\documentclass[10pt]{article}

\usepackage{amsmath,amssymb}
\usepackage{mathtools}
\usepackage{geometry} 
\usepackage{authblk} 

\usepackage[amsmath,amsthm,thmmarks]{ntheorem} 
\usepackage[nottoc]{tocbibind} 
\usepackage{appendix}

\usepackage{stackrel} 
\usepackage{leftidx} 
\usepackage{scalerel} 

\usepackage{amsfonts}
\usepackage{euscript}
\usepackage{pxfonts}
\usepackage{bbold}

\usepackage[shortlabels]{enumitem}
\usepackage{hyperref}
\usepackage{cleveref}
\usepackage[normalem]{ulem}

\usepackage[dvipsnames]{xcolor}
\usepackage{tcolorbox}
\usepackage{colortbl}

\usepackage{graphicx}
\usepackage{subfigure}
\usepackage{adjustbox}
\usepackage{float} 

\usepackage[all,2cell,arrow,matrix]{xy} \UseAllTwocells \SilentMatrices
\usepackage{tikz-cd}
\usepackage{tikz}
\usetikzlibrary{arrows}
\usetikzlibrary{decorations.pathreplacing,decorations.markings,decorations.pathmorphing}
\usepackage{commutative-diagrams} 

\usepackage{multirow}
\usepackage{array}
\usepackage{makecell} 

\usepackage{preamble/rongge_macro}

\begin{document}

\title{2-Morita Equivalent Condensable Algebras and Domain Walls in 2+1D Topological Orders}
\author[a,b,c]{Rongge Xu \thanks{Email: \href{mailto:xurongge@westlake.edu.cn}{\tt xurongge@westlake.edu.cn}}}
\author[d]{Holiverse Yang \thanks{Email: \href{mailto:holiversey@gmail.com}{\tt holiversey@gmail.com}}}
\affil[a]{Department of Physics, Fudan University, Shanghai, 200433, China}
\affil[b]{School of Science, Westlake University, Hangzhou, 310030, China}
\affil[c]{Institute of Natural Sciences, Westlake Institute of Advanced Study, Hangzhou 310024, China}
\affil[d]{Department of Physics, The Chinese University of Hong Kong,\authorcr Shatin, New Territories, Hong Kong, China}
\date{\vspace{-5ex}}

\maketitle

\begin{abstract}
    We classify \(E_2\) condensable algebras in a modular tensor category \(\EuScript{C}\) up to 2-Morita equivalence. 
From a physical perspective, this is equivalent to providing a criterion for when different \(E_2\) condensable algebras result in the same condensed topological phase in a 2d anyon condensation process. 
By considering the left and right centers of \(E_1\) condensable algebras in \(\EuScript{C}\), we exhaust all 2-Morita equivalent \(E_2\) condensable algebras in \(\EuScript{C}\) and provide a method to recover \(E_1\) condensable algebras from 2-Morita equivalent \(E_2\) condensable algebras. 
We also prove that intersecting Lagrangian algebras in \(\EuScript{C} \boxtimes \overline{\EuScript{C}}\) with its left and right components generates all 2-Morita equivalent \(E_2\) condensable algebras in \(\EuScript{C}\). 
This paper establishes a complete interplay between \(E_1\) condensable algebras in \(\EuScript{C}\), 2-Morita equivalent \(E_2\) condensable algebras in \(\EuScript{C}\), and Lagrangian algebras in \(\EuScript{C} \boxtimes \overline{\EuScript{C}}\). 

The relations between different condensable algebras can be translated into their module categories, which correspond to domain walls in topological orders. 
We introduce a two-step condensation process and study the fusion of domain walls. We also show that an automorphism of an \(E_2\) condensable algebra may lead to a nontrivial braided autoequivalence in the condensed phase. 
As concrete examples, we interpret the categories of quantum doubles of finite groups. We also discuss examples beyond group symmetries. 
Moreover, our results can be generalized to Witt-equivalent modular tensor categories.

\end{abstract}

\tableofcontents

\section{Introduction} \label{sec:introduction}

Classical Morita theory \cite{Morita58} provides a powerful tool in many aspects of mathematics especially in representation theory \cite{Ost03}. 
Recent years, people find it also important in studying quantum many-body systems and quantum field theories (such as open-closed conformal field theory \cite{KR08,KR09} and topological quantum field theory \cite{FSV13,CRS18,CRS18defect,Turzillo20}).
Many significant results including boundary-bulk relation and anyon condensation theory in the study of topological orders \cite{KK12,Kon14,KWZ15,KZZZ24}, have been developed using categorical Morita theory.
However, most of these results use only 1-Morita theory, which is applicable in systems associated with $E_1$ algebras.
The study of higher Morita theory of $E_n$ algebra \cite{Hau17} in higher dimensional physics and mathematics is still in its infancy.
With the rapid development of topological orders and categorical tools in physics, a well studied higher Morita theory is becoming more and more in demand when we encounter higher dimensions.
In the last few years, 2-Morita equivalence of braided fusion categories has been studied in the context of fusion 2-categories \cite{BJS21,BJSS21,Dec22center}.
It is natural to consider the 2-Morita equivalence of $E_2$ algebras in the context of fusion 1-categories.
As front-runners, we classify 2-Morita equivalent class of $E_2$ condensable (connected commutative separable) algebras in 2d\footnote{We use $nd$ to represent $n$ spatial dimension and $(n+1)D$ to represent $n+1$ spacetime dimension.} topological orders.

From physical perspective, it has been well-known that a 2d topological order can be described by a (unitary)\footnote{We do not consider the unitary structure in this paper.} modular tensor category (MTC) \cite{Kit06}, usually denoted by $\CC$.
And an anyon condensation process, which is a selecting of energy-favorable subspaces of the original Hilbert space, may happen in $\CC$. An $E_2$ (or 2d) condensable algebra\footnote{Through this paper, we use the terminology "$E_2$ ($E_1$) condensable algebra" in mathematical context and use the terminology "2d (1d) condensable algebra" in physical context.} $A$ in $\CC$ is viewed as the new vacuum in the energy-favorable subspace. 
This new subspace is also a topological order, which can be described by an MTC $\CC^{loc}_A$, the category of local $A$-modules in $\CC$ \cite{Kon14}.
We will explain more details in \hyperref[section:preliminary]{preliminary}.

However, characterization of when two 2d condensable algebras, say $A_1$ and $A_2$ in $\CC$, produce the same condensed phase $\CC^{loc}_{A_1} \simeq \CC^{loc}_{A_2}$, remains incomplete.
And this is indeed the same question of classifying the 2-Morita equivalent $E_2$ condensable algebras in an MTC, namely, $\CC^{loc}_{A_1} \simeq \CC^{loc}_{A_2} \Leftrightarrow A_1 \widesim[3]{2-Morita}{} A_2$.

On the other hand, classifications of $E_1$ condensable (indecomposable separable) algebra have been developed using algebra centers.
Two algebras $A$, $B$ in a monoidal category $\CC$ are said to be 1-Morita equivalent if their categories of (right) modules are equivalent as module categories over $\CC$.
It is known that by computing full centers of $E_1$ condensable algebras, we can classify $E_1$ condensable algebras up to 1-Morita equivalence \cite{KR08}.


Motivated by algebraic centers appearing in 1-Morita theory, we study $E_2$ condensable algebras in a MTC $\CC$ from the perspective of higher Morita theory.
According to \cite{FFRS06}, given an $E_1$ condensable algebra $B$ in $\CC$, there is an equivalence $\CC_{Z_l(B)}^{loc}\simeq \CC_{Z_r(B)}^{loc}$ of MTCs, where $Z_l(B)$ is the left center of $B$ and $Z_r(B)$ is the right center of $B$.
This result provides a method to generate some $E_2$-Morita equivalent condensable algebras.
In this paper, we further prove that any pair of $E_2$-Morita equivalent condensable algebras $(A_1,A_2)$ in $\Alg_{E_2}^{cond}(\CC)$, there exists an $E_1$ condensable algebra $B$ such that $A_1 \simeq Z_l(B)\widesim[3]{2-Morita}{} Z_r(B)\simeq A_2 $.
In this way we can classify all 2-Morita equivalent condensable algebras $\{A_i\} \in \CC$ that result in the same topological phases after anyon condensation.


To be more specific, if we 
consider a trivial 1d domain wall (i.e. a 1d subregion) of a 2d topological order described by MTC $\CC$, in which this 1d domain wall is still described by $\CC$ (viewing as a fusion category\footnote{The fusion categories appearing in this article are assumed to have spherical structures.}). 
Then all gapped domain walls within MTC $\CC$ can be described by categories ${}_{B_i}{\CC}_{B_i}$ of bimodule over 1d condensable algebras $\{B_i\}$\footnote{Strictly speaking, the gapped domain walls are classified by 1-Morita classes of 1d condensable algebras $B_i$. However, since only the 1-Morita class of $B_i$ is physically detectable, we would abuse 'a 1d condensable algebra $B$' as $B$'s 1-Morita class unless emphasized otherwise.} in fusion category $\CC$ \cite{Kon14}, see the following figure.

\begin{figure}[H]
    \centering
    \begin{tikzpicture}[scale=0.7]
        \scriptsize
        \filldraw[fill=gray!40, draw=none] (-5,0) rectangle (0,5);
        \filldraw[fill=gray!40, draw=none] (0,0) rectangle (5,5);
        \draw[very thick] (0,0) -- (0,4);
        \draw[very thick, dashed] (0,4) -- (0,5);
        \draw[very thick] (-0.1,1) -- (0.1,1);
        \draw[very thick] (-0.1,2) -- (0.1,2);
        \draw[very thick] (-0.1,3) -- (0.1,3);
        \draw[very thick] (-0.1,4) -- (0.1,4);

        \node at(-2.5,2.5){\large $\CC$};
        \node at(3.5,2.5){\large $\CC$};
        \node at(0.5,0.5){$\cdots$};
        \node at(0.8,1.5){${}_{B_4}\CC_{B_4}$};
        \node at(0.8,2.5){${}_{B_3}\CC_{B_3}$};
        \node at(0.8,3.5){${}_{B_2}\CC_{B_2}$};
        \node at(0.4,4.5){$\CC$};
    \end{tikzpicture}
    \caption{Gapped domain walls within a topological order $\CC$ can be classified by the category of bimodules ${}_{B_i}{\CC}_{B_i}$ of 1d condensable algebras $\{B_i\}$ in the trivial domain wall $\CC$ (drawn in the dashed line). In particular, ${}_{B_1}{\CC}_{B_1} \simeq \CC$ for $B_1$ being the tensor unit $\one$ of the fusion category $\CC$.}
\end{figure}
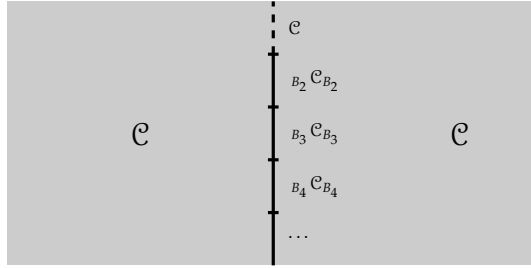

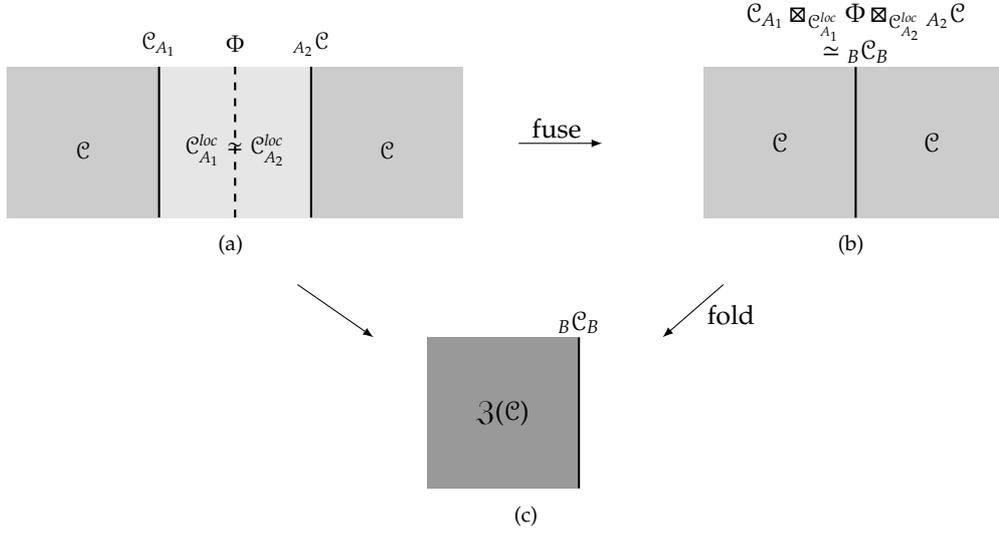
\begin{figure}[H]
    \centering    
    \subfigure[]{
        \begin{minipage}[t]{0.45\linewidth}
        \centering
        \begin{tikzpicture}
            \centering
            \filldraw[fill=gray!40, draw=none] (-3,-1) rectangle (-1,1);
            \filldraw[fill=gray!20, draw=none] (-1,-1) rectangle (1,1);
            \filldraw[fill=gray!40, draw=none] (1,-1) rectangle (3,1);
            \draw[thick](-1,-1)--(-1,1);
            \draw[thick](1,-1)--(1,1);
            \draw[dashed,thick](0,-1)--(0,1);
            \node at(-2,-0.1){\small $\CC$};
            \node at(2,-0.1){\small $\CC$};
            \node at(0,-0.1){\small $\CC^{loc}_{A_1}\simeq \CC^{loc}_{A_2}$};
            \node at(-1,1.3){\small $\CC_{A_1}$};
            \node at(1,1.3){\small ${}_{A_2} \CC$};
            \node at(0,1.3){\small $\Phi$};
        \end{tikzpicture}
        \end{minipage}
    }
    \begin{tikzpicture}
        \node at(0,1.2){fuse};
        \draw[-latex] (-0.5,1) -- (0.6,1);
        \draw[draw=none] (-0.5,0) -- (0.5,0.5);
    \end{tikzpicture}
    \subfigure[]{
        \begin{minipage}[t]{0.4\linewidth}
        \centering
        \begin{tikzpicture}
            \filldraw[fill=gray!40,draw=none] (-2,0) rectangle(2,2);
            \draw[thick] (0,0) -- (0,2);
            \node at(0,2.2){$\simeq {}_{B}\CC_{B}$};
            \node at(0,2.6){$\CC_{A_1}\btd_{\CC_{A_1}^{loc}}\Phi \btd_{\CC_{A_2}^{loc}}{}_{A_2} \CC$};
            \node at(-1,1){$\CC$};
            \node at(1,1){$\CC$};
        \end{tikzpicture}
        \end{minipage}
    }
    \\
    \subfigure[]{
        \begin{minipage}[t]{0.45\linewidth}
        \centering
        \begin{tikzpicture}
            \filldraw[fill=gray!80,draw=none] (-1.5,0.3) rectangle(0.5,2.3);
            \draw[thick] (0.5,0.3) -- (0.5,2.3);
            \draw[-latex] (-3.2,3) -- (-2.2,2.3);
            \draw[-latex] (2.4,3) -- (1.6,2.3);
            \draw[draw=none] (2,1.3) -- (2.5,1.3);
            \node at(2.5,2.6){fold};
            \node at(0.5,2.5){$ {}_{B}\CC_{B}$};
            \node at(-0.5,1.3){$\FZ(\CC)$};
        \end{tikzpicture}
        \end{minipage}
    }
    \caption{The fusion process from (a) to (b) shows that for a gapped domain wall ${}_B{\CC}_B \simeq \CC_{A_1}\btd_{\CC_{A_1}^{loc}}\Phi \btd_{\CC_{A_2}^{loc}}{}_{A_2} \CC$ in $\CC$, we have equivalent condensed phases $\CC_{A_1}^{loc} \simeq \CC_{A_2}^{loc}$ 'hidden inside' this wall ${}_B{\CC}_B$. This process gives an intuitive way to understand why $B$ can recover 2-Morita equivalent 2d condensable algebras $A_1$ and $A_2$ in $\CC$.
    Moreover, using folding trick from (b) to (c), the correspondence between 1d condensable algebras in $\CC$ and lagrangian algebras in $\fZ(\CC)$ can be characterized by the correspondence between domain walls in $\CC$ and gapped boundaries of $\fZ(\CC)$.
    } \label{general_picture}
\end{figure}

On the other hand, consider condensing two 2-Morita equivalent $E_2$ condensable algebras $A_1 \widesim[3]{2-Morita}{\phi} A_2 \in \CC$ where the equivalence is given by $\phi:\CC_{A_1}^{loc} \xrightarrow{\sim} \CC_{A_2}^{loc}$. 
Then two 1d domain walls $\CC_{A_1}$ and ${}_{A_2} \CC$ together with an invertible domain wall $\Phi$ induced by $\phi$ are generated through the condensation process, see figure \ref{general_picture} (a).
Now we 
fuse $\CC_{A_1}$ and ${}_{A_2} \CC$ through interlayer phase, 
we obtain $\CC_{A_1}\btd_{\CC_{A_1}^{loc}}\Phi \btd_{\CC_{A_2}^{loc}}{}_{A_2} \CC$ (fig. \ref{general_picture} (b)).
Since topological order only reveals observables at fixed point, after rescaling the system up to a proper correlation length,
$\CC_{A_1}\btd_{\CC_{A_1}^{loc}}\Phi \btd_{\CC_{A_2}^{loc}}{}_{A_2} \CC$ can be viewed as a 1d domain wall between $\CC$ and itself. 
Therefore, $\CC_{A_1}\btd_{\CC_{A_1}^{loc}}\Phi \btd_{\CC_{A_2}^{loc}}{}_{A_2} \CC$ should be monoidal equivalent to ${}_{B_i}{\CC}_{B_i}$ for some 1d condensable algebra $B_i$ in $\CC$.
This process is also called dimensional reduction \cite{KWZ15,AKZ17}.

Motivated by the fusion process from figure \ref{general_picture} (a) to figure \ref{general_picture} (b), we show in the main body that there exists a specific 1d condensable algebra $B$ such that $Z_l(B)\cong A_1$ and $Z_r(B)\cong A_2$.
And this procedure exhausts all 2-Morita equivalent classes in $\CC$. 
A detailed proof is given in section \ref{section:proof}.

Moreover, we can fold $\CC$ through a gapped domain wall ${}_B \CC_B$ (arrow from fig. \ref{general_picture} (b) to fig. \ref{general_picture} (c)). After folding, we get a new topological order described by the Drinfeld center of $\FZ(\CC) \simeq \CC \boxtimes \bar{\CC}$  with ${}_B \CC_B$ becomes the gapped boundary.
It is known that the gapped boundaries of $\FZ(\CC)$ are classified by lagrangian algebras (a specific 2d condensable $A^L$ in which $\CC^{loc}_{A^L} \simeq \vect$) in $\FZ(\CC)$ \cite{DMNO13,Kon14}.
By this folding trick, gapped domain walls in $\CC$ are one-to-one corresponding to gapped boundaries of $\fZ(\CC)$.
So there is also a bijection between the set of 1-Morita classes of 1d condensable algebras in $\CC$ and the set of isomorphic classes of lagrangian algebras in $\fZ(\CC)$ \cite{KR08}.
Indeed, taking the full center $Z(B)$ of a 1d condensable algebra $B$ will produce a lagrangian algebra in $\fZ(\CC)$ \cite{KR09}.
As a consequence, the 2-Morita equivalent condensable algebras in $\CC$ can also be classified by lagrangian algebras in $\FZ(\CC)$.

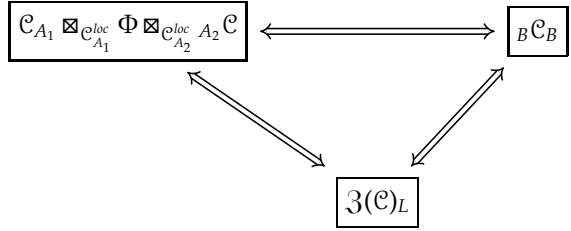
\begin{figure}[H]
    \centering
    \begin{tikzcd}[arrows=Leftrightarrow]
        \fbox{\begin{tabular}{@{}c@{}c@{}}$\CC_{A_1}\btd_{\CC_{A_1}^{loc}}\Phi \btd_{\CC_{A_2}^{loc}}{}_{A_2} \CC$
        \end{tabular}}\ar[ddr] && \fbox{\begin{tabular}{@{}c@{}c@{}}${}_B \CC_B$ \end{tabular}} \ar[ll] \ar[ddl]\\
        &&\\
        &\fbox{\begin{tabular}{@{}c@{}}$\FZ(\CC)_L$\end{tabular}}&
    \end{tikzcd}
    \caption{We have proven the bijections between these three kind of domain wall topological orders.}\label{fig:walls_cycle}
\end{figure}

The arrow from figure \ref{general_picture} (a) to figure \ref{general_picture} (c) is not obvious. By a method called the {\it 2-step condensation}, we prove that the domain walls $\CC_{A_1}$ and ${}_{A_2} \CC$ together produce a boundary of $\fZ(\CC)$ (see section \ref{section:proof} for details). 
Indeed, these three arrows are all invertible. 
Namely, we can unfold a stable gapped boundary of $\fZ(\CC) \simeq \CC\boxtimes \overline{\CC}$ to be a gapped domain wall ${}_B{\CC}_B$ in $\CC$ (\ref{general_picture} (c) to \ref{general_picture} (b)), and any stable gapped domain wall ${}_B{\CC}_B$ can be opened to contain a condensed interlayer MTC (\ref{general_picture} (b) to \ref{general_picture} (a)). We illustrate all of them in our main body, and eventually get figure \ref{fig:walls_cycle}. 

\begin{rem}
    Using language of higher condensation \cite{GJF19,KZZZ24}, we show that condensable 2-codimensional defects (i.e. particle-like defects) can be used to classify 1-codimensional defects (i.e. 1d gapped domain walls) in a 2+1D topological order.
\end{rem}

Above physical pictures depict 
1d gapped domain wall (1-codimensional defect) classifications in $\CC$ through three different perspectives.
The correspondence between these modules categories can reflect the correspondence in algebraic levels, which lead us to prove  classification theorems of 2-Morita equivalent condensable algebras:

\begin{thm} \label{main_theorem}
    Given a modular tensor category $\CC$ with tensor unit $\one$ and consider all indecomposable separable algebras $\{B_i\}$ in $\CC$, $L_i\simeq Z(B_i)$, $:A_{l_i} := Z_l(B_i)$ and $A_{r_i} :=Z_r(B_i)$ represent the full, left, right centers respectively, 
    \begin{itemize}
        \item all the pairs of 2-Morita equivalent condensable algebras in $\CC$ can be obtained by taking left and right centers of $B_i$, 
        in which the resulted category of local modules $\CC_{Z_l(B)}^{loc}$ and $\CC_{Z_r(B)}^{loc}$ are equivalent as modular tensor categories.  

        \item Or, all the pairs of 2-Morita equivalent condensable algebras in $\CC$ can be obtained by lagrangian algebras $L_i$'s in $\CC\boxtimes \overline{\CC}$, precisely speaking, $L_i\cap(\CC\boxtimes \one)=:A_{l_i}\widesim[3]{2-Morita}{} A_{r_i}:= L_i\cap (\one \boxtimes \overline{\CC})$.
        
    \end{itemize}
\end{thm}


        

By proving the above theorem, we show the power of algebraic centers in classifying condensed phases in a MTC, and the physical correspondence of left/right/full center in MTC appears simultaneously.
In addition to this proof, our paper indeed give more fruitful results --- A complete cycle of above condensable algebras! In which we summarize their relations in the following Trinity:

\begin{figure}[H]
    \centering
    \begin{tikzcd}[arrows=Rightarrow]
        \fbox{\begin{tabular}{@{}c@{}c@{}}2-Morita equivalent\\ condensable algebras in $\CC$\\
        \end{tabular}} \ar[rr, shift left=-0.5ex, "\text{6. Extended tensor}"'] \ar[ddr,"\text{1. Extension}" ] && \fbox{\begin{tabular}{@{}c@{}c@{}}1-Morita class of 1d \\
            condensable algebras in $\CC$\end{tabular}} \ar[ll, shift left=-0.5ex,"\text{3. Left and right center}"'] \ar[ddl,"\text{4. Full center}"' ]\\
        &&\\
        &\fbox{\begin{tabular}{@{}c@{}}Lagrangian algebras\\ in $\fZ(\CC)\simeq \CC\boxtimes \overline{\CC}$\end{tabular}} \ar[uul, shift left=1.5ex, start anchor={[yshift=0.8ex,xshift=-1ex]},"\text{2. $\cap$ with components}"] \ar[uur, shift left=-1.5ex, start anchor={[yshift=0.9ex,xshift=0.95ex]},end anchor={[yshift=0.8ex,xshift=1ex]}, "\text{5. Forget}"']&
    \end{tikzcd}
    \caption{Results of this paper can be summarized by this Trinity, all arrows appear here are reversible. 
    \Ar{1} was first discussed in \cite{DNO12}, an alternative proof using 2-step condensation is provided in section \ref{sec:pf_lag_alg}; 
    \Ar{2} was first stated by Davydov \cite[Theorem 2.5.1]{Dav10a}, and we prove it using results in \cite{DNO12};
    \Ar{3} is proved in \cite{FFRS06}; 
    \Ar{4} is proved by Kong and Runkel in \cite{KR09}; 
    \Ar{5} has long been a folklore without enough discussions, we reformulate this 'forget' process in section \ref{sec:centers};
    \Ar{6} is first discussed in this work.}\label{fig:alg_cycle}
\end{figure}
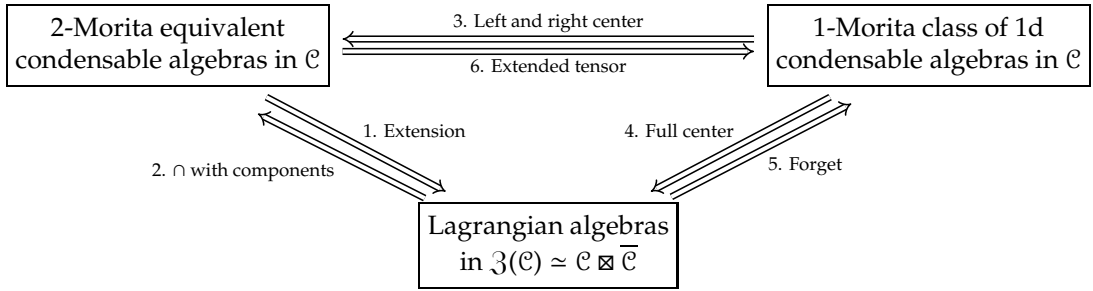

Figure \ref{general_picture} corresponds to the inner commutative part of this Trinity.
In order to get the whole cycle works,
an important ingredient one should consider is symmetry $\phi$ (braided autoequivalence) appearing in the condensed phase $\CC_A^{loc}$,  we discuss this in section \ref{sec:any_cond} and section \ref{sec:sym}. 
With this ubiquitous symmetry revealing, we can give a complete relationship between lagrangian algebras in $\CC$'s Drinfeld center $\FZ(\CC)$, 1-Morita class of 1d condensable algebras in $\CC$ and 2-Morita equivalent condensable algebras in $\CC$, the interplay of these three 
also motive us to introduce a new concept called \emph{extended tensor}, 
that helps us to recover 1d 1-Morita equivalent condensable algebras explicitly through the 2-Morita equivalent 2d condensable algebras.
These results can be packed into six Arrows drawn above.
We use all these six Arrows frequently throughout our paper to show the universality of this Trinity.
 

Our analysis is universal and model-independent, which can be applied to many physical systems.
We compute some examples in section \ref{section:lattice_model}, some of them are the case of group-theoretical categories $\FZ(\rep(G))$ including $\bZ_2$, $\bZ_4$ and $S_3$ gauge symmetries. Our method meets with the known classification result of condensed phases in $\FZ(\rep(G))$ \cite{Dav10a, DS17}. 
We also develop a method to realize 1d condensation on toric code model, and an interpretation of left/right center on lattice, which can be generalized easily.

Actually, Figure \ref{general_picture} can be transported to a more general environment, in which $\CC_1$ and $\CC_2$ are two different MTCs that are connected by another MTC $\CD$ in between (Or to say, we can treat the case for $\CC_1$ and $\CC_2$ are Witt equivalent). 

Here we explain the layout of this paper.
In the next section, we show how these categorical and algebraic structures emerge from natural physical requirements and explain each of the Arrows in Trinity \ref{fig:alg_cycle} explicitly:
\Ar{1} corresponds to Lemma \ref{lem:1d_cond_alg_lag_alg}; \Ar{2} corresponds to Corollary \ref{crl:lag_alg_to_2-Morita_algs}; \Ar{3} corresponds to Theorem \ref{thm:1d_cond_alg_to_2_Morita_cond_alg}; \Ar{4} and \Ar{5} correspond to Lemma \ref{lem:1d_cond_alg_lag_alg}; \Ar{6} corresponds to Algorithm \ref{cnj:Arrow6}.
In section \ref{section:proof}, we carry out our proof of Theorem \ref{main_theorem} and discuss how the automorphisms of condensable algebras affect the condensation process, this section might be more suitable for reader with mathematical backgrounds. 
In section \ref{section:lattice_model}, we treat the case of group-theoretical categories, in which we classify 2d condensable algebra up to 2-Morita equivalence for any finite group $G$ and realize 1d condensable algebras and left/right center on toric code model.
We also discuss examples beyond common group symmetries.
Discussions on Witt equivalences and other generalizations are performed in section \ref{section:outlook}.
Mathematical background including higher Morita theory, center of algebras, and condensable algebra classifications are given in the \hyperref[appendix:morita]{appendices}.

\bigskip
\noindent \textbf{Acknowledgements}: 
We would like to thank Liang Kong for inspiring us to dive into this interesting topic. We thank Zhi-Hao Zhang for helpful discussions and Hao Xu for comments on references. HY thanks useful discussions with Tian Lan and Gen Yue. RX thanks Jian Li for funding and travel supporting, and acknowledges Shenzhen Institute of Quantum Science and Engineering for hospitality during the visits. 
HY is supported by Research Grants Council (RGC), University Grants Committee (UGC) of Hong Kong (ECS No. 24304722).

\section{Main Story} \label{section:preliminary}
In this section, we go through the preliminaries and main story of classification of 2-Morita equivalent condensable algebras, in which we give our main results by string all six Arrows of the Trinity \ref{fig:alg_cycle} clearly.
For sake of conciseness, we leave detailed proof in section \ref{section:proof}. 

\subsection{Anyon condensations and 2-Morita equivalence}\label{sec:any_cond}
A 2d anomaly-free stable\footnote{In topological orders, a phase is called {\it anomaly-free} if it does not admit a non-trivial higher dimensional bulk.
And a phase is {\it stable} means the macroscopic observables in it are invariant under small perturbations, 
stable corresponds to indecomposable in the categorical language.
In this paper, 2d topological orders that we discuss are assumed to be anomaly-free and stable.} topological order (a system of anyons) can be described by a (unitary) modular tensor category (MTC) $\CC$ with a central charge $c\in\bC$ \cite{Kit06,KZ22a}.
Two adjacent topological orders are separated by a domain wall, which is mathematically described by a (spherical) fusion category.

\begin{figure}[H]
    \centering
        \begin{tikzpicture}
            \filldraw[fill=gray!40,draw=white] (-2,0) rectangle(2,2);
            \draw[thick] (0,0) -- (0,2);

            \node at (0,-0.3) {$\CM$};
            \node at (0,2.3) {$\CN$};
            \node at (0.2,1) {$\CX$};
            \node at(-1,1){$\CC_1$};
            \node at(1,1){$\CC_2$};
            \filldraw[fill=white] (-0.05, 0.95)rectangle(0.05,1.05);
        \end{tikzpicture}
    \caption{$\CX$ is the category of 0d domain wall conditions, which is a $\CM$-$\CN$-bimodule category. And a 1d gapped domain wall $\CM$ (or $\CN$) is described by a closed monoidal $\CC_1$-$\CC_2$-bimodule.
    } \label{fig:modules}
\end{figure}
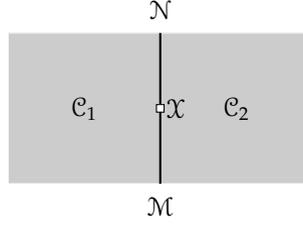

Intuitively, a 0d gapped domain wall (2-codimensional defect) between two 1d gapped phases described by fusion categories $\CM$ and $\CN$ consists of wall conditions and form a category $\CX$ (figure \ref{fig:modules}).
The category $\CX$ of wall conditions naturally has an $\CM$ and $\CN$ action.
This action makes $\CX$ also a $\CM$-$\CN$-bimodule category.
Similarly, the 1d gapped domain wall (1-codimensional defect) $\CM$ (or $\CN$) between 2d topological orders $\CC_1$ and $\CC_2$ should be described by a $\CC_1$-$\CC_2$ bimodule category.
However, since the 2d phases $\CC_1$ and $\CC_2$ have braiding structures and $\CM$ also admits a fusion structure on itself, these structures should be compatible under the action of 2d bulk on 1d domain wall.
So a 1d gapped domain wall is indeed described by a {\bf closed monoidal} $\CC_1$-$\CC_2$-bimodule, i.e. $\fZ(\CM)\simeq \CC_1\btd \overline{\CC_2}$ \cite{KYZ21}.
Here $\fZ(\CM)$ is the {\bf Drinfeld center} of $\CM$, which gives the 2d (folded) bulk $\CC_1\btd \overline{\CC_2}$ of the 1d phase $\CM$ \cite{EGNO15,KWZ15}.
See appendix \ref{appendix:module_cat} for definitions of monoidal modules.

One source of gapped domain walls comes from anyon condensation \cite{Kon14}.
The original phase, a topological order described by a MTC $\CC$, is sitting on one side of the wall, while a 2d condensable algebra (or $E_2$ condensable algebra in mathematical term) $A \in \CC$ is condensed on the other side \cite{BS09,BSH09,Kon14}.
Then a new topological order, described by the category $\CC_A^{loc}$ of local $A$-modules in $\CC$, is obtained, in which the condensed algebra plays the role of the vacuum in the condensed phase $\CC_A^{loc}$.
For precise mathematical definitions of a 2d condensable algebra and local modules, see Appendix \ref{appendix:condensable_algebras}.
Domain wall excitations consist of anyons in $\CC$ that are confined from going to the condensed phase $\CC_A^{loc}$ together with anyons that move transparently. 
The wall excitations together with their fusion structures form a fusion category $\CC_A$, the category of right $A$-modules in $\CC$ \cite{Kon14, DMNO13}.

\begin{figure}[H]
    \centering
    \begin{tikzpicture}[scale=0.7]
        \scriptsize
        \filldraw[fill=gray!40, draw=none] (-5,0) rectangle (0,6);
        \filldraw[fill=gray!20, draw=none] (0,0) rectangle (5,6);
        \draw[very thick] (0,0) -- (0,6);
        \draw[very thick] (-0.1,1.25) -- (0.1,1.25);
        \draw[very thick] (-0.1,2.5) -- (0.1,2.5);
        \draw[very thick] (-0.1,3.75) -- (0.1,3.75);

        \node at(-2.5,3.1){\large $\CC$};
        \node at(3.2,3.1){\large $\CC^{loc}_{A}$};
        \node at(1.2,0.5){$\cdots \widesim[2]{1-Morita}{} \CC_{A}$};
        \node at(0.9,1.75){${}_{B_2}(\CC_A)_{B_2}$};
        \node at(0.9,3){${}_{B_1}(\CC_A)_{B_1}$};
        \node at(0.5,4.6){\large $\CC_{A}$};
        \draw [>=latex][<->, thick] (-0.5,4) -- (-0.5,1);
        \draw[-latex, thick] (-2,5.2) -- (2,5.2);
        \node[rotate=90] at(-0.8,2.5){1d condensation};
        \node[] at(0,5.4){2d condensation};
    \end{tikzpicture}
    \caption{This figure shows the directions of 1d and 2d anyon condensations. For any $B_i$, ${}_{B_i}(\CC_A)_{B_i}$ is 1-Morita equivalent as fusion categories to $\CC_A$ since the Drinfeld center of ${}_{B_i}(\CC_A)_{B_i}$ is equivalent to the Drinfeld center of $\CC_A$, i.e. $ \fZ(\CC_A) \simeq \fZ({}_{B_i}(\CC_A)_{B_i}) \simeq \CC \btd \overline{\CC^{loc}_A} $  \cite{Sch01}.} 
    \label{fig:1d_cond}
\end{figure}
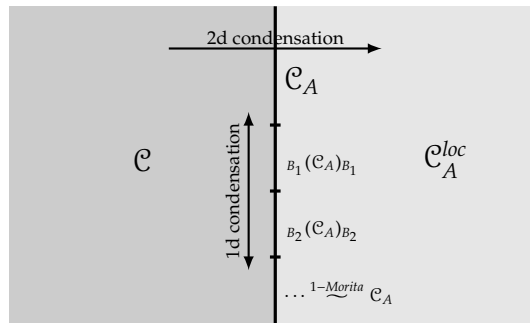

Despite the domain wall generated by anyon condensation, there exist other types of domain walls between $\CC$ and $\CC_A^{loc}$, which are described by the fusion category ${}_B(\CC_A)_{B}$, the category of $B$-$B$-bimodules in $\CC_A$.
The anyons in original phase and condensed phase have action on wall excitation which gives ${}_B(\CC_A)_{B}$ the monoidal $\CC$-$\CC_A^{loc}$-bimodule structure.
The vacuum on the wall ${}_B(\CC_A)_B$ is given by an 1d condensable algebra $B$ in $\CC_A$, $B$'s properties are similar to that of $A$ except that it is not necessarily commutative.  
It defines a new type of condensation but confined to the $1d$ domain wall between $\CC$ and $\CC^{loc}_A$, which is called the $1d$ {\it condensation}. $\{{}_{B_i}(\CC_A)_{B_i}\mid B_i\in\Alg_{E_1}^{cond}(\CC_A)\}$ exhausts all stable gapped domain walls between $\CC$ and $\CC_A^{loc}$ \cite{Kon14}. Different from 2d condensation, 1d condensation is invertible\footnote{The reason for 2d condensation is not invertible is that the $E_2$ condensable algebra is a codimension 2 defect. Actually, codimension 1 condensations are all reversible \cite{KZZZ24}.}, namely one can also find a 1d condensable algebra $B'$ in ${}_B(\CC_A)_B$ such that ${}_{B'}({}_B(\CC_A)_B)_{B'}\simeq \CC_A$ (\mynote{Does $B'$ just $A \in \CC$ when putting into ${}_B(\CC_A)_B$?}).
Some 1d condensed phases ${}_{B_i}(\CC_A)_{B_i}$ can be written as $\CC_{A_i}$ for some 2d condensable algebra $A_i$, in which the condensed phase $\CC_{A_i}^{loc}$ via $A_i$ is equivalent to $\CC_A^{loc}$.
See figure \ref{fig:1d_cond} for an illustration of 1d and 2d condensation.

Using the folding trick, we can fold the 2d phase $\CC_A^{loc}$ to another side through the 1d gapped domain wall, and obtain a time reversal phase $\overline{\CC_A^{loc}}$ whose braiding is defined by anti-braiding in $\CC_A^{loc}$. 
The folded 2d phase $\CC \btd \overline{\CC^{loc}_A}$ can be regarded as a blank stacking of phases $\CC$ and $\overline{\CC_A^{loc}}$ with boundaries $\{{}_{B_i}(\CC_A)_{B_i}\mid B_i\in\Alg^{cond}_{E_1}(\CC_A)\}$.
By boundary-bulk relation \cite{DMNO13,KWZ15}, we have $\fZ({}_{B}(\CC_A)_{B})\simeq \fZ(\CC_A) \simeq \CC \btd \overline{\CC^{loc}_A} $ as MTCs. Two fusion categories $\CM$ and $\CN$ are {\bf 1-Morita equivalent} if and only if their Drinfeld centers $\FZ(\CM) \simeq \FZ(\CN)$ as braided fusion categories \cite{ENO11}.  
That is to say, for any $1d$ condensable algebra $B$ in $\CC_A$, ${}_{B}(\CC_A)_{B}$ is 1-Morita equivalent to $\CC_A$ \cite{Sch01}.

A natural question in anyon condensation theory arises, given two 2d condensable algebras $A_1$ and $A_2$ in a 2d topological order $\CC$, how to tell if they condense to a same phase or not? i.e. Which pair of $A_1$ and $A_2$ leads to $\CC_{A_1}^{loc}\simeq \CC_{A_2}^{loc}$?
To answer this question, we need to briefly review the definition of 1-Morita equivalent algebras, and introduce the notion of 2-Morita equivalence of $E_2$-algebras.

\begin{defn}[\cite{Morita58}]
    Let $\CC$ be a monoidal category.
    Two $E_1$-algebras $B_1$, $B_2\in\CC$ are $1$-{\bf Morita equivalent}\footnote{Note the difference between algebraic Morita equivalent and categorical Morita equivalent, which is explained in appendix \ref{morita_equvi}.}, denoted by $B_1\widesim[3]{1-Morita}{} B_2$, if there is an equivalence of categories $\CC_{B_1}\simeq \CC_{B_2}$.
\end{defn}

Based on above definition, we define 2-Morita equivalence of $E_2$-algebras iteratively.
\begin{defn}\label{defn:2-Morita}
    Let $\CC$ be a braided fusion category.
    Two $E_2$-algebras $A_1$, $A_2\in\CC$ are {\bf 2-Morita equivalent}, denoted by $A_1\widesim[3]{2-Morita}{} A_2$, if $\CC_{A_1}\widesim[3]{1-Morita}{} \CC_{A_2}$. 
\end{defn}

By \cite{ENO11}, $\CC_{A_1}\widesim[3]{1-Morita}{} \CC_{A_2}$ if and only if $\fZ(\CC_{A_1})\simeq \fZ(\CC_{A_2})$. 
Since $\fZ(\CC_A)\simeq \CC\btd \overline{\CC_A^{loc}}$ for a MTC $\CC$ \cite{DMNO13}, then $\fZ(\CC_{A_1})\simeq \fZ(\CC_{A_2})$ implies $\CC_{A_1}^{loc}\simeq \CC_{A_2}^{loc}$ and vice versa.
Therefore, $A_1\widesim[3]{2-Morita}{} A_2$ is equivalent to $\CC_{A_1}^{loc}\simeq \CC_{A_2}^{loc}$\footnote{Since local modules are $E_2$ modules over $E_2$ algebras, it is natural to characterize 2-Morita equivalence of $E_2$-algebras by $E_2$-monoidal equivalence between their $E_2$-module categories.}. Hence, we can translate the question of classifying equivalent condensed phases to the question of classifying 2-Morita equivalent algebras. 
Equivalent definitions of 2-Morita equivalence can be summarized as follows:
\begin{align*}
    A_1 \widesim[3]{2-Morita}{} A_2 \, \, \, \Leftrightarrow\,\,\,  \CC_{A_1} \widesim[3]{1-Morita}{}\CC_{A_2}\,\,\, \Leftrightarrow\,\,\,  \CC_{A_1}^{loc}\simeq \CC_{A_2}^{loc}\\
    \text{($E_2$ algebras) }\,\,\,\,\,\,\,\,\,\,\text{(fusion categories) } \,\,\,\,\,\, \,\text{ (MTCs) }
\end{align*}

\begin{expl}
    For the special case when $\CC^{loc}_A$ is trivial, i.e. $\CC^{loc}_A = \vect$, in which $\vect$ is the category of finite dimensional $\bC$-vector spaces.
    $\CC_A$ is now a gapped boundary of $\CC$. 
    A 2d condensable algebra $A$ in this case is called the {\bf lagrangian algebra}. 
    The gapped boundaries $\{{}_{B_i}(\CC_A)_{B_i}\}$ of the $\CC$-phase are classified by the lagrangian algebras $\{A_i^L \in \Alg_{E_2}^{cond}(\CC)\}$\footnote{We use $A^L$ to denote lagrangian algebras in $\CC$ while $L$ to denote lagrangian algebras in $\CC \boxtimes \overline{\CC}$.}. 
    All lagrangian algebras in $\CC$ are 2-Morita equivalent since $\CC^{loc}_{A_i^L} \simeq \vect$, $\forall A_i^L$.
\end{expl}

\begin{rem}\label{rem:full_closed_CFT}
    Lagrangian algebras play a central role in rational 1+1D conformal field theory (CFT).
    A rational closed 1+1D CFT can be mathematically described by a rational vertex operator algebra (VOA) $V$ \cite{FHL93} and a lagrangian algebra $L\in \mathrm{Mod}_V\boxtimes \overline{\mathrm{Mod}_V}\simeq \fZ(\mathrm{Mod}_V)$, where the MTC $\mathrm{Mod}_V$ is the category of modules over $V$ \cite{MS89,Kong07}.
    Here the lagrangian algebra $L$ represents the Hilbert space of this CFT and determines the partition functions $Z_L$ correspondingly.
    Indeed, the modular invariance of the partition function is equivalent to the condensable algebra $L$ is lagrangian \cite{Kon08,KR09}.
    Since all lagrangian algebras are 2-Morita equivalent, then all full 2D closed CFTs over $V$ are equivalent up to 2-Morita equivalence.
\end{rem}

\subsection{Invertible domain walls in anyon condensation}

\begin{figure}[H]
    \centering
    
    \subfigure[]{
    \begin{minipage}[t]{0.45\linewidth}
        \centering
        \begin{tikzpicture}[scale=1.4]
            \filldraw[draw=none, fill=gray!40] (-4, 0)rectangle(-2,2);
            \filldraw[draw=none, fill=gray!20] (-2, 0)rectangle(0,2);
            \draw[thick](-2, 0)--(-2,2);

            \draw[thick, dashed](-4, 1)--(0,1);
            \draw[dashed](-2, 0.7)--(0,0.7);
                
            \node[]at(-2, -0.2){$\CC_{A_1}$};
            \node[]at(-2, 2.2){$\CC_{A_2}$};
            \node[]at(-3, 1.3){$\CC$};
            \node[]at(-1, 1.5){$\CC_{A_2}^{loc}$};
            \node[]at(-1, 0.4){$\CC_{A_1}^{loc}$};
            \node[]at(0.15, 1){$\Phi$};
            \node[]at(0.22, 0.68){$\Phi_{\varphi}$};
            \node[]at(-4.2, 1){$\Phi'$};

            \node[]at(-2.15, 0.7){\footnotesize$A_1$};
            
            \draw[-latex] (-2.25,0.8) arc(25:320:0.17);
        
            \node[]at(-2.65, 0.65){\footnotesize $\varphi$};
            \filldraw[fill=white] (-2.05, 0.95)rectangle(-1.95,1.05);
        \end{tikzpicture}
    \end{minipage}    
    }
    \begin{tikzpicture}
        \draw[-latex, thick] (-0.5,0) -- (0.5,0);
        \node at(0,0.3) {bend};
        \draw[draw=none] (0,0)--(0,-2);
    \end{tikzpicture}
    \subfigure[]{
        \begin{minipage}[t]{0.45\linewidth}
        \centering
        \begin{tikzpicture}[scale=1.4]
            \filldraw[fill=gray!40, draw=none] (-2,0) rectangle (2,2);
            \filldraw[fill=gray!20, draw=none] (-1,0)--(1,0)--(0,1.8)--(-1,0);
            \draw[thick, dashed] (0,0) -- (0,2);
            \draw[thick](0,1.7)--(0,1.85);
            \draw[thick](-1,0)--(0,1.8);
            \draw[thick](1,0)--(0,1.8);
            \node at (0,0.6) {$\CC^{loc}_{A_1} \simeq \CC^{loc}_{A_2}$};
            \node at (0.75,0.9) {${}_{A_2} \CC$};
            \node at (-0.75,0.9) {$\CC_{A_1}$};
            \node at (1.5,0.9) {$\CC$};
            \node at (-1.5,0.9) {$\CC$};
            \node at (0,-0.2) {$\Phi$};
            \draw[draw=none] (0,0) -- (0,-0.4);
            \filldraw[fill=white] (-0.05, 1.7)rectangle(0.05,1.8);
        \end{tikzpicture}
        \label{fig:a}
        \end{minipage}
    }
    \caption{An interchange of anyons between $\CC_{A_2}^{loc}$ and $\CC_{A_1}^{loc}$ can be regarded as an invertible domain wall $\Phi$, and fusing $\Phi$ with $\CC_{A_1}$ through $\CC_{A_1}^{loc}$ leads to a gapped domain wall characterized by ${}_{B}(\CC_{A_2})_B$.
 } \label{fig:open_to_phi}
\end{figure}
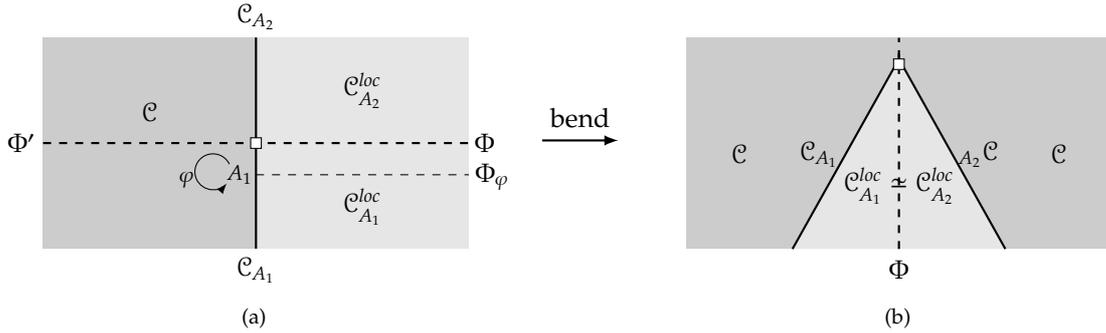

Some gapped domain wall ${}_{B}(\CC_{A_2})_{B}$ in figure \ref{fig:1d_cond} can be viewed as the fused phase $\CC_{A_1} \boxtimes_{\CC^{loc}_{A_1}} \Phi$ for some 2d condensable algebra $A_1$ in $\CC$\footnote{Not all domain walls ${}_{B_i}(\CC_A)_{B_i}$ can be written as $\CC_{A_1} \boxtimes_{\CC^{loc}_{A_1}} \Phi$ for some 2d condensable algebra $A_1$ in $\CC$, 
we discuss some algorithms related to $B$ in section \ref{section:outlook}.}, such that $A_1$ and $A_2$ are 2-Morita equivalent (figure \ref{fig:open_to_phi} (a)).
Here $\Phi$ is created implicitly by an interchange of anyons  $\phi:\CC_{A_1}^{loc}\xrightarrow{\sim} \CC_{A_2}^{loc}$ 
between the condensed topological order $\CC_{A_2}^{loc}$ and $\CC_{A_1}^{loc}$, which can be regarded as an {\it invertible} domain wall. 
The group $\Aut_{E_2}(\CC_{A_1}^{loc})$ of braided auto-equivalence of $\CC_{A_1}^{loc}$ has a natural action on the set $\{\phi:\CC_{A_1}^{loc}\to\CC_{A_2}^{loc}\}$ of all braided equivalence between $\CC_{A_1}^{loc}$ and $\CC_{A_2}^{loc}$ by composition.
Indeed, this set is a $\Aut_{E_2}(\CC_{A_1}^{loc})$-torsor, which means we can use the group $\Aut_{E_2}(\CC_{A_1}^{loc})$ (or $\Aut_{E_2}(\CC_{A_2}^{loc})$ equivalently) to describe
all invertible domain walls between $\CC_{A_2}^{loc}$ and $\CC_{A_1}^{loc}$.

\begin{rem}\label{Pic_iso}
    Any stable invertible domain wall in $\CC$ is an indecomposable invertible monoidal $\CC$-$\CC$-bimodule.
    Under the Deligne's tensor product $\btd_{\CC}$, they form a group $\mathrm{BrPic}_{E_1}(\CC)$ (see Definition \ref{dfn:E_1_BrPic}).
    There is an isomorphism between $\mathrm{BrPic}_{E_1}(\CC)$ and the Picard group ${\rm Pic}(\CC)$ of all invertible $\CC$-modules (see Theorem \ref{thm:E_1_BrPic}).
    By \cite{ENO10}, the group $\Aut_{E_2}(\CC)$ of braided auto-equivalences of $\CC$ is also isomorphic to ${\rm Pic}(\CC)$.
    Therefore, we obtain 
    \begin{equation*}
        \mathrm{BrPic}_{E_1}(\CC) \stackrel{\cong}{\longrightarrow} \Aut_{E_2}(\CC),
    \end{equation*}
    which shows that braided autoequivalences in $\CC$ actually characterize the invertible monoidal $\CC$-$\CC$-bimodules.
\end{rem}

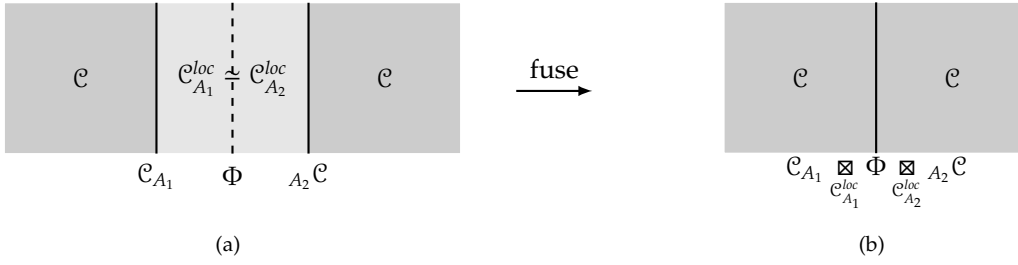
\begin{figure}[H]
    \centering
    \subfigure[]{
        \begin{minipage}[t]{0.45\linewidth}
            \centering
        \begin{tikzpicture} 
            \filldraw[fill=gray!40, draw=white] (-6,1) rectangle (-4,-1);
            \filldraw[fill=gray!20, draw=white] (-4,1) rectangle (-2,-1);
            \filldraw[fill=gray!40, draw=white] (-2,1) rectangle (0,-1);
            \draw[thick, dashed] (-3,-1) -- (-3,1);
            \draw[thick] (-4,1) -- (-4,-1);
            \draw[thick] (-2,1) -- (-2,-1);
            \node at (-3,-1.3) {$\Phi$};
            \node at (-3,0) {$\CC_{A_1}^{loc}\simeq \CC_{A_2}^{loc}$}; 
            \node at (-5,0) {$\CC$};
            \node at (-1,0) {$\CC$};
            \node at (-4,-1.3) {$\CC_{A_1}$};
            \node at (-2,-1.3) {${}_{A_2} \CC$};
            \draw[draw=none] (0,-1) -- (0,-1.87);
        \end{tikzpicture}
        \end{minipage}
    }
    \begin{tikzpicture}
        \draw[-latex, thick] (-0.5,0) -- (0.5,0);
        \node at(0,0.3) {fuse};
        \draw[draw=none] (0,0)--(0,-1.7);
    \end{tikzpicture}
    \subfigure[]{
        \begin{minipage}[t]{0.45\linewidth}
        \centering
        \begin{tikzpicture}
            \filldraw[fill=gray!40,draw=white] (-2,0) rectangle(2,2);
            \draw[thick] (0,0) -- (0,2);
            \node at (0,-0.4) {$\CC_{A_1}\btd\limits_{\CC_{A_1}^{loc}}\Phi \btd\limits_{\CC_{A_2}^{loc}}{}_{A_2} \CC$};
            \node at(-1,1){$\CC$};
            \node at(1,1){$\CC$};
        \end{tikzpicture}
        \end{minipage}
    }
    \caption{    
    A gapped domain wall within a 2d topological order $\CC$ can be described by $\CC_{A_1}\btd_{\CC_{A_1}^{loc}}\Phi \btd_{\CC_{A_2}^{loc}}{}_{A_2} \CC$. 
     Such a fused wall is non-invertible in most cases, except when 2d condensable algebras are both trivial, i.e. $A_1 = \one = A_2$.
    And (b) here is actually figure \ref{fig:1d_cond} with the special case $A= \one$ and $\CC^{loc}_{A} \simeq \CC$.
    } \label{fig:bimodule_C}

\end{figure}

Now we bend gapped domain walls $\CC_{A_2}$ down in figure \ref{fig:open_to_phi} (a), such that the condensed MTC $\CC^{loc}_{A_1}$ is sandwiching in between the gapped domain walls $\CC_{A_1}$, $\Phi$ and ${}_{A_2} \CC$\footnote{For a commutative algebra $A$, the category of right $A$-modules $\CC_A$ is canonically isomorphic to the category of left $A$-modules ${}_A \CC$. However, the $\CC$-module category structure depends on the position relative to $\CC$, so we use ${}_{A_2} \CC$ after we bend $\CC_{A_2}$ to the right-hand side of $\CC_{A_2}^{loc}$}, as figure \ref{fig:open_to_phi} (b) displays. 
And if we completely close these two domain walls described by $\CC_{A_1}$ and ${}_{A_2} \CC$ through the interlayer phase, we get a gapped domain wall within $\CC$ described by $\CC_{A_1}\btd_{\CC_{A_1}^{loc}}\Phi \btd_{\CC_{A_2}^{loc}}{}_{A_2} \CC$ in $\CC$, as figure \ref{fig:bimodule_C} (b) shows. 

\begin{expl}
    ~
    \begin{itemize}
        \item If $\CC_{A_1}^{loc} \simeq \CC$, then $\CC$, as a trivial domain wall is apparently a kind of invertible domain wall. 
        \item If $\CC_{A^L_1}^{loc} \simeq \vect$, there does not exist non-trivial invertible domain walls in $\vect$, and $\CC_{A^L_i}\btd {}_{A^L_j} \CC$ for lagrangian algebras $\{A^L_i\} \in \Alg_{E_2}^{cond} (\CC)$ are indeed stable gapped domain walls within $\CC$.
    \end{itemize}
\end{expl}

However, some invertible domain walls in condensed phase $\CC_{A_1}^{loc}$ (or $\CC_{A_2}^{loc}$) does not generate distinguishable gapped domain walls in original phase $\CC$, namely, we can have $\CC_{A_1} \boxtimes_{\CC^{loc}_{A_1}} \Phi = \CC_{A_1}$. 
In section \ref{sec:sym} we prove that invertible domain walls induced from algebra automorphisms of $A_1$ does not affect the classification of gapped domain walls in $\CC$ (however, it would have non-trivial impact in $\CC_{A_1}^{loc}$).
So for the sake of classification, we distinguish all invertible domain walls in the condensed phase as generated by two kinds of invertible walls (as figure \ref{fig:open_to_phi} (a) depicts):


\begin{itemize}
    \item $\Phi$ is induced from the auto-equivalences $\Phi'$ in the original phase;
    \item $\Phi_{\varphi}$ is induced from the algebra automorphism $\varphi$ of 2d condensable algebra $A_1$.  
\end{itemize}


Note that the braided auto-equivalences induced by algebra automorphisms of $A_1$ form a subgroup of $\Aut_{E_2}(\CC_{A_1}^{loc})$. 
We denote $[\phi]$ as a braided equivalence $\CC_{A_1}^{loc}\xrightarrow{\sim} \CC_{A_2}^{loc}$ that mod this redundant subgroup i.e. we define $\phi\sim \phi'$ if there is algebra isomorphism $\varphi_1:A_1\to A_1$, such that $\phi\circ \phi_1=\phi'$ where $\phi_1$ is the braided autoequivalence induced by algebra isomorphisms $\varphi_1$. 
Then $(A_1, A_2, [\phi])$ determines the classification of domain walls in $\CC$.

\begin{rem}
     Since $\CC_{A_1}^{loc}\simeq \CC_{A_2}^{loc}$, we have $\Aut_{E_2}(\CC_{A_1}^{loc})\cong \Aut_{E_2}(\CC_{A_2}^{loc})$. And $\{\phi:\CC_{A_1}^{loc}\to \CC_{A_2}^{loc} \}$ is both $\Aut_{E_2}(\CC_{A_1}^{loc})$-torsor and $\Aut_{E_2}(\CC_{A_2}^{loc})$-torsor. So the braided equivalence induced by $\varphi \in \Aut(A_1)$ and $\varphi' \in \Aut(A_2)$ are equivalent.
    Hence, we only need to choose one side as redundancy.
\end{rem}

\begin{rem}
    In a more general multistep condensation picture, both $\Phi$ and $\Phi_{\varphi}$ should come from some 2d condensable algebra's automorphisms in a bigger MTC $\CB$ that can condense to $\CC$. So we believe that by excluding the second kind of invertible domain walls $\Phi_{\varphi}$ in $\CC_{A_1}^{loc}$, the left ones are just the first kind of $\Phi$. In order to clarify what happens through $\Phi'$ to $\Phi$, we need to appeal to 0d defects, in which we omit in this paper. A comprehensive study will be performed in our future works.
\end{rem}



\subsection{2-Morita equivalence through lagrangian algebras}

We can also use lagrangian algebras in $\CC \btd \overline{\CC}$ to classify 2-Morita equivalent condensable algebras or gapped domain walls in $\CC$.

Recall figure \ref{general_picture} (c), by using folding trick to figure \ref{fig:bimodule_C} (b), the 1d gapped domain wall $\CC_{A_1}\btd_{\CC_{A_1}^{loc}}\Phi \btd_{\CC_{A_2}^{loc}}{}_{A_2} \CC$ becomes a boundary of the folded 2d bulk $\fZ(\CC)\simeq \CC\btd \overline{\CC}$.
Here $\Phi$ becomes a boundary of ${\CC_{A_1}^{loc}\btd \overline{\CC_{A_2}^{loc}}}$ determined by lagrangian algebra $L_{\phi}$ in the folded condensed phase (see figure below).

\begin{figure}[H]
    \centering
        \begin{minipage}[t]{0.45\linewidth}
            \centering
        \begin{tikzpicture} 
            \filldraw[fill=gray!40, draw=white] (-6,1) rectangle (-4,-1);
            \filldraw[fill=gray!20, draw=white] (-4,1) rectangle (-2,-1);
            \filldraw[fill=gray!40, draw=white] (-2,1) rectangle (0,-1);
            \draw[thick, dashed] (-3,-1) -- (-3,1);
            \draw[thick] (-4,1) -- (-4,-1);
            \draw[thick] (-2,1) -- (-2,-1);
            \node at (-3,-1.3) {$\Phi$};
            \node at (-3,0) {$\CC_{A_1}^{loc}\simeq \CC_{A_2}^{loc}$}; 
            \node at (-5,0) {$\CC$};
            \node at (-1,0) {$\CC$};
            \node at (-4,-1.3) {$\CC_{A_1}$};
            \node at (-2,-1.3) {${}_{A_2} \CC$};
        \end{tikzpicture}
        \end{minipage}
    \begin{tikzpicture}
        \draw[-latex, thick] (-0.5,0) -- (0.5,0);
        \node at(0,0.3) {fold};
        \draw[draw=none] (0,0)--(0,-1.7);
    \end{tikzpicture}
        \begin{minipage}[t]{0.45\linewidth}
        \centering
        \begin{tikzpicture}
            \centering
            \filldraw[fill=gray!40, draw=none] (-3,-1) rectangle (-1,1);
            \filldraw[fill=gray!20, draw=none] (-1,-1) rectangle (0,1);
            \filldraw[fill=gray!0, draw=none] (0,-1) rectangle (3,1);
            \draw[thick](-1,-1)--(-1,1);
            \draw[thick](0,-1)--(0,1);
            \node at(-2,0){\small $\CC \boxtimes\overline{\CC}$};
            \node at(-1,-1.3){\small $(\CC\boxtimes \overline{\CC})_{A_1\boxtimes A_2}$};
            \node at(-0.15,0){\scriptsize $L_{\phi}$};
        \end{tikzpicture}
        \end{minipage}
\end{figure}

By \cite{DMNO13,Kon14}, the stable gapped boundaries of $\fZ(\CC)$ are classified by lagrangian algebras $L$ in $\fZ(\CC)$.
We show in the next section that the triple ($A_1, A_2, L_{\phi}$) (or $A_1, A_2, \Phi$ equivalently) corresponds to a stable gapped boundary of $\fZ(\CC)$ after fusion by appealing to a notion call 2-step condensation \ref{sec:two_step}:

\begin{lem}\label{lem:2-Morita_E2_alg_to_lag_alg}
    Let $\CC$ be a MTC.
    Given any pair of $E_2$-Morita equivalent condensable algebras $A_1 \widesim[3]{2-Morita}{\phi} A_2$, $\CC_{A_1}\btd_{\CC_{A_1}^{loc}}\Phi \btd_{\CC_{A_2}^{loc}}{}_{A_2} \CC$ is equivalent to $\FZ(\CC)_L$ as monoidal $\CC$-$\CC$-bimodule for some lagrangian algebra $L\in\fZ(\CC)$.
\end{lem}

Thus, the fused domain walls $\CC_{A_1}\btd_{\CC_{A_1}^{loc}}\Phi \btd_{\CC_{A_2}^{loc}}{}_{A_2} \CC$ after folding can be written as $\fZ(\CC)_{L}$ for some lagrangian algebra $L\in\fZ(\CC)$, hence are stable. The above lemma gives \Ar{1} of the Trinity \ref{fig:alg_cycle} from "2-Morita equivalent condensable algebras" to "lagrangian algebras in $\fZ(\CC)$".
Note that two different equivalence $\phi$ and $\phi'$ might produce the same lagrangian algebras in $\fZ(\CC)$ (see for example $S_3$ topological order in section \ref{sec:expl_S3}).


This process can also be reversed.
Namely, given a lagrangian algebra $L \in \Alg_{E_2}^{lag}(\fZ(\CC))$, we can intersect $L$ with its left and right components 
to obtain subalgebras $A_l:=L\cap(\CC\btd \one)$ and $A_r:=(\one\btd \overline{\CC})\cap L$ in $\CC$ and $\overline{\CC}$ respectively.
By Corollary 3.3 in \cite{DNO12}, $A_l$ and $A_r$ are both 2d condensable algebras in $\CC$.
Moreover, by Proposition 3.7 and Theorem 3.6 in \cite{DNO12}, we have
\begin{cor}\label{crl:lag_alg_to_2-Morita_algs}
    $A_l$ and $A_r$ are 2-Morita equivalent.
\end{cor}

Taking intersections of $L$ to obtain $A_l \widesim[3]{2-Morita}{} A_r$ gives \Ar{2} from "lagrangian algebras in $\fZ(\CC)$" to "2-Morita equivalent condensable algebras" in the Trinity.
However, this process throw away the information given by $L_{\phi}$ (or $\Phi$). In order for \Ar{2} and \Ar{1} to be invertible to each other, we need to add back $L_{\phi}$ by computing the condensation of $L$ by its subalgebra $A_1\btd A_2$.

In summary, given a lagrangian algebra $L\in\fZ(\CC)$, its components $A_l$ and $A_r$ together with $L_{\phi}$ can reproduce $L$.
And, given a pair of 2-Morita equivalent condensable algebras $A_1 \widesim[3]{2-Morita}{\phi} A_2$ in $\CC$, the left and right components of the lagrangian algebra $L:=\Ext^R_{A_1\btd A_2}(L_{\phi})$ are again themselves, in the sense that $L$ is the extension (Lemma \ref{lem:ext}) of $L_{\phi}$ over $A_1\btd A_2$, see section \ref{sec:pf_lag_alg} for detail proof. 
To summarize, we give a proof that 
\begin{thm}\label{thm:classify_by_lag_alg}
    There is a one-to-one correspondence between the set of equivalent triples $(A_1,A_2,[\phi])$ where $A_1 \widesim[3]{2-Morita}{[\phi]} A_2$ in $\CC$, and the set of isomorphic classes of lagrangian algebras $L$ in $\fZ(\CC)$.
\end{thm}

\begin{rem}\label{rmk:A_1boxA_2}
    $\CC \btd \overline{\CC}$ can have more lagrangian algebras $L$ than \{$A^L_i \btd A^L_j$\} for $A^L_i \in \Alg_{E_2}^{lag}(\CC)$ and $A^L_j \in \Alg_{E_2}^{lag}({\overline{\CC}})$ due to $L_{\phi}$ hidden in the interlayer $\CC_{A_1}^{loc}$.
    By taking intersections, we terminate the entanglement between $\CC$ and $\overline{\CC}$, thus we are only left with 2-Morita equivalent condensable algebras from separate layers.
\end{rem}

Theorem \ref{thm:classify_by_lag_alg} upgrades the classification of lagrangian algebras in $\fZ(\CC)$ in \cite[Proposition 3.7]{DNO12} by modifying $\phi$ to $[\phi]$ (see also Remark \ref{rem:DNO}).
Our result provides a geometric comprehension of DNO's theorem and can be easily generated to the case $\CC_1\btd \overline{\CC_2}$.

\begin{rem}
    As Remark \ref{rem:full_closed_CFT} mentioned, in closed 1+1D CFT, lagrangian algebras in $\fZ(\mathrm{Mod}_V)$ determines a closed CFT over $V$ and corresponds to a modular invariant.  
    So Theorem \ref{thm:classify_by_lag_alg} also provides a classification of CFTs (or modular invariants) for a given vertex operator algebra $V$ by 2-Morita equivalent condensable algebras (together with the braided autoequivalence $\phi$) in $\mathrm{Mod}_V$.
\end{rem}



\subsection{Centers and 1d condensable algebras}\label{sec:centers}
The classification Theorem \ref{thm:classify_by_lag_alg} tells that we can obtain all 2-Morita equivalent condensable algebras in $\CC$ by classifying lagrangian algebras in $\fZ(\CC)$.
However, for cases other than $\fZ(\vect_{G})$, we do not have a systematical method to classify lagrangian algebras in $\fZ(\CC)$ without knowing a priori classification of 2-Morita equivalent condensable algebras.
To resolve this issue, we develop another method to classify 2-Morita equivalent condensable algebras via 1d condensable algebras.

Recall that lagrangian algebras in $\fZ(\CC)$ are one-to-one corresponding to gapped boundaries of $\fZ(\CC)$, and gapped boundaries of $\fZ(\CC)$ are one-to-one corresponding to gapped domain walls within $\CC$.
Hence, the 2-Morita equivalence of 2d condensable algebras in $\CC$ is also encoded in gapped domain walls each described by
${}_{B_i} \CC_{B_i}$ for some 1d condensable algebra ${B_i}\in\CC$.
Therefore, 2-Morita equivalent condensable algebras can also be classified by 1d condensable algebras.

One important method was developed in finding 2-Morita equivalent algebras in MTCs based on 1d condensable algebras --- the left/right center \cite{Ost03,FFRS06,Dav10}.

\begin{defn}[\cite{KYZ21}]\label{defn:center}
    Let $\CM$ be a monoidal left $\CC$-module and let $M\in\Alg_{E_1}(\CM)$.
    The {\bf left center} of $M$ in $\CC$ is a pair $(Z_l(M),u_l)$, where $Z_l(M)\in\Alg_{E_1}(\CC)$ and $u_l:Z_l(M)\odot M\to M$ is a unital $Z_l(M)$-action (see appendix \ref{appendix:algebra_center} for definition of unital action) on $M$, such that it is terminal among all such pairs.
    \begin{equation}\label{eq:uni_prop}
        \begin{codi}[hexagonal=vertical side 4.5em angle 60]
            \obj {
                 & |(Z)| Z_l(M)\odot M & \\
                |(1)|\one_{\CC}\odot M & |(X)|X\odot M & M \\
                };
            \mor :[bend left] 1 -> Z u_l:-> M;
            \mor * -> X -> *;
            \mor [swap] * \sim:-> *;
            \mor [mid]:[dashed] X \exists !:-> Z;
        \end{codi}
    \end{equation}   
    For $N \in \Alg_{E_1}(\CN)$ where $\CN$ is a monoidal right $\CC$-module, the {\bf right center} of $N$ in $\CC$ is defined to
    be the left center of $N$ in $\overline{\CC}$ by regarding $\CN$ as a monoidal left $\overline{\CC}$-module.   
\end{defn}

\begin{rem}\label{rmk:Dav_center}
    There is also another definition of right/left center $C_r(B)/C_l(B)\in\CC$ for an algebra $B$ in a braided monoidal category $\CC$ introduced by Davydov \cite{Dav10,Dav10a} (see Appendix \ref{appendix:algebra_center} for details).
    When $\CC$ is viewed as a $\CC$-$\CC$-bimodule category, the left/right center in Definition \ref{defn:center} coincides with the Davydov's right/left center, i.e. $Z_l(B)\cong C_r(B)$ and $Z_r(B)\cong C_l(B)$ for any algebra $B\in\CC$. 
\end{rem}

Taking left and right centers of a 1d condensable algebra $B$ would produce a pair of 2-Morita equivalent condensable algebras $(Z_l(B),Z_r(B))$:
\begin{thm}[\cite{FFRS06}]\label{thm:1d_cond_alg_to_2_Morita_cond_alg}
    Let $\CC$ be a MTC, $B$ be a 1d condensable algebra in $\CC$.
    Then there is an equivalence of MTCs:
    \begin{align*}
        \CC_{Z_l(B)}^{loc}\simeq \CC_{Z_r(B)}^{loc}
    \end{align*}
\end{thm}

\begin{cor}
    For any 1d condensable algebra $B$ in $\CC$, $Z_l(B)\widesim[3]{2-Morita}{} Z_r(B)$.
\end{cor}
 
\begin{expl}
    Let $B$ be a commutative 1d condensable algebra, which can be naturally regarded as a 2d condensable algebra.
    Then since the left/right center of a commutative algebra is still itself, we have $Z_l(B)\cong B\cong Z_r(B)$.
    This provides a trivial pair of 2-Morita equivalent condensable algebras, i.e. $B\widesim[3]{2-Morita}{}B$.
\end{expl}

This procedure gives \Ar{3} from "1-Morita class of 1d condensable algebras" to "2-Morita equivalent condensable algebras" in Trinity \ref{fig:alg_cycle}.
However, the bijectivity of \Ar{3} is not provided according to \cite{FFRS06}, namely, theorem \ref{thm:1d_cond_alg_to_2_Morita_cond_alg} does not tell whether all 2-Morita equivalent condensable algebras can be obtained by taking left and right centers.

On the other hand, let $(A_1,A_2)$ be a pair of 2-Morita equivalent condensable algebras in $\CC$ with $\phi:\CC_{A_1}^{loc}\simeq \CC_{A_2}^{loc}$.
Recall that $L_{\phi}$ is the lagrangian algebra in $\CC_{A_1}^{loc}\btd\overline{\CC_{A_2}^{loc}}$ corresponding to the $\Phi$.
This $L_{\phi}$ can also be regarded as a lagrangian algebra in $\CC_{A_1}^{loc}\btd\overline{\CC_{A_1}^{loc}}$ or in $\CC_{A_2}^{loc}\btd\overline{\CC_{A_2}^{loc}}$.
By applying tensor functor $\ot_{A_1}:\CC_{A_1}^{loc}\btd\overline{\CC_{A_1}^{loc}}\to \CC_{A_1}^{loc}$ on $L_{\phi}$, we obtain a direct sum of 1-Morita equivalent 1d condensable algebras in $\CC_{A_1}^{loc}$. 
Let us choose an indecomposable one as $B_{\phi}$, then we can have $\Ext^R_{A_1}(B_{\phi})\in \CC$ by extending this $B_{\phi}$ over $A_1$.
A similar procedure in $\CC_{A_2}^{loc}$ results in $\Ext^L_{A_2}(B_{\phi})\in\CC$.
We claim that $\Ext^R_{A_1}(B_{\phi})\ot \Ext^L_{A_2}(B_{\phi})$ would give the 1d condensable algebra $B$ in $\CC$ corresponding to $(A_1,A_2)$ and $\phi$.

\begin{Algr}\label{cnj:Arrow6}
    An indecomposable subalgebra $B\hookrightarrow\Ext^R_{A_1}(B_{\phi})\ot \Ext^L_{A_2}(B_{\phi})$ is the 1d condensable algebra corresponding to the 2-Morita equivalent pair $(A_1,A_2)$, i.e. $Z_l(B)\cong A_1$ and $Z_r(B)\cong A_2$.
\end{Algr}

This procedure gives \Ar{6} from "2-Morita equivalent condensable algebras” to "1-Morita class of 1d condensable algebras” in Trinity \ref{fig:alg_cycle}.

\begin{expl}\label{expl:lag_ext}
    When $A_1^L$ and $A_2^L$ are lagrangian algebras in $\CC$.
    Then the extended algebra is a subalgebra $B\hookrightarrow A_1^L\ot A^L_2$ since $\Ext^R_{A_1^L}(\one)=A_1^L$.
    Note that this algebra also determines the category of 0d domain wall conditions $\CC_B$ between two boundaries $\CC_{A_1^L}$ and $\CC_{A_2^L}$.
\end{expl}

We can also prove the bijectivity of \Ar{3} by proving bijective of \Ar{4} and \Ar{5} in Trinity ($5\circ 4 \simeq \id$), i.e. the bijection between the set of 1-Morita class of 1d condensable algebras in $\CC$ and the set of lagrangian algebras in $\fZ(\CC)$.
Since we have shown \Ar{1} and \Ar{2} are invertible, i.e. there is a bijection between lagrangian algebras in $\fZ(\CC)$ and pairs of 2-Morita equivalent condensable algebras in $\CC$, it is clear that the composed \Ar{$2\circ 4$} and \Ar{$5\circ 1$} between "pairs of 2-Morita equivalent condensable algebras in $\CC$" and "1-Morita class of 1d condensable algebras in $\CC$" should also be bijective.

So in order to show the bijection between 1-Morita class of 1d condensable algebras in $\CC$ and lagrangian algebras in $\fZ(\CC)$, we need to use another important algebraic center called 'full center' \cite{FFRS08,Dav10,DKR11}.

\begin{defn}\label{defn:full_center}
    If left (right) $\CC$-module $\CM$ satisfies $\CC=\fZ(\CM)$, then the left (right) center of $M\in\Alg_{E_1}(\CM)$ is called the {\bf full center} of $M$, denoted by $Z(M)$.
\end{defn}

\begin{rem}
    Left and right centers are actually dual concepts.
    Left center $Z_l(B) \in \CC$ is equivalent to the right center $Z_r(B) \in \overline{\CC}$ by regarding the fusion category $\CC$ as a monoidal right $\CC$-module.
    In the folded case (figure \ref{general_picture} (c)), the usual left/right center for $B$ coincides with the full center $Z(B)$ that results in a lagrangian algebra in $\FZ(\CC)$. 
\end{rem}

Let $B$ be a 1d condensable algebra in $\CC$, its full center $Z(B)$ is a lagrangian algebra in $\fZ(\CC)$ \cite{KR09}.
Hence, the procedure of "taking full center" gives \Ar{4} from "1-Morita classes of 1d condensable algebras in $\CC$" to "isomorphic classes of lagrangian algebras in $\fZ(\CC)$" in Trinity \ref{fig:alg_cycle}.
\Ar{4} is injective since two 1d condensable algebras $B_1$ and $B_2$ are 1-Morita equivalent if and only if $Z(B_1)\cong Z(B_2)$ \cite{KR08}.

\Ar{4} is also surjective, i.e. given any lagrangian algebras $L$ in $\fZ(\CC)$, there is a 1d condensable algebra $B$ such that $Z(B)\cong L$. 
Under the forgetful functor $U:\fZ(\CC)\to\CC$, $L$ becomes a separable algebra $U(L)$ in $\CC$.
However, $U(L)$ may not be indecomposable since it is a direct sum of matrix algebras in $\CC$.
A 1d condensable algebra $B$ can only be found as an indecomposable subalgebra in $U(L)$ in the sense of 1-Morita equivalence \cite{KZ17}.
This forgetting and picking process gives \Ar{5} in Trinity.

To see $Z(B)\cong L$, consider the indecomposable left $\CC$-module $\CC_B$.
By Proposition 4.8 in \cite{DMNO13}, indecomposable left $\CC$-modules are one-to-one corresponding to isomorphic classes of lagrangian algebras in $\fZ(\CC)$, i.e. $\Fun_{\CC}(\CC_B,\CC_B)\simeq {}_B\CC_B\simeq \fZ(\CC)_L$ (see Appendix \ref{appendix:algebra_center} for the definition of $\Fun_{\CC}(\CC_B,\CC_B)$).
And since ${}_B\CC_B\simeq \fZ(\CC)_{Z(B)}$, we have $Z(B)\cong L$.

Therefore, we have shown \Ar{4} and \Ar{5} are inverse to each other:
\begin{lem}\label{lem:1d_cond_alg_lag_alg}
    There is a bijection between set of 1-Morita classes of 1d condensable algebras in $\CC$ and  set of isomorphic classes of lagrangian algebras in $\fZ(\CC)$.
\end{lem}

\begin{rem}
    Similar to the forgetful functor $U$, we can also act tensor functor $\otimes$ on the lagrangian algebra $L$ in $\CC\btd\overline{\CC}$ to obtain a separable algebra $\otimes(L)$ which consists of 1d condensable algebras that are 1-Morita equivalent in $\CC$.
    This is due to the equivalence $\fZ(\CC)\simeq \CC\btd \overline{\CC}$, in which the tensor product functor $\otimes:\CC\btd \overline{\CC}\to\CC$ is a central functor, i.e. the following diagram commutes:
$$
\begin{tikzcd}
   \CC\btd \overline{\CC} \ar[r, "\sim"] \ar[dr, "\otimes"']& \FZ(\CC) \ar[d, "\text{Forget}"]\\
   & \CC
\end{tikzcd}
$$
Example \ref{expl:lag_ext} can also be explained directly by acting tensor functor $\ot$ on $A_1^L\btd A_2^L$.
\end{rem}

\begin{rem}
    Full open-closed 2D CFT are classified by a lagrangian algebra $A$ in $\fZ(\Mod_V)$ which determines the closed (bulk) CFT, and a simple special Frobenius algebra $B$ in $\Mod_V$ such that $A\cong Z(B)$ which determines the open (boundary) CFT \cite{FRS02,FFRS08,KR09}.
    Different open CFT $B_i$ that share the same bulk CFT $A$ are 1-Morita equivalent.
\end{rem}

We are left to show \Ar{3} is the composition of \Ar{2} and \Ar{4} in Trinity \ref{fig:alg_cycle}, i.e. for a 1d condensable algebra $B$, taking left/right center $Z_l(B)$/$Z_r(B)$, is equivalent to first taking full center $Z(B)$ then intersect with the left/right components of $\CC\btd \overline{\CC}$:

\begin{lem}\label{lem:left_right_full_centers}
    $Z_l(B)\cong Z(B)\cap(\CC\btd \one_{\CC})$ and $Z_r(B)\cong Z(B)\cap (\one_{\CC}\btd \overline{\CC})$ as algebras.
\end{lem} 
Above Lemma was first stated in the language of Davydov's center \cite[Section 2.5]{Dav10a}.
We give a proof in section \ref{sec:pf_center} using Definition \ref{defn:center}.
Since \Ar{2} and \Ar{4} are both bijections, thus \Ar{3} is also a bijection.
In other words, for two 2-Morita equivalent condensable algebras $A_1$, $A_2$ in $\CC$, there exists a 1d condensable algebra $B\in \CC$ such that $Z_l(B)\cong A_1$ and $Z_r(B)\cong A_2$.

\begin{rem}
    The above lemma can be generalized to a closed monoidal $\CC_1$-$\CC_2$-bimodule $\CM$: let $B$ be a 1d condensable algebra in $\CM$, we have $Z_l(B)\cong Z(B)\cap(\CC_1\btd \one_{\CC_2})$ and $Z_r(B)\cong Z(B)\cap (\one_{\CC_1}\btd \overline{\CC_2})$ as algebras.
\end{rem}



Then we finish the proof of classification by 1d condensable algebras in $\CC$.
\begin{thm}\label{thm:classify_by_1d_cond_alg}
    There is a one-to-one correspondence between the set of equivalent triples $(A_1,A_2,[\phi])$ where $A_1\widesim[3]{2-Morita}{[\phi]} A_2$ in $\CC$, and the set of 1-Morita classes of 1d condensable algebras in $\fZ(\CC)$.
\end{thm}


All Arrows that connect "2-Morita equivalent condensable algebras", "1-Morita class of condensable algebras", and "Lagrangian algebras" in Trinity \ref{fig:alg_cycle} are now been illustrated. 
Once we know a corner of the Trinity, we can have the other two. Different condensable algebras are unified through this Trinity.
Next section we give the detailed proof left in preliminary, namely Lemma \ref{lem:2-Morita_E2_alg_to_lag_alg} and Lemma \ref{lem:left_right_full_centers}. 
We also invent a method using internal hom to find 1-Morita equivalent condensable algebras $B$, see section \ref{sec:expl_S3}.

\section{Proof of Main Results} \label{section:proof}
In this section, we first introduce a process called 2-step condensation and prove some equivalence on fusion of domain walls.
Based on these, we further prove
that given a pair of 2-Morita equivalent condensable algebras $(A_1,A_2)$ in $\CC$, there is a lagrangian algebra $L\in\CC\boxtimes\overline{\CC}$ such that $L\cap (\CC\boxtimes \one)\cong A_1$ and $L\cap (\one \boxtimes \overline{\CC})\cong A_2$ (Lemma \ref{lem:2-Morita_E2_alg_to_lag_alg} in preliminary). Then we prove that taking full center of a 1d condensable algebra $B\in\CC$ then intersect with components of $\CC\boxtimes \overline{\CC}$ is equivalent to taking left/right centers of $B$ directly, i.e. Lemma \ref{lem:left_right_full_centers}.
In the last part of this section, we discuss the capability of the algebra automorphisms of $A$ in producing non-trivial symmetries in the condensed phase $\CC_A^{loc}$.


\subsection{Domain walls in two-step condensations}\label{sec:two_step}
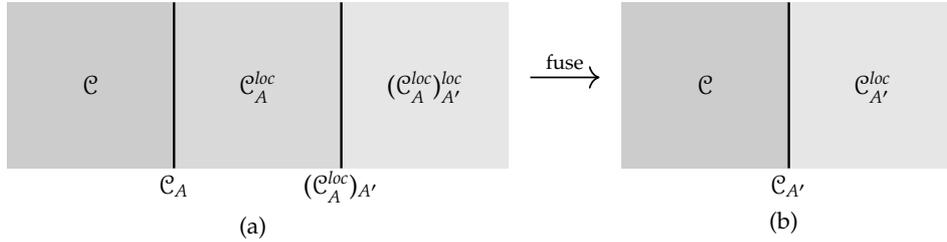
\begin{figure}[H]
    \centering
    \adjustbox{scale=1.1}{
    \begin{tikzcd}
        \subfigure[]{
        \begin{tikzpicture}
            \centering
            \filldraw[fill=gray!40, draw=none] (-3,-1) rectangle (-1,1);
            \filldraw[fill=gray!30, draw=none] (-1,-1) rectangle (1,1);
            \filldraw[fill=gray!20, draw=none] (1,-1) rectangle (3,1);
            \draw[thick](-1,-1)--(-1,1);
            \draw[thick](1,-1)--(1,1);
            \node at(-2,-0.1){\small $\CC$};
            \node at(2,-0.1){\small $(\CC^{loc}_{A})^{loc}_{A'}$};
            \node at(0,-0.1){\small $\CC^{loc}_{A}$};
            \node at(-1,-1.3){\small $\CC_{A}$};
            \node at(1,-1.3){\small $(\CC^{loc}_{A})_{A'}$};
        \end{tikzpicture}
        }
        \ar[r, "\text{fuse}"] &
        \subfigure[]{
        \begin{tikzpicture}
            \centering
            \filldraw[fill=gray!40, draw=none] (-2,-1) rectangle (0,1);
            \filldraw[fill=gray!20, draw=none] (0,-1) rectangle (2,1);
            \draw[thick](0,-1)--(0,1);
            \node at(-1,-0.1){\small $\CC$};
            \node at(1,-0.1){\small $\CC^{loc}_{A'}$};
            \node at(0,-1.3){\small $\CC_{A'}$};
        \end{tikzpicture}
        }
    \end{tikzcd}
    }
    \caption{For two 2d condensable algebras $A$ and $A'$ with $A\hookrightarrow A'$ in $\CC$, first condensing $A$ in $\CC$ then $A'$ condensing $A'$ in $\CC_A^{loc}$ gives the same phase as condensing $A'$ in $\CC$ directly, i.e. $(\CC^{loc}_{A})^{loc}_{A'}\simeq \CC^{loc}_{A'}$. However, whether the gapped domain walls generated in the two-step condensation after fusion (i.e. $\CC_{A} \boxtimes_{\CC^{loc}_{A}} (\CC^{loc}_{A})_{A'}$) is equivalent to $\CC_{A'}$ or not has not been discussed before.}
    \label{fig:2_step_cond}
\end{figure}

Let $A$ and $A'$ be two condensable algebras in MTC $\CC$ with an inclusion $A\hookrightarrow A'$, i.e. $A$ is a subalgebra of $A'$.
If we condense $A$ to obtain a condensed phase $\CC_A^{loc} $ and a domain wall $\CC_A$ between $\CC$ and $\CC_A^{loc}$,
then $A'$ would still be a condensable algebra in the condensed phase $\CC_A^{loc}$ \cite{DNO12}. 
Next we can condense $A'$ in $\CC_{A}^{loc}$ to produce a new condensed phase $(\CC_A^{loc})_{A'}^{loc}$ and a gapped domain wall $(\CC_{A}^{loc})_{A'}$ between $\CC_{A}^{loc}$ and $(\CC_A^{loc})_{A'}^{loc}$ (see figure \ref{fig:2_step_cond} (a)). This step-by-step condensation process to obtain $(\CC_A^{loc})_{A'}^{loc}$ from $\CC$ is called a \emph{two-step condensation}.

On the other hand, we can condense $A'$ in $\CC$ directly, which results in a condensed phase $\CC_{A'}^{loc}$ and a gapped domain wall $\CC_{A'}$ between $\CC$ and $\CC_{A'}^{loc}$, see figure \ref{fig:2_step_cond} (b).
It is known that the phase generated by a 2-step condensation with $A\hookrightarrow A'$ is equivalent to the direct condensed phase generated by $A'$, namely $(\CC^{loc}_{A})^{loc}_{A'}\simeq \CC^{loc}_{A'}$ \cite{FFRS06}.

\begin{rem}\label{rem:ext_voa}
   Consider the MTC $\mathrm{Mod}_V$ for a VOA $V$, a 2d condensable algebra $A'$ in $\mathrm{Mod}_V$ corresponds to an extension $V\hookrightarrow V'$ of VOA over $V$ \cite[Theorem 3.6]{HKL15} i.e., $(\mathrm{Mod}_V)_{A'}^{loc}\simeq \mathrm{Mod}_{V'}$.
   Two-step condensation $A'\hookrightarrow A''$ in $\mathrm{Mod}_V$ corresponds to a two-step conformal embedding $V\hookrightarrow V' \hookrightarrow V''$.
\end{rem}

However, under this equivalence, there is a hidden fusion process between the gapped domain walls $\CC_A$ and $(\CC_A^{loc})_{A'}$ through the intermediate condensed phase $\CC_A^{loc}$. Intuitively, The fused wall $\CC_{A} \boxtimes_{\CC^{loc}_{A}} (\CC^{loc}_{A})_{A'}$ should be equivalent to $\CC_{A'}$ as fusion categories.
But this equivalence has not been discussed before.
We fill this loophole here by proving the following theorem.
\begin{thm}\label{thm:2-step_cond}
    $\CC_{A} \boxtimes_{\CC^{loc}_{A}} (\CC^{loc}_{A})_{A'} \simeq \CC_{A'} $ as monoidal $\CC$-$\CC_{A'}^{loc}$-bimodule.
\end{thm} 

To prove theorem \ref{thm:2-step_cond}, we need to first prove a useful Lemma \ref{lem:absorb}. 

    Let $A$ be an algebra in a braided fusion category $\CD$. 
    Let $\CE$ be a monoidal right $\CD$-module with module action $\odot:\CE\times\CD\to\CE$.
Similar to the category $\CD_A$ of right $A$-modules in $\CD$, we have the category of right $A$-modules in $\CE$, denoted by $\CE_A$ (see appendix \ref{appendix:module_cat} for the precise definition of right $A$-modules in $\CE$, see also \cite{KYZ21}).

\begin{prop}
    Let $A$ be an $E_2$-algebra (commutative algebra) in $\CD$, then $\CE_A$ admits a monoidal structure.
\end{prop}
\begin{proof}
    Notice that $\one_{\CE}\odot A$ is an algebra in $\CE$, where the multiplication is given by 
    \begin{align*}
        (\one_{\CE}\odot A)\ot_{\CE}(\one_{\CE}\odot A)\simeq (\one_{\CE}\ot_{\CE}\one_{\CE})\odot (A\ot_{\CD}A)\xrightarrow{\Id\odot m_A}\one_{\CE}\odot A
    \end{align*}
    Moreover, consider $M,N\in\CE_A$, we have $M\odot A\simeq (M\ot_{\CE}\one_{\CE})\odot(\one_{\CD}\ot_{\CD}A)\simeq  (M\odot \one_{\CD})\otimes_{\CE} (\one_{\CE}\odot A)\simeq M\ot_{\CE}(\one_{\CE}\odot A)$.
    So $A$-action on $M$ can be equivalently characterized by a right $(\one_{\CE}\odot A)$-module structure on $M$.
    Indeed, an $A$-action on $N$ can also be characterized by a left $\one_{\CE}\odot A$-module structure on $\CN$, since $N\odot A\simeq (\one_{\CE}\ot_{\CE}N)\odot(A\ot_{\CD}\one_{\CD})\simeq  (\one_{\CE}\odot A)\otimes_{\CE} (N\odot \one_{\CD})\simeq (\one_{\CE}\odot A)\otimes_{\CE} N$.
    Hence, we can define left and right actions of $\one_{\CE}\odot A$ on $M$ and $N$.
    Then the relative tensor product of the algebra $\one_{\CE}\odot A$ 
    \begin{align*}
        \xymatrix{
            M\ot_{\CE}(\one_{\CE}\odot A)\ot_{\CE} N\ar@<1ex>[r]\ar@<-1ex>[r] & M\ot_{\CE}N \ar[r]\ar[dr] & M\otimes_{\one_{\CE}\odot A} N\ar@{-->}[d]^{\exists !}\\
             & & X 
        }
    \end{align*}
    in $\CE$ is well-defined and induces a monoidal structure in $\CE_A$.
    For simplicity, we denote the monoidal structure $M\otimes_{\one_{\CE}\odot A} N$ as $M\odotd_A N$.
\end{proof}

\begin{lem}\label{lem:absorb}
    Let $\CD$ be a braided fusion category.
    Let $\CE$ be a monoidal right $\CD$-module and let $A$ be an $E_2$ algebra in $\CD$.
    Then we have
    \begin{align*}
        \CE\btd_{\CD}\CD_A\simeq \CE_A
    \end{align*}
    as monoidal categories.
\end{lem}
\begin{proof}
    By \cite{KZ18}, there is an equivalence
    \begin{align*}
        F:\CE\btd_{\CD}\CD_A&\to \CE_A\\
        X\btd_{\CD}M&\mapsto X\odot M
    \end{align*}
    of categories, where $X\in \CE$ and $M\in\CD_A$.
    The right $A$-module action on $X\odot M$ is induced by the right $A$-module action on $M$.
    What remains is to show this equivalence is monoidal.

    Let $X_1\btd_{\CD}M_1$ and $X_2\btd_{\CD} M_2$ be two objects in $\CE\btd_{\CD}\CD_A$, using the following diagram, we can obtain the natural isomorphism $\nabla_{-,-}: F(-)\odot_A F(-)\Rightarrow F(-\otimes -)$ in monoidal functor.
    \begin{align*}
        \xymatrix@C=5em{
            (X_1\btd_{\CD}M_1)\ot(X_2\btd_{\CD}M_2)\ar@{|->}[d]\ar[r]^= & (X_1\otd_{\CE}X_2)\btd_{\CD} (M_1\otd_{A}M_2)\ar@{|->}[d] \\
            (X_1\odot M_1)\odotd_A (X_2\odot M_2)\ar[r]_{\nabla_{X_1\boxtimes_{\CD}M_1,X_2\boxtimes_{\CD}M_2}} & (X_1\otd_{\CE}X_2)\odot (M_1\otd_A M_2)
        }
    \end{align*}
    $(X_1\odot M_1)\odotd_A (X_2\odot M_2)$ is the coequalizer of $(X_1\odot M_1)\otd_{\CE}(\one_{\CE}\odot A)\otd_{\CE} (X_2\odot M_2)$.
    This can also be regarded as the coequalizer of $(X_1\otd_{\CE} \one_{\CE}\otd_{\CE}X_2)\odot(M_1\otd_{\CD} A\otd_{\CD} M_2)$, which obviously is $(X_1\otd_{\CE}X_2)\odot (M_1\otd_A M_2)$.
    So there is a natural isomorphism between $(X_1\odot M_1)\odotd_A (X_2\odot M_2)$ and $(X_1\otd_{\CE}X_2)\odot (M_1\otd_A M_2)$, which is the natural isomorphism $\nabla_{-,-}$ of monoidal functor we need.
\end{proof}

Lemma \ref{lem:absorb} has a graphical explanation in which $\CD$ can be regarded as a 2d phase and $\CE$, $\CD_A$ can be regarded as 1d phases at left and right side of $\CD$ respectively.
The equivalence in Lemma \ref{lem:absorb} tells us the fusion of $\CE$ and $\CD_A$ through $\CD$ is equivalent to $\CE_A$.
See the following figure.
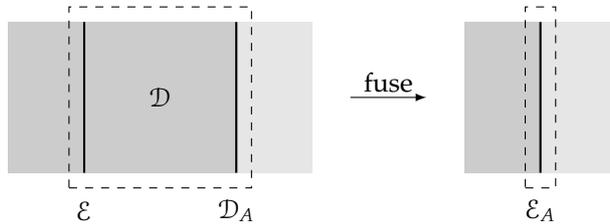
\begin{figure}[H]
    \centering
    \begin{tikzpicture}
        \filldraw[fill=gray!40,draw=none] (-3,0) rectangle(0,2);
        \filldraw[fill=gray!20,draw=none] (0,0) rectangle(1,2);
        \filldraw[fill=gray!40,draw=none] (3,0) rectangle(4,2);
        \filldraw[fill=gray!20,draw=none] (4,0) rectangle(5,2);
        \draw[thick] (-2,0) -- (-2,2);
        \draw[thick] (0,0) -- (0,2);
        \draw[thick] (4,0) -- (4,2);
        \node at(2,1.2){fuse};
        \draw[-latex] (1.5,1) -- (2.5,1);
        \draw[dashed] (-2.2, -0.2) rectangle (0.2, 2.2);
        \draw[dashed] (3.8, -0.2) rectangle (4.2, 2.2);
        \node at(-1,1){$\CD$};
        \node at(-2,-0.5){$\CE$};
        \node at(0,-0.5){$\CD_A$};
        \node at(4,-0.5){$\CE_A$};
    \end{tikzpicture}
\end{figure}

Manifestly, figure \ref{fig:2_step_cond} is a just special case of the above figure, in which we can substitute $\CE$ with $\CC_A$, $\CD_A$ with $(\CC_A^{loc})_{A'}$, and $\CD$ with $\CC_A^{loc}$ to recover the fusion of domain walls in the two-step condensation.

On the other hand, it is well known that $(\CC_A)_{A'}\simeq \CC_{A'}$. So based on this Lemma, we can easily prove Theorem \ref{thm:2-step_cond} 
\begin{proof}[Proof of Theorem \ref{thm:2-step_cond}]
    $\CC_{A} \boxtimes_{\CC^{loc}_{A}} (\CC^{loc}_{A})_{A'} \widesimeq[5]{Lemma\, \ref{lem:absorb}} (\CC_A)_{A'} \simeq \CC_{A'}$.
\end{proof}

We also prove a useful lemma which shows that for two 1d phases attached to a same 2d phase, the operations of folding and fusing commutes. 
See the following figure.
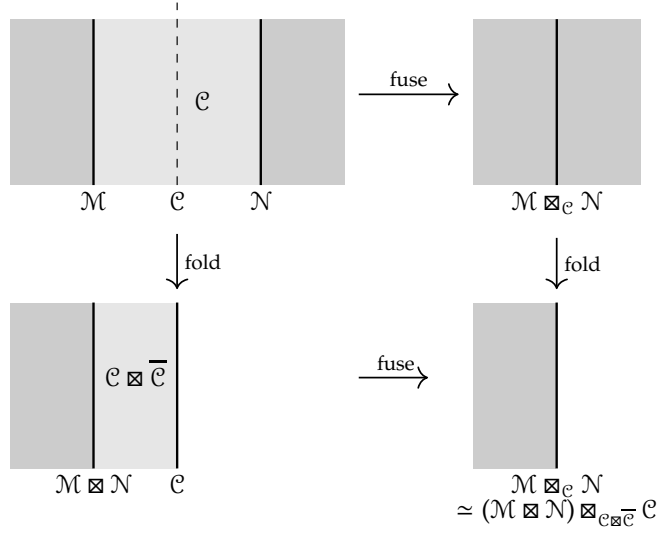
\begin{figure}[H]
    \centering
    \adjustbox{scale=1.1}{
    \begin{tikzcd}
        \begin{tikzpicture}
            \centering
            \filldraw[fill=gray!40, draw=none] (-2,-1) rectangle (-1,1);
            \filldraw[fill=gray!20, draw=none] (-1,-1) rectangle (1,1);
            \filldraw[fill=gray!40, draw=none] (1,-1) rectangle (2,1);
            \draw[thick](-1,-1)--(-1,1);
            \draw[thick](1,-1)--(1,1);
            \draw[dashed](0,-1)--(0,1.2);
            \node at(0.3,-0.1){\small $\CC$};
            \node at(-1,-1.3){\small $\CM$};
            \node at(1,-1.3){\small $\CN$};
            \node at(0,-1.3){\small $\CC$};
        \end{tikzpicture}
        \ar[r, "\text{$\quad$ fuse $\quad$}"] \ar[d, "\text{fold}"]&
        \begin{tikzpicture}
            \centering
            \filldraw[fill=gray!40, draw=none] (-1,-1) rectangle (0,1);
            \filldraw[fill=gray!40, draw=none] (0,-1) rectangle (1,1);
            \draw[thick](0,-1)--(0,1);
            \node at(0,-1.3){\small $\CM\boxtimes_{\CC}\CN$};
        \end{tikzpicture} \ar[d, "\text{fold}"] \\
        \begin{tikzpicture}[scale=1]
            \centering
            \filldraw[fill=gray!40, draw=none] (-2,-1) rectangle (-1,1);
            \filldraw[fill=gray!20, draw=none] (-1,-1) rectangle (0,1);
            \filldraw[fill=gray!0, draw=none] (1,-1) rectangle (2,1);
            \draw[thick](-1,-1)--(-1,1);
            \draw[thick](0,-1)--(0,1);
            \node at(-0.5,0){\small $\CC \boxtimes\overline{\CC}$};
            \node at(-1,-1.3){\small $\CM\boxtimes \CN$};
            \node at(0,-1.3){\small $\CC$};
        \end{tikzpicture}
        \ar[r, "\text{ $\quad$ fuse  $\quad$}"] &
        \begin{tikzpicture}[scale=1]
            \centering
            \filldraw[fill=gray!40, draw=none] (-1,-1) rectangle (0,1);
            \filldraw[fill=gray!0, draw=none] (0,-1) rectangle (1,1);
            \draw[thick](0,-1)--(0,1);
            \node at(0,-1.3){\small $\CM\boxtimes_{\CC}\CN$};
            \node at(0,-1.6){\small $\simeq (\CM\boxtimes\CN)\btd_{\CC\boxtimes \overline{\CC}}\CC$};
        \end{tikzpicture}
    \end{tikzcd}
    }
\end{figure}

\begin{lem}\label{lem:fold_fuse}
    Let $\CC$ be a braided fusion category.
    Let $\CM$ be a monoidal right $\CC$-module and $\CN$ be a left $\CC$-module.
    Then there is an equivalence 
    $\CM\btd_{\CC}\CN\simeq (\CM\boxtimes \CN)\btd_{\CC\boxtimes \overline{\CC}}\CC$
    of monoidal categories.
\end{lem}
\begin{proof}
    By Lemma 3.1.1 in \cite{KZ18}, there is an equivalence of categories
    \begin{align*}
        F:(\CM\boxtimes \CN)\btd_{\CC\boxtimes \overline{\CC}}\CC&\xrightarrow{\sim}\CM\btd_{\CC}\CN\\ 
        (x\boxtimes y)\btd_{\CC\boxtimes \overline{\CC}}c&\mapsto x\btd_{\CC} (c\odot_{\CN}y)
    \end{align*}
    where $\odot_{\CN}:\CC\times \CN\to\CN$ is the left $\CC$-module action on $\CN$.

    To show this equivalence is monoidal, we need to find a natural isomorphism $\nabla_{-,-}$ between functors $F(-)\ot F(-)$ and $F(-\ot-)$.
    Now let $(x_1\btd y_1)\btd_{\CC\btd\overline{\CC}}c_1$ and $(x_2\btd y_2)\btd_{\CC\btd\overline{\CC}}c_2$ be two objects in $(\CM\btd \CN)\btd_{\CC\btd\overline{\CC}}\CC$.
    Their images under $F$ are $x_1\btd_{\CC} (c_1\odot_{\CN}y_1)$ and $x_2\btd_{\CC} (c_2\odot_{\CN}y_2)$, which should be tensored to 
    \begin{align}\label{eq:FF}
        (x_1\otd_{\CM} x_2)\btd_{\CC} ((c_1\odotd_{\CN}y_1)\otd_{\CN}(c_2\odotd_{\CN}y_2)).
    \end{align}
    On the other hand, we have 
    \begin{align*}
        ((x_1\btd y_1)\btd_{\CC\btd\overline{\CC}}c_1)\ot ((x_2\btd y_2)\btd_{\CC\btd\overline{\CC}}c_2)=((x_1\otd_{\CM} x_2)\btd (y_1\otd_{\CN}y_2))\btd_{\CC\btd\overline{\CC}}(c_1\otd_{\CC}c_2).
    \end{align*}
    Its image under $F$ is $(x_1\otd_{\CM} x_2)\btd_{\CC} ((c_1\otd_{\CC}c_2)\odot_{\CN}(y_1\otd_{\CN}y_2))$.
    So the natural isomorphism $\nabla$ should be induced by the interchanging isomorphism in the definition of monoidal modules.
\end{proof}

\begin{rem}
Lemma \ref{lem:absorb} and lemma \ref{lem:fold_fuse} can be used to prove many results about fusion of 1d phases. For example, they can prove a often mentioned conclusion which provides a method to compute fusions of any 1d domain walls \cite{ENO10, DNO12,HBJP23}. We formulate this conclusion as follows:
\begin{thm}
    Let $\CB$, $\CD$ be braided fusion categories, and $\CC$ be a MTC.
    Let $\CM$ be a monoidal $\CB$-$\CC$-bimodule and $\CN$ be a monoidal $\CC$-$\CD$-bimodule.
    Then there is a monoidal equivalence
    \begin{align*}
        \CM\btd_{\CC}\CN\simeq (\CM\boxtimes \CN)_{\ot_{\CC}^R(\one_{\CC})}
    \end{align*}
    as monoidal $\CB$-$\CD$-bimodules, where $\ot_{\CC}^R$ is the right adjoint of the tensor functor $\ot_{\CC}:\CC\boxtimes\overline{\CC}\to \CC$. 
\end{thm}
\begin{proof}
    Folding the entire phase through the trivial domain wall in $\CC$. We have a new phase with boundary $\CC$ and a domain wall $\CM \boxtimes \CN$.
    Since $\CC$, when viewed as a boundary of $\fZ(\CC) \simeq \CC\boxtimes \overline{\CC} $, can be written as $\fZ(\CC)_{\otimes^R(\one_{\CC})}$, in which $\otimes^R(\one_{\CC})$ is the canonical lagrangian algebra.
    Then the folded phase, after fusing $\CM\boxtimes \CN$ with  $\fZ(\CC)_{\otimes^R(\one_{\CC})}$, becomes $(\CM\boxtimes\CN)\btd_{\fZ(\CC)}\fZ(\CC)_{\ot_{\CC}^R(\one_{\CC})}$. 
    On the other hand, we have $\CM\btd_{\CC}\CN$ after fusing $\CM$ with $\CN$ in the unfolded phase.
    Thus, by Lemma \ref{lem:absorb} and Lemma \ref{lem:fold_fuse}, we have 
    \begin{align*}
        \CM\btd_{\CC}\CN \widesimeq[5]{Lemma\, \ref{lem:fold_fuse}} (\CM\boxtimes\CN)\btd_{\fZ(\CC)}\fZ(\CC)_{\ot_{\CC}^R(\one_{\CC})} \widesimeq[5]{Lemma\, \ref{lem:absorb}} (\CM\boxtimes \CN)_{\ot_{\CC}^R(\one_{\CC})}
    \end{align*}
\end{proof}

\end{rem}

\subsection{Classification of 2-Morita equivalent condensable algebras}
In this subsection, we show that 2-Morita equivalent condensable algebras in a modular tensor category $\CC$ can be classified 
through two different ways: one way is to use lagrangian algebras, another way is to use 1d condensable algebras. 
In other words, we finish the proof of our main theorem \ref{main_theorem}.

\subsubsection{Lagrangian algebras}\label{sec:pf_lag_alg}
We first prove Lemma \ref{lem:2-Morita_E2_alg_to_lag_alg} in the preliminary using Theorem \ref{thm:2-step_cond}, which we restate below. 


\setcounter{section}{2}
\setcounter{thm}{0}
\begin{lem}
    Let $\CC$ be a MTC.
    Given any pair of $E_2$-Morita equivalent algebras $A_1 \widesim[3]{2-Morita}{\phi} A_2$, $\CC_{A_1}\btd_{\CC_{A_1}^{loc}}\Phi \btd_{\CC_{A_2}^{loc}}{}_{A_2} \CC$ is equivalent to $\FZ(\CC)_L$ as monoidal $\CC$-$\CC$-bimodule for some lagrangian algebra $L\in\fZ(\CC)$.
\end{lem}
\begin{figure}[H]
    \centering
    \adjustbox{scale=1.1}{
    \begin{tikzcd}
        \begin{tikzpicture}
            \centering
            \filldraw[fill=gray!40, draw=none] (-3,-1) rectangle (-1,1);
            \filldraw[fill=gray!20, draw=none] (-1,-1) rectangle (1,1);
            \filldraw[fill=gray!40, draw=none] (1,-1) rectangle (3,1);
            \draw[thick](-1,-1)--(-1,1);
            \draw[thick](1,-1)--(1,1);
            \draw[dashed,thick](0,-1)--(0,1);
            \node at(-2,-0.1){\small $\CC$};
            \node at(2,-0.1){\small $\CC$};
            \node at(0,-0.1){\small $\CC^{loc}_{A_1}\simeq \CC^{loc}_{A_2}$};
            \node at(-1,-1.3){\small $\CC_{A_1}$};
            \node at(1,-1.3){\small ${}_{A_2} \CC$};
            \node at(0,-1.3){\small $\Phi$};
        \end{tikzpicture}
        \ar[r, "\text{$\quad$ fuse $\quad$}"] \ar[d, "\text{fold}"]&
        \begin{tikzpicture}
            \centering
            \filldraw[fill=gray!40, draw=none] (-2,-1) rectangle (0,1);
            \filldraw[fill=gray!40, draw=none] (0,-1) rectangle (2,1);
            \draw[thick](0,-1)--(0,1);
            \node at(-1,-0.1){\small $\CC$};
            \node at(1,-0.1){\small $\CC$};
            \node at(0,-1.3){\small $\CC_{A_1}\btd_{\CC_{A_1}^{loc}}\Phi \btd_{\CC_{A_2}^{loc}}{}_{A_2} \CC$};
        \end{tikzpicture} \ar[d, "\text{fold}","\simeq"'] \\
        \begin{tikzpicture}[scale=1]
            \centering
            \filldraw[fill=gray!40, draw=none] (-3,-1) rectangle (-1,1);
            \filldraw[fill=gray!20, draw=none] (-1,-1) rectangle (0,1);
            \filldraw[fill=gray!0, draw=none] (0,-1) rectangle (3,1);
            \draw[thick](-1,-1)--(-1,1);
            \draw[thick](0,-1)--(0,1);
            \node at(-2,0){\small $\CC \boxtimes\overline{\CC}$};
            \node at(1.5,0){\small $\vect$};
            \node at(-0.37,0){\tiny $(\CC \boxtimes\overline{\CC})^{loc}_{A_1\boxtimes A_2}$};
            \node at(-1,-1.3){\small $(\CC\boxtimes \overline{\CC})_{A_1\boxtimes A_2}$};
            \node at(0.1,-1.3){\small $\Phi$};
        \end{tikzpicture}
        \ar[r, "\text{ $\quad$ fuse  $\quad$}"] &
        \begin{tikzpicture}[scale=1]
            \centering
            \filldraw[fill=gray!40, draw=none] (-1,-1) rectangle (1,1);
            \filldraw[fill=gray!0, draw=none] (1,-1) rectangle (3,1);
            \draw[thick](1,-1)--(1,1);
            \node at(-0.1,0){\small $\CC \boxtimes\overline{\CC}$};
            \node at(2,0){\small $\vect$};
            \node at(1.3,-1.3){\small $\fZ(\CC)_{L}$};
        \end{tikzpicture}
    \end{tikzcd}
    }
    \caption{This figure depicts the logic flow of the proof of Lemma \ref{lem:2-Morita_E2_alg_to_lag_alg}. By Lemma \ref{lem:fold_fuse}, the whole diagram commutes.
    Based on two-step condensation, we prove $\CC_{A_1}\btd_{\CC_{A_1}^{loc}}\Phi \btd_{\CC_{A_2}^{loc}}{}_{A_2} \CC \simeq \FZ(\CC)_L$ as monoidal $\CC$-$\CC$-bimodules.} \label{fig:proof}
\end{figure}
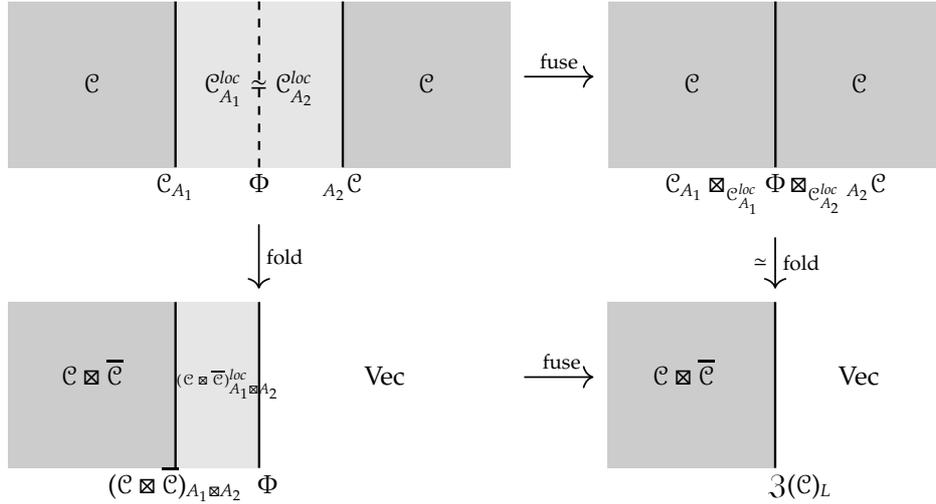
\begin{proof}
    The sub-figure in the upper left corner of fig. \ref{fig:proof} is just figure \ref{fig:bimodule_C} (a) in preliminary. 
    Starting from this sub-figure, we can perform two operations that lead to the bottom right figure. 
    One way is to first fuse $\CC_{A_1}$, $\Phi$ and ${}_{A_2} \CC$,
    another way is to first fold the whole phase through the invertible domain wall $\Phi$. 
    According to Lemma \ref{lem:fold_fuse}, we have $(\CC\boxtimes \overline{\CC})_{A_1\boxtimes A_2}\btd_{(\CC\boxtimes \overline{\CC})_{A_1\boxtimes A_2}^{loc}}\Phi \simeq \CC_{A_1}\btd_{\CC_{A_1}^{loc}}\Phi \btd_{\CC_{A_2}^{loc}}{}_{A_2} \CC$.
    
    Since $\Phi$ is an indecomposable monoidal $\CC_{A_1}^{loc}$-$\CC_{A_2}^{loc}$-bimodule, it should also be an indecomposable monoidal left $(\CC\boxtimes \overline{\CC})_{A_1\boxtimes A_2}^{loc}$-module when viewed as a boundary.
    According to \cite{DMNO13}, there exists a lagrangian algebra $L_{\phi}$ in $(\CC\boxtimes \overline{\CC})_{A_1\boxtimes A_2}^{loc}$, such that $\Phi\simeq ((\CC\boxtimes \overline{\CC})_{A_1\boxtimes A_2}^{loc})_{L_{\phi}}$.
    On the other hand, $L_{\phi}$ can also be extended to a lagrangian algebra $L:=\Ext^R_{A_1\boxtimes A_2}(L_{\phi})$ in $\CC\boxtimes\overline{\CC}$, where $A_1\boxtimes A_2$ is a subalgebra of $L$ \cite{DNO12}.
    Thus, fusing $(\CC\boxtimes \overline{\CC})_{A_1\boxtimes A_2}$ and $\Phi$ through $(\CC\boxtimes \overline{\CC})_{A_1\boxtimes A_2}^{loc}$ can be regarded as fusing the domain walls in a 2-step condensation $A_1\boxtimes A_2\hookrightarrow L$.
    By Theorem \ref{thm:2-step_cond}, we have $(\CC\boxtimes \overline{\CC})_{A_1\boxtimes A_2}\btd_{(\CC\boxtimes \overline{\CC})_{A_1\boxtimes A_2}^{loc}}((\CC\boxtimes \overline{\CC})_{A_1\boxtimes A_2}^{loc})_{L_{\phi}}\simeq (\CC\boxtimes \overline{\CC})_{L}$ as monoidal left $\fZ(\CC)$-module categories.
    Thus, $\CC_{A_1}\btd_{\CC_{A_1}^{loc}}\Phi \btd_{\CC_{A_2}^{loc}}{}_{A_2} \CC \simeq \FZ(\CC)_L$ as monoidal $\CC$-$\CC$-bimodules.

\end{proof}
\setcounter{section}{3}
\setcounter{thm}{5}

\begin{rem}
    Here $L_{\phi}:=\bigoplus_{x\in\mathrm{Irr}(\CC_{A_1}^{loc})}x\btd \phi(x^*)$ in $(\CC\btd \overline{\CC})_{A_1\btd A_2}^{loc}$ can be computed by internal hom $[\one,\one]$ in the left $(\CC\btd \overline{\CC})_{A_1\btd A_2}^{loc}$-module category $\Phi$.
\end{rem}

Lemma \ref{lem:2-Morita_E2_alg_to_lag_alg} together with Corollary \ref{crl:lag_alg_to_2-Morita_algs} show that there is a one-to-one correspondence between the set of pairs of 2-Morita equivalent condensable algebras in $\CC$ and the set of isomorphic classes of lagrangian algebras in $\fZ(\CC)$, which finishes the proof of Theorem \ref{thm:classify_by_lag_alg}.

\subsubsection{Left, right and full centers}\label{sec:pf_center}
Instead of using lagrangian algebra $L \in \FZ(\CC)$, we can also use the left/right centers of 1d condensable algebras $B \in \CC$ to classify 2-Morita equivalent condensable algebras.
Indeed, by Lemma \ref{lem:1d_cond_alg_lag_alg}, we have $Z(B)\cong L$ for some 1d condensable algebra $B$ in $\CC$.
Recall corollary \ref{crl:lag_alg_to_2-Morita_algs} that $Z(B) \cap (\CC\boxtimes \one)$ and $Z(B)\cap (\one \boxtimes \overline{\CC})$ are 2-Morita equivalent. At the same time, we have a pair of 2-Morita equivalent condensable algebras $(Z_l(B),Z_r(B))$ by taking the left and right centers of $B$ according to Theorem \ref{thm:1d_cond_alg_to_2_Morita_cond_alg}.
We now show these two ways are the same, i.e. for a 1d condensable algebra $B$, taking left/right center $Z_l(B)$/$Z_r(B)$, is equivalent to first taking full center $Z(B)$ then intersect with the left/right components of $\CC\boxtimes \overline{\CC}$:

\setcounter{section}{2}
\setcounter{thm}{6}
\begin{lem}
    $Z_l(B)\cong Z(B)\cap (\CC\boxtimes \one)$ and $Z_r(B)\cong Z(B)\cap (\one\boxtimes \overline{\CC})$.
\end{lem}
\begin{proof}
    Consider $Z(B)\cap (\CC\boxtimes \one)$, we show it satisfies the universal property of left center $Z_l(B)$, i.e. the commutative diagram (a). 
    Unital action $u_l:(Z(B)\cap (\CC\boxtimes \one))\otimes B \to B$ in diagram of left center is induced by the unital action $u:Z(B)\odot B\to B$ of $Z(B)$ in diagram (b).
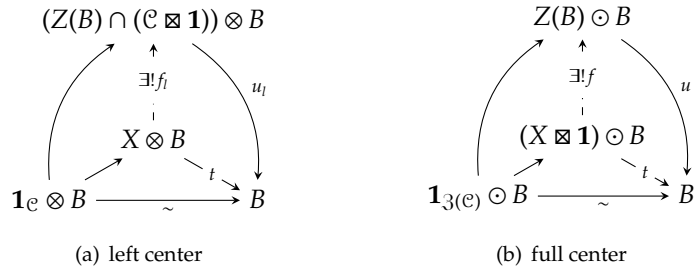
    \begin{figure}[H]
        \centering
        \subfigure[][left center]{
            \begin{codi}
                \obj [hexagonal=vertical side 4.5em angle 60] {
                     & |(Z)| (Z(B)\cap (\CC\boxtimes \one))\otimes B & \\
                    |(1)|\one_{\CC}\otimes B & |(X)|X\otimes B & B \\
                    };
                \mor :[bend left] 1 -> Z u_l:-> B;
                \mor [mid] * -> X t: -> *;
                \mor [swap] * \sim:-> *;
                \mor [mid]:[dashed] X {\exists ! f_l} :-> Z;
            \end{codi}
        }
        \qquad\qquad
        \subfigure[][full center]{
            \begin{codi}
                \obj [hexagonal=vertical side 4.5em angle 60] {
                     & |(Z)| Z(B)\odot B & \\
                    |(1)|\one_{\fZ(\CC)}\odot B & |(X)|(X\boxtimes \one)\odot B & B \\
                    };
                \mor :[bend left] 1 -> Z u:-> B;
                \mor [mid]* -> X t:-> *;
                \mor [swap] * \sim:-> *;
                \mor [mid]:[dashed] X {\exists ! f} :-> Z;
            \end{codi}
        }
    \end{figure}

    For any $X$ together with a unital action $t:X\otimes B\to B$ such that the lower triangle in the diagram of left center commutes, we consider $X\boxtimes \one$ in the diagram (b) which satisfies the universal property of full center $Z(B)$.
    Since $(X\boxtimes \one)\odot B:=\otimes(X\boxtimes \one)\otimes B
    \simeq X\otimes B$,
    the morphism $t':(X\boxtimes\one)\odot B\to B$ is actually given by $t$, so the lower triangle in right diagram must commutes.
    Hence, by universal property of the full center $Z(B)$, there exists a unique morphism $f:(X\boxtimes \one)\to Z(B)$ such that the right triangle in diagram (b) commutes.
    Since $Z(B)\cap (\CC\boxtimes \one)$ consists of objects only in $\CC$, we restrict $f$ to the left component of $\CC\boxtimes \overline{\CC}$, in which we obtain $f_l: X\to Z(B)\cap(\CC\boxtimes \one)$.
    And $f_l$ makes the right triangle in diagram (a) commutes since $f$ does.

    $Z(B)\cap (\one\boxtimes \overline{\CC})$ can be proven to satisfy the universal property of right center $Z_r(B)$ by a similar process.
\end{proof}
\setcounter{section}{3}
\setcounter{thm}{5}

\begin{rem}
    Lemma \ref{lem:left_right_full_centers} was first stated by Davydov in \cite{Dav10,Dav10a} without proof. 
    Our process also proves his statement since Davydov's center is equivalent to Definition \ref{defn:center} for $B \in \CC$.  
\end{rem}

To summarize, for two 2-Morita equivalent condensable algebras $A_1 \widesim[3]{2-Morita}{\phi} A_2$, we have a lagrangian algebra $L$ such that $L\cap (\CC\boxtimes\one)\cong A_1$ and $L\cap (\one \boxtimes \overline{\CC})\cong A_2$.
Since $L\cong Z(B)$ for some 1d condensable algebra $B$, we have $A_1=Z(B)\cap(\CC\boxtimes \one)$ and $A_2=Z(B)\cap (\one \boxtimes \overline{\CC})$.
Now by Lemma \ref{lem:left_right_full_centers}, we have $A_1=Z_l(B)$ and $A_2=Z_r(B)$.
Consequently, for any 2-Morita equivalent condensable algebras $A_1$ and $A_2$, there exists a 1d condensable algebra $B$ such that $A_1=Z_l(B)$ and $A_2=Z_r(B)$,
which finishes the proof of Theorem \ref{thm:classify_by_1d_cond_alg}.

Combine Theorem \ref{thm:classify_by_lag_alg} and Theorem \ref{thm:classify_by_1d_cond_alg}, we prove our main Theorem \ref{main_theorem}.

\vspace{2em}
We can translate above algebraic results to their module categories.
The indecomposable monoidal $\CC$-$\CC$-bimodule $\CC_{A_1}\btd_{\CC_{A_1}^{loc}}\Phi \btd_{\CC_{A_2}^{loc}}{}_{A_2} \CC$ can be written as ${}_B\CC_B$ for some 1d condensable algebra $B$.
Since $A_1=Z_l(B)$ and $A_2=Z_r(B)$, we can directly derive following theorem:

\begin{thm}\label{thm:pull_open}
    Let $\CC$ be a MTC.
    An indecomposable $\CC$-$\CC$-bimodule in $\CC$ can be written as ${}_B\CC_B$ for some 1d condensable algebra $B\in\CC$. We have ${}_B\CC_B \simeq \CC_{Z_l(B)}\boxtimes_{\CC_{Z_l(B)}^{loc}}\Phi \boxtimes_{\CC_{Z_r(B)}^{loc}}{}_{Z_r(B)} \CC$ where $\Phi$ is the invertible domain wall induced by the equivalence $\phi: \CC_{Z_l(B)}^{loc}\xrightarrow{\sim}\CC_{Z_r(B)}^{loc}$ of MTCs.
\end{thm}

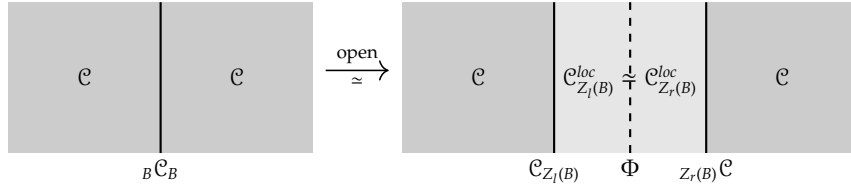
\begin{figure}[H]
    \centering
    \begin{tikzcd}
        \begin{tikzpicture}
            \centering
            \filldraw[fill=gray!40, draw=none] (-2,-1) rectangle (0,1);
            \filldraw[fill=gray!40, draw=none] (0,-1) rectangle (2,1);
            \draw[thick](0,-1)--(0,1);
            \node at(-1,-0.1){\small $\CC$};
            \node at(1,-0.1){\small $\CC$};
            \node at(0,-1.3){\small ${}_B\CC_B$};
        \end{tikzpicture}
        \ar[r, "\text{$\quad$ open $\quad$}","\simeq"'] &
        \begin{tikzpicture}
            \centering
            \filldraw[fill=gray!40, draw=none] (-3,-1) rectangle (-1,1);
            \filldraw[fill=gray!20, draw=none] (-1,-1) rectangle (1,1);
            \filldraw[fill=gray!40, draw=none] (1,-1) rectangle (3,1);
            \draw[thick](-1,-1)--(-1,1);
            \draw[thick](1,-1)--(1,1);
            \draw[dashed,thick](0,-1)--(0,1);
            \node at(-2,-0.1){\small $\CC$};
            \node at(2,-0.1){\small $\CC$};
            \node at(0,-0.1){\small $\CC^{loc}_{Z_l(B)}\simeq \CC^{loc}_{Z_r(B)}$};
            \node at(-1,-1.3){\small $\CC_{Z_l(B)}$};
            \node at(1,-1.3){\small ${}_{Z_r(B)} \CC$};
            \node at(0,-1.3){\small $\Phi$};
        \end{tikzpicture}
    \end{tikzcd}
\end{figure}

In the language of topological orders, Theorem \ref{thm:pull_open} tells us any stable gapped domain wall in $\CC$ can be 'pulled open'. On the other hand, by Lemma \ref{lem:2-Morita_E2_alg_to_lag_alg}, any fused domain wall whose inner part is obtained by condensation, is stable. See section \ref{section:outlook} for more discussions on the criterion of gapped domain wall stability. Or to say, we classify all simple 1-codimensional defects of a 2+1D topological order through 2-codimensional condensations.

Theorem \ref{thm:pull_open} can be further generalized to any stable gapped domain wall between 2d topological orders $\CC$ and $\CD$. See Generalization \ref{sec:generalize}.

\subsection{Symmetries induced by algebra automorphisms}\label{sec:sym}
Note that an $E_2$ condensable algebra $A$ in a braided fusion category $\CC$ may have non-trivial algebra automorphisms. In this subsection, we show a non-trivial algebra automorphism $\varphi$ leads to a braided autoequivalence $\phi$ in $\CC_A^{loc}$.

Now suppose $A$ admits a non-trivial algebra automorphism, say $\varphi:A\to A$, which satisfies
\begin{align}\label{diag:alg_homo}
    \xymatrix{
        A\ot A\ar[r]^{\varphi\ot\varphi}\ar[d]_{m_A} & A\ot A\ar[d]^{m_A}\\
        A\ar[r]_{\varphi} &A
    }
\end{align}

For a right $A$-module $(M,r_M)$, by pre-composing the right $A$-module action $r_M$ with this algebra automorphism $\varphi$, we obtain a new morphism from $M\ot A$ to $M$: $M\ot A \xrightarrow{\id_M\ot\varphi} M\ot A\xrightarrow{r_M} M$.
\begin{prop}
    $(M,r_M\circ(\id_M\ot \varphi))$ is also a right $A$-module.
\end{prop}
\begin{proof}
    Consider the following diagram,
    \begin{center}
        \begin{codi}
            \obj[tetragonal=base 9em height 3.5em]{ 
                |(MAA)| M\ot A \ot A & |(MAA1)|M\ot A\ot A & |(MAr1)|M\ot A\\
                 & |(MAA2)|M\ot A\ot A & |(MAr2)|M\ot A\\
                |(MAl1)| M\ot A & |(MAl2)| M\ot A &M \\
            };
            \mor: MAA {\id_M\ot \varphi\ot\id_A}:-> MAA1 {r_M\ot \id_A}:-> MAr1 {\id_M\ot\varphi}:-> MAr2 r_M:-> M;
            \mor[swap]: * "\id_M\ot m_A":-> MAl1 "\id_M\ot\varphi":-> MAl2 r_M :-> *;
            \mor: MAA1 "\id_M\ot \id_A\ot \varphi":-> MAA2 "r_M\ot \id_A" :-> MAr2;
            \mor[swap]: MAA2 "\id_M\ot m_A" :-> MAl2; 
        \end{codi}
    \end{center}
    The left sub-diagram commutes since it is diagram \ref{diag:alg_homo} tensoring with $M$, the rest sub-diagrams commute, apparently.
    So the outer diagram commutes.
\end{proof}

\begin{cor}
    This $\varphi$ induces an autoequivalence in $\CC_A$,
    \begin{align*}
        \phi:\CC_A&\to \CC_A\\
        (M,r_M)&\mapsto (M,r_M\circ (\id_M\ot\varphi))
    \end{align*}
\end{cor}

\begin{lem}\label{lem:phi_is_monoidal}
    Let $A$ be a commutative algebra in $\CC$ and let $\varphi$ be an automorphism of $A$.
    Then autoequivalence $\phi$ induced by $\varphi$ is monoidal.  
\end{lem}
\begin{proof}
    First, since $A$ can be regarded as the free $A$-module $\one\ot A$, by Proposition \ref{prp:fix_free_mod}, we see $\phi$ preserves tensor unit $A$ of $\CC_A$. 
    Then we prove $\phi$ preserves monoidal structure.
    Given two right $A$-modules $(M,r_M)$ and $(N,r_N)$.
    We consider the natural isomorphism $\nabla$ between $\phi((M,r_M)\ot_A(N,r_N))$ and $\phi(M,r_M)\ot_A \phi(N,r_N)$.
    By definition, 
    \begin{align*}
        \phi((M,r_M)\ot_A(N,r_N))=\phi(M\ot_A N,r_{M\ot_A N})=(M\ot_A N, r_{M\ot_A N}\circ (\id\ot \varphi))\\
        \phi(M,r_M)\ot_A \phi(N,r_N)=(M,r_M\circ(\id\ot \varphi))\ot_A(N,r_N\circ(\id\ot \varphi)).
    \end{align*}
    Since the following diagram commutes,
    \begin{center}
        \begin{codi}
            \obj[tetragonal=base 11em height 3.5em]{
                |(1)|M\ot A\ot N \ot A & |(2)|M\ot A\ot N \ot A\\
                |(3)|M\ot A\ot N \ot A & |(4)|M\ot A\ot N \ot A\\
                |(AN)|M\ot A\ot N & |(AN')|M\ot A\ot N \\ 
            };
            \mor: 1 "\id_M\ot\varphi\ot\id_{AN}":-> 2 "\id_{MAN}\ot\varphi":-> 4 "\id_{MA}\ot r_N":-> AN';
            \mor[swap]: * "\id_{MAN}\ot\varphi":-> 3 "\id_{MA}\ot r_N":-> AN "\id_M\ot \varphi\ot\id_{N}":-> *;
            \mor: 3 "\id_M\ot \varphi\ot\id_{NA}":-> 4;
        \end{codi}
    \end{center}
    then their coequalizer diagram must commute, which is the diagram of right $A$-module isomorphisms between $\phi((M,r_M)\ot_A(N,r_N))$ and $\phi(M,r_M)\ot_A \phi(N,r_N)$.
    Note that the natural isomorphism $\nabla=\id$.
\end{proof}

\begin{prop}\label{prp:preserves_locality}
    Let $(M,r_M)$ be a right $A$-module.
    If $M$ is local, then $(M,r_M\circ (\id_M\ot\varphi))$ is still local.
\end{prop}
\begin{proof}
    The following diagram commutes naturally.
    \begin{center}
        \begin{codi}
            \obj[tetragonal=base 6em height 3.5em]{
                |(MA)|M\ot A & & |(AM)|A\ot M\\
                |(MA')|M\ot A & |(AM')|A\ot M & \\
                M & |(MA'1)| M\ot A & |(MA1)|M\ot A\\
            };
            \mor: MA "\beta_{M,A}":-> AM "\beta_{A,M}":-> MA1 "\id\ot\varphi":-> MA'1 "r_M":-> M;
            \mor[swap]: * "\id\ot\varphi":-> MA' "r_M":-> *;
            \mor: MA' "\beta_{M,A}":-> AM' "\beta_{A,M}":-> MA'1;
            \mor: AM "\varphi\ot\id":-> AM'; 
        \end{codi}
    \end{center}
\end{proof}

\begin{lem}\label{lem:phi_is_br_auto}
    Let $A$ be a commutative algebra in $\CC$ and let $\varphi$ be an automorphism of $A$.
    The monoidal autoequivalence $\phi$ induced by $\varphi$ is a braided autoequivalence when restrict to $\CC_A^{loc}$.
    \begin{align*}
        \phi:\CC_A^{loc}&\to \CC_A^{loc}\\
        (M,r_M)&\mapsto (M,r_M\circ (\id_M\ot\varphi))    
    \end{align*}
\end{lem}
\begin{proof}
    By Proposition \ref{prp:preserves_locality}, restrict $\phi:\CC_A\to \CC_A$ to $\CC_A^{loc}$, we obtain an autoequivalence.
    The monoidal structure of $\CC_A^{loc}$ is inherited from that of $\CC_A$, by Lemma \ref{lem:phi_is_monoidal}, we obtain a monoidal autoequivalence $\phi:\CC_A^{loc}\to \CC_A^{loc}$.
    Since the natural isomorphism $\nabla_{-,-}$ of monoidal functor $\phi$ is $\id$, so it is automatically a braided autoequivalence.
\end{proof}

In other words, Lemma \ref{lem:phi_is_monoidal} and Lemma \ref{lem:phi_is_br_auto} tells us $\Aut(A)$ has an action on $\Aut_{E_1}(\CC_A)$ and on $\Aut_{E_2}(\CC_A^{loc})$.
However, the $\Aut(A)$-action on $\Aut_{E_2}(\CC_A^{loc})$ is not free in general. In other words, $\phi \in \CC_A^{loc}$ induced by $\varphi \in \Aut(A)$ may not be non-trivial.

\begin{prop}\label{prp:fix_free_mod}
    For any free\footnote{Note that the {\bf free module} is not the same as the {\bf free action}.} right $A$-module $(X\ot A, \id_X\ot m_A)$, any $\varphi\in\Aut(A)$ fixes it.
\end{prop}
\begin{proof}
    The following diagram commutes 
    \begin{center}
        \begin{codi}
            \obj[tetragonal=base 8.5em height 3.5em]{
                |(AA)|X\ot A\ot A & |(AA1)|X\ot A\ot A\\
                & |(AA2)| X\ot A\ot A\\
                |(A1)| X\ot A & |(XA)|X\ot A\\
            };
            \mor: AA "\id\ot\varphi\ot \id":-> AA1 "\id\ot \id\ot\varphi":-> AA2 "\id\ot m":-> XA;
            \mor[swap]: * "\id\ot m":-> A1 "\id\ot\varphi":-> *;
        \end{codi}
    \end{center}
    which is the diagram \ref{diag:alg_homo} tensoring with $X$.
\end{proof}

\begin{cor}\label{crl:free_local_are_fixed}
    If $\CC_A^{loc}$ all consists of free local $A$-modules, then for any $\varphi \in \Aut(A)$, $\phi=\Id$.
\end{cor}

\begin{expl}
    Here we provide some trivial examples.
    \begin{itemize}
        \item The lagrangian algebra $\one\oplus \bfe$ in $\fZ(\vect_{\bZ_2})$ has a non-trivial automorphism $\varphi$.
        But the condensed phase $\fZ(\vect_{\bZ_2})_{\one\oplus \bfe}^{loc}\simeq \vect$ consists of all free local modules, by Corollary \ref{crl:free_local_are_fixed}, $\varphi$ induces the trivial braided autoequivalence $\Id$ in $\vect$. In general, for $G$ abelian, all condensed phases of $\fZ(\vect_G)$ consist of free modules, so any algebra automorphism $\varphi \in \Aut(A)$ for $A \in \Alg_{E_2}^{cond}(\fZ(\vect_G))$ induce trivial
        braided autoequivalence in $\fZ(\vect_G)_A^{loc}$.

        \item By \cite{LY23}, the $\bZ_{2k+1}$ double parafermion $\fZ(\mathrm{PF}_{2k+1})$ can condense to $\bZ_{2k+1}$ topological order, i.e. $\fZ(\mathrm{PF}_{2k+1})_{A_0}^{loc}\simeq \fZ(\vect_{\bZ_{2k+1}})$ for some 2d condensable algebra $A_0$, in which all simple objects in $\fZ(\mathrm{PF}_{2k+1})_{A_0}^{loc}$ are free modules.
        Thus, by Corollary \ref{crl:free_local_are_fixed}, any automorphism of $A_0$ will trivially act in $\fZ(\vect_{\bZ_{2k+1}})$.
        Or to say, the braided autoequivalence in $\fZ(\vect_{\bZ_{2k+1}})$ is not induced by automorphism of $A_0$.
    \end{itemize}
    
\end{expl}



\begin{expl}
    Consider the case when Double Ising $\fZ(\Is)$ condense to $\Zb_2$ topological order \cite{CJKYZ20} (see also section \ref{sec:expl_double_is}), in which the $E_2$ condensable algebra can be written as $A_2:=(\one \boxtimes \overline{\one}) \oplus (\psi \boxtimes \overline{\psi})$,
    i.e. $\fZ(\Is)_{A_2}^{loc}\simeq \fZ(\vect_{\bZ_2})$.
    For the four simple objects in $\fZ(\Is)_{A_2}^{loc}$: two local $A_2$-modules $(\one\boxtimes \overline{\one})\ot A_2\mapsto \one$ and $(\psi\boxtimes \overline{\psi})\ot A_2\mapsto \bff $ are free; the other two local $A_2$-modules $\sigma\boxtimes \overline{\sigma}:=(\sigma\boxtimes \overline{\sigma},r)\mapsto \bfe$ and $(\sigma\boxtimes\overline{\sigma})^{tw}:= (\sigma\boxtimes\overline{\sigma}, r \circ (\id\ot \varphi))\mapsto \bfm $ are not.

 There is a non-trivial algebra automorphism $\varphi$ of $A_2$ given by 
    \begin{align*}
        (\one \boxtimes \overline{\one}) \oplus (\psi \boxtimes \overline{\psi})\xrightarrow{1\oplus -1}(\one \boxtimes \overline{\one}) \oplus (\psi \boxtimes \overline{\psi})
    \end{align*}
    It is clear that $\varphi^2=\id_{A_2}$. 
    By Lemma \ref{lem:phi_is_br_auto}, this non-trivial automorphism induce a braided autoequivalence $\phi$ in $\fZ(\vect_{\bZ_2})$.
    Based on Proposition \ref{prp:fix_free_mod}, two free local $A_2$ modules $\one$ and $\bff$ are invariant under the action of $\phi$.
    However, the two non-free local $A_2$-modules $\bfe$ and $\bfm$ exchange.
\end{expl}

The automorphism of algebras also affects two-step condensations.
Let $i:A\hookrightarrow A'$ be an inclusion of 2d condensable algebras in a MTC $\CC$.
This inclusion determines a two-step condensation process. 
However, when $A$ admits non-trivial automorphisms, by composing with one automorphism $\varphi:A\to A$, we will obtain another inclusion $i\circ \varphi=i':A\hookrightarrow A'$.
Since $A$-module action on $A'$ may change due to $\varphi$ such that $A'$ becomes a different object in $\CC_A^{loc}$, $i'$ may lead to a different two-step condensation process. 
However, when $\varphi$ would lead to different 2-step condensation is not clear, which is worthy to be explored.
The example below gives a situation that $\varphi$ generates different condensable algebras in condensed phase. In Sec. \ref{sec:expl_S3}, we also provide an example of $S_3$ topological order in which $\varphi$ generates the same condensable algebra.



\begin{expl}\label{proof:exp_DIS}
    Again considering Double Ising condense to $\bZ_2$ topological order.
    The obvious inclusion $i:A_2\hookrightarrow A_L$ determines a 2-step condensation process.
    The lagrangian algebra $A_L:=(\one \boxtimes \overline{\one}) \oplus (\sigma\boxtimes \overline{\sigma})\oplus (\psi \boxtimes \overline{\psi})$ should become a lagrangian algebra in $\fZ(\Is)_{A_2}^{loc}\simeq \fZ(\vect_{\bZ_2})$. It is clear its components $(\one \boxtimes \overline{\one})\oplus (\psi \boxtimes \overline{\psi})$ corresponds to $\one\in \fZ(\vect_{\bZ_2})$ and the component $(\sigma\boxtimes \overline{\sigma})$ corresponds to $\bfe$ (or $\bfm$, depending on the convention).
    So $A_L$ becomes the lagrangian algebra $\one\oplus \bfe$ (or $\one\oplus\bfm$) in $\fZ(\vect_{\bZ_2})$.
    After composing $i:A_2\hookrightarrow A_L$ with $\varphi$, we obtain a new two-step condensation $i':A_2\hookrightarrow A_L$.
    The component $(\one \boxtimes \overline{\one})\oplus (\psi \boxtimes \overline{\psi})$ is invariant and still corresponds to $\one$, but the component $(\sigma\boxtimes \overline{\sigma})$ becomes $(\sigma\boxtimes \overline{\sigma})^{tw}$, which corresponds to $\bfm$ (or $\bfe$) now.
    Hence, $A_L$ becomes another lagrangian algebra $\one \oplus \bfm$ (or $\one\oplus \bfe$) in $\fZ(\vect_{\bZ_2})$ under this condensation process. 
\end{expl}

Conversely, two-step condensation tells us that different condensable algebras in the condensed phase $\CC_{A}^{loc}$ may have the same extension. More precisely, $(A',i)$ and $(A',i')$ in $\CC_A^{loc}$ induced by $i$ and $i'=i \circ \varphi$ comes from the same $A' \in \CC$.
This can also be explained by the automorphisms of $A$.
We rigorously explain this phenomena by proving the following theorem:


\begin{thm}\label{thm:phi_alg_iso}
    Let $\CC$ be a MTC, and $A$ be a 2d condensable algebra in $\CC$.
    Let $\phi\in\Aut_{E_2}(\CC_{A}^{loc})$ be a braided autoequivalence in $\CC_{A}^{loc}$ induced by an automorphism $\varphi\in\Aut(A)$ of $A$.
    If two 2d condensable algebras $(A_1,m_1:A_1\ot_A A_1\to A_1)$ and $(A_2,m_2:A_2\ot_A A_2\to A_2)$ in $\CC_A^{loc}$ are connected by $\phi$, i.e. $\phi(A_1)\cong A_2$ or $\phi(A_2)\cong A_1$.
    Then the extensions of $A_1$ and $A_2$ over $A$ are isomorphic in $\CC$.
\end{thm}
\begin{proof}
    Let $A_0$ denote the image of $A_1$ and $A_2$ under forgetful functor $U:\CC_A^{loc}\to\CC$.
    Suppose $A_1=(A_0,r_1)$, then we have $\phi(A_1)=(A_0,r_1\circ (\id\ot \varphi))$.
    Let $f:\phi(A_1)\to A_2$ be the algebra isomorphism in $\CC_A^{loc}$, i.e. the following diagram commutes in $\CC_A^{loc}$.
    \begin{align*}
        \xymatrix{
            \phi(A_1)\ot_A\phi(A_1)\ar[r]^{\phi(m_1)}\ar[d]_{f\ot_A f} & \phi (A_1)\ar[d]^{f}\\
            A_2\ot_A A_2\ar[r]_{m_2} &A_2
        }
    \end{align*}
    Note that $\phi(m_1)=m_1$, so by applying forgetful functor $U$ on this diagram, we obtain the following commutative diagram
    \begin{align*}
        \xymatrix{
            A_0\ot_A A_0\ar[r]^{m_1}\ar[d]_{f\ot_A f} & A_0\ar[d]^{f}\\
            A_0\ot_A A_0\ar[r]_{m_2} &A_0
        }
    \end{align*}    
    The extension $\Ext^R_{A}(A_i)$ of $A_i$ over $A$ is given by $(A_0,m_i^{ext}:A_0\ot A_0\xrightarrow{p_{A_0,A_0}}A_0\ot_A A_0\xrightarrow{m_i}A_0)$ \cite{Dav10a}. 
    To show that $\Ext^R_{A}(A_1)$ is isomorphic to $\Ext^R_{A}(A_2)$, we prove the following diagram commutes 
    \begin{align*}
        \xymatrix{
            A_0\ot A_0\ar[r]^{p_{A_0,A_0}}\ar[d]_{f\ot f} &A_0\ot_A A_0\ar[r]^{m_1}\ar[d]_{f\ot_A f} & A_0\ar[d]^{f}\\
            A_0\ot A_0\ar[r]_{p_{A_0,A_0}} &A_0\ot_A A_0\ar[r]_{m_2} &A_0
        }
    \end{align*}  
    where the left sub-diagram commutes by the universal property of coequalizer.  The case for $\phi(A_2)\cong A_1$ is the same.
\end{proof}

\begin{expl}
    The extension of two lagrangian algebras $\one \oplus \bfe $ and $\one \oplus \bfm$ over $A_2$ are both the unique lagrangian algebra $A_L$ in Double Ising. It is natural because the incarnation of $A_L$ which results $\one \oplus \bfe$ or $\one \oplus \bfm$ is due to the non-trivial automorphism $\varphi$.
\end{expl}

\begin{cor}
    Let $A$ be a condensable algebra in $\CC$.
    Let $A_1$ and $A_2$ be two condensable algebra in $\CC_{A}^{loc}$ that are connected by a braided autoequivalence $\phi$ induced by an automorphism on $A$, i.e. $\phi(A_1)\cong A_2$.
    Then we have $\CC_{A}\btd_{\CC_{A}^{loc}}(\CC_{A}^{loc})_{A_1}\simeq \CC_{A}\btd_{\CC_{A}^{loc}}(\CC_{A}^{loc})_{A_2}$ as left monoidal $\CC$-module categories.
\end{cor}
This tells us that the fusion of $\CC_{A}$ with $(\CC_{A}^{loc})_{A_1}$ and $(\CC_{A}^{loc})_{A_1}$ results in a same domain wall $\CC_{A'}$.

In particular, let $A_1\boxtimes A_2$ be a condensable algebra in $\CC\boxtimes \overline{\CC}$. Then the autoequivalence (or invertible domain wall $\Phi_{\varphi}$) induced by the automorphism $\varphi \in \Aut(A_1)$ (or $\Aut(A_2)$) does not affect the classification of lagrangian algebras $L \in \Alg_{E_2}^{lag}(\CC\boxtimes \overline{\CC})$. More precisely, we have $\phi (L_{\phi_1}) \simeq L_{\phi_2}$ in which $L_{\phi_2}$ corresponds to $\Phi_{\varphi} \boxtimes_{\CC\boxtimes \overline{\CC}_{A_1\boxtimes A_2}^{loc}} \Phi_1$, and we got 
\begin{align*}
    (\CC\boxtimes \overline{\CC})_{A_1\boxtimes A_2}\btd_{(\CC\boxtimes \overline{\CC})_{A_1\boxtimes A_2}^{loc}}((\CC\boxtimes \overline{\CC})_{A_1\boxtimes A_2}^{loc})_{L_{\phi_1}}\simeq (\CC\boxtimes \overline{\CC})_{A_1\boxtimes A_2}\btd_{(\CC\boxtimes \overline{\CC})_{A_1\boxtimes A_2}^{loc}}((\CC\boxtimes \overline{\CC})_{A_1\boxtimes A_2}^{loc})_{L_{\phi_2}}\simeq (\CC\btd \overline{\CC})_L.
\end{align*}

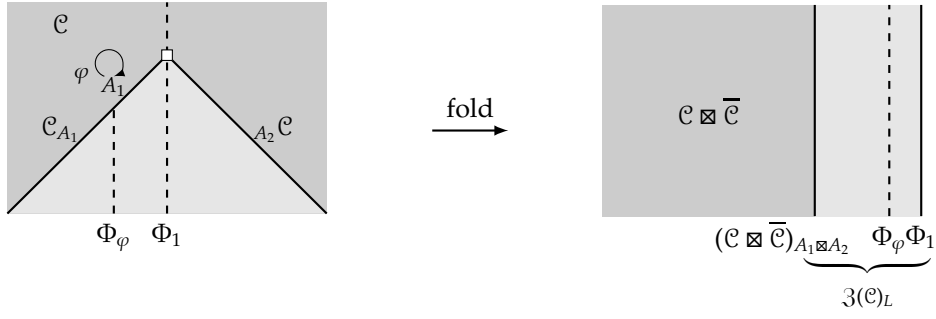
\begin{figure}[H]
    \centering
        \begin{minipage}[t]{0.43\linewidth}
        \centering
        \begin{tikzpicture}[scale=1.4]
            \filldraw[fill=gray!40, draw=none] (-1.5,0) rectangle (1.5,2);
            \filldraw[fill=gray!20, draw=none] (-1.5,0)--(1.5,0)--(0,1.5)--(-1.5,0);
            \draw[thick, dashed] (0,0) -- (0,2);
            \draw[thick](-1.5,0)--(0,1.5);
            \draw[thick](1.5,0)--(0,1.5);
            \node at (1,0.8) {${}_{A_2} \CC$};
            \node at (-0.5,1.2) {\footnotesize $A_1$};
            \draw[thick, dashed] (-0.5,0) -- (-0.5,1);
            \node at (-1,0.8) {$\CC_{A_1}$};
            \draw[latex-] (-0.5,1.3) arc(-70:250:0.13);
            \node[]at(-0.8, 1.3){\footnotesize $\varphi$};
            \node at (-1,1.8) {$\CC$};
            \node at (0,-0.2) {$\Phi_1$};
            \node at (-0.5,-0.23) {$\Phi_{\varphi }$};
            \draw[draw=none] (0,0) -- (0,-0.4);
            \filldraw[fill=white] (-0.05, 1.45)rectangle(0.05,1.55);
            \draw[thick, draw=none](-1, -1)--(-0.5,-1);
        \end{tikzpicture}
        \end{minipage}
    \begin{tikzpicture}
        \draw[-latex, thick] (-0.5,0) -- (0.5,0);
        \node at(0,0.3) {fold};
        \draw[draw=none] (0,0)--(0,-2.5);
    \end{tikzpicture}
    \begin{minipage}[t]{0.43\linewidth}
        \centering
        \begin{tikzpicture}[scale=1.4]
            \filldraw[draw=none, fill=gray!40] (-3, 0)rectangle(-1,2);
            \filldraw[draw=none, fill=gray!20] (-1, 0)rectangle(0,2);
            \draw[thick](-1, 0)--(-1,2);
            \draw[thick](0, 0)--(0,2);
            \draw[thick, dashed](-0.3, 0)--(-0.3,2);
            \node[]at(-2, 1){$\CC\boxtimes \overline{\CC}$};
            \node[]at(0, -0.2){$\Phi_1$};
            \node[]at(-0.3, -0.23){$\Phi_{\varphi}$};
            \node[]at(-1.3, -0.2){$(\CC\boxtimes \overline{\CC})_{A_1\boxtimes A_2}$};
            \node[]at(-0.5, -0.6){\large $\underbrace{\qquad \qquad }_{\fZ(\CC)_L}$};
        \end{tikzpicture}
    \end{minipage}    
    \caption{Invertible domain wall $\Phi_{\varphi}$ induced by the automorphism $\varphi \in \Aut(A_1)$  does not affect the classification of gapped boundaries given by $L$ in $\FZ(\CC) \simeq \CC\boxtimes \overline{\CC}$, so does the gapped domain walls given by 1d condensable algebras $B$ in $\CC$.}
\end{figure}

Now suppose $\Phi$ is an invertible domain wall in $\CC_{A}^{loc}$ connecting two lagrangian algebras $L_1$, $L_2\in\Alg_{E_2}^{lag}(\CC_A^{loc})$, i.e. $\phi(L_1)\cong L_2$, such that the fused boundaries $\CC_{A}\btd_{\CC_A^{loc}}(\CC_A^{loc})_{L_1}$ and $\CC_{A}\btd_{\CC_A^{loc}}(\CC_A^{loc})_{L_2}$ are the same.
Or to say, the extension $\Ext^R_A(L_1)$ and $\Ext^R_A(L_2)$ are algebraically isomorphic in $\CC$.
We denote their extensions by $L\in \Alg_{E_2}^{lag}(\CC)$, the algebra isomorphism $f: \Ext^R_A(L_1)\to \Ext^R_A(L_2)$ induces an algebra automorphism on $L$.
It is clear that $A$ is a subalgebra of $L$, so we can restrict $f$ on $A$ to obtain an algebra automorphism $\varphi$ on $A$.
Thus, this invertible domain wall $\Phi$ should be induced by $\varphi$.
We have argued that for a 2-step condensation $A\hookrightarrow A'$, all the redundant invertible domain walls in condensed phase should be induced by an algebra automorphism $\varphi\in\Aut(A)$.
Therefore, it is reasonable to obtain the complete classification of domain walls $\CC_{A_1}\btd_{\CC_{A_1}^{loc}}\Phi \btd_{\CC_{A_2}^{loc}}{}_{A_2} \CC$ by modding out $\Phi_{\varphi}$ given by the subgroup generated by $\Aut(A_1)$ in $\Aut_{E_2}(\CC_{A_1}^{loc})$.

In section \ref{sec:expl_S3}, we find an issue related to $\phi$ of DNO's classification \cite[Theorem 3.6]{DNO12} of 2d condensable algebras in $\CC\boxtimes \CD$.
We will use Theorem \ref{thm:phi_alg_iso} to the $S_3$ topological order to explain why DNO's statement is not rigorous enough.


\section{Examples}\label{section:lattice_model}
In this section we give some physical examples to exhibit the utility of Arrows in Trinity \ref{fig:alg_cycle}.
In section \ref{section:toric_code}, we explicitly compute left and right centers of 1d condensable algebras in $\bZ_2$ topological order $\fZ(\vect_{\bZ_2})$ and lagrangian algebras in $\fZ(\vect_{\bZ_2})\btd\overline{\fZ(\vect_{\bZ_2})} \simeq \fZ(\vect_{\bZ_2 \times \bZ_2})$, in which we demonstrate that by using \Ar{2} and \Ar{3}, we can obtain 2-Morita equivalent condensable algebras.
We also illustrate all other Arrows precisely in this example.
Realizations of 1d condensable algebras and their left/right centers together with invertible domain walls are also constructed on toric code model.
In section \ref{sec:Z4} with $\Zb_4$ topological order, besides computing left/right centers of 1d condensable algebras to obtain all 2-Morita equivalent condensable algebras, we also illustrate non-trivial examples of \Ar{6}.
A general computation method of left/right centers for abelian group cases called Kreuzer-Schellekens bicharacters is also mentioned in this subsection.
In section \ref{sec:expl_S3}, we review the characters in $\FZ(\vect_{G})$ and use them to compute lagrangian algebras in $\FZ(\vect_{G})\btd\overline{\FZ(\vect_{G})}$ to obtain 2-Morita equivalent condensable algebras, this method can be applied to any finite group $G$. We illustrate this method in case of $\S_3$ topological order.
In the end of this section, we discuss some examples beyond group-like gauge symmetries, in which we illustrate the impact of algebra automorphism through Double Ising topological order.

\subsection{Group gauge symmetries \texorpdfstring{$\FZ(\vect_{G})$}{Z(VecG)}}\label{section:ZVec_G}

We first discuss MTCs with traditional gauge symmetries, namely, these topological orders that can be described by $\FZ(\vect_{G}) \simeq \FZ(\rep(G))$ with some finite group $G$.
This kind of $2$d bulk phase can also be realized by the Kitaev quantum double model \cite{Kit03}.
In \cite{Dav10a}, Davydov classifies condensable algebras in $\FZ(\vect_{G})$:

\begin{thm}\label{thm:classification_lag_alg_ZVecG}
  	A $2$d-condensable algebra $A(H, F,\omega, \epsilon)$ in $\FZ(\vect_{G})$ is determined by a subgroup $H\subset G$, a normal subgroup $F\lhd H$, a 2-cocycle $\omega \in Z^2(F,\mathbb{C}^{\times})$ and $\epsilon : H \times F \rightarrow  \mathbb{C}^{\times}$ satisfying some axioms. (More precisely, $A(H, F,\omega, \epsilon ) = \fun(G)\otd_{\bC[H]}\bC[F,\omega,\epsilon]$, see appendix \ref{appendix:condensable_algebras} for the detail conditions). 

	The algebra $A(H,F,\omega, \epsilon)$ is lagrangian if and only if $F=H$. 
	In this case, $\epsilon$ is uniquely determined by $\omega$. 
	Or to say, a lagrangian algebra in $\FZ(\vect_{G})$ is determined by a pair $(H \subset G,\omega)$, where $\omega\in \mathrm{H}^2(H,\bC^{\times})$ is a 2-cohomology class.
\end{thm}


Using this classification theorem of 2d condensable algebras in $\FZ(\vect_{G})$, we can obtain all 2-Morita equivalent condensable algebras in $\FZ(\vect_{G})$ through A$\overrightarrow{\text{rrow}}$ 2 in Trinity \ref{fig:alg_cycle}:
By Theorem \ref{thm:classify_by_lag_alg}, first we need to know the classification of lagrangian algebras in $\fZ(\FZ(\vect_{G}))$.
Since $\FZ(\vect_{G})$ is an MTC, we have $\fZ(\FZ(\vect_{G}))\simeq \FZ(\vect_{G})\btd \overline{\FZ(\vect_{G})}\simeq \fZ(\vect_{G\times G})$ \cite{Dav10a}\footnote{This equivalence is related to which boundary of $\FZ(\vect_{G})\btd \overline{\FZ(\vect_{G})}$ is viewed as canonical.}. 
Then we can use Theorem \ref{thm:classification_lag_alg_ZVecG} to find all lagrangian algebra in $\fZ(\vect_{G\times G})$.
After that, we can obtain all 2-Morita equivalent condensable algebras in $\FZ(\vect_{G})$ by intersecting all lagrangian in $\fZ(\vect_{G\times G})$ with left/right components of $\FZ(\vect_{G})\btd \overline{\FZ(\vect_{G})}$ (Corollary \ref{crl:lag_alg_to_2-Morita_algs}).

In some situations, it is easier to find all 1d condensable algebras in $\FZ(\vect_{G})$. 
For abelian group $G$, 1d condensable algebras in $\fZ(\vect_G)$ can be written explicitly as a twisted group algebra $\bC[H,\omega]$, where $H$ is a subgroup of $G\times G$ and $\omega\in \mathrm{H}^2(H,\bC^{\times})$ \cite{Ost03}.
Hence, for abelian group $G$, we can compute left/right centers of these twisted group algebras to obtain 2-Morita equivalent condensable algebras directly, i.e. we can classify the 2d condensable algebras in $\FZ(\vect_{G})$ using A$\overrightarrow{\text{rrow}}$ 3 in Trinity \ref{fig:alg_cycle}. 

We illustrate some explicit examples using both methods in this subsection.

\subsubsection{Toric code model \texorpdfstring{$\fZ(\vect_{\bZ_2})$}{Z(vecZ2)}}\label{section:toric_code}
We start from the most well known topological order, the 2d toric code model \cite{Kit03} described categorically by MTC $\FZ(\vect_{\Zb_2}) \simeq \FZ(\rep(\Zb_2))$ (or we simply denoted by $\TC$). Here we can translate the abstract process in the Trinity \ref{fig:alg_cycle} into concrete lattice models.

We consider a square lattice where each edge has a 1/2 spin.
The local Hilbert space for each edge $e$ is $\cH_e=\bC^2$ and the total Hilbert space $\cH_{tot}=\bigotimes_{e}\cH_e$.
For each vertex $v$ we define a vertex operator  $\sA_v := \prod_e \sigma_x^e $ acting on adjacent edges; For each plaquette $p$, we define a dual operator $\sB_p := \prod_{e'} \sigma^{e'}_z$. Here $\sigma_x^e$ and $\sigma_z^e$ are Pauli matrices acting on edge $e$.

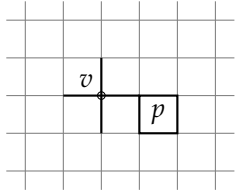
\begin{figure}[H]
  \centering
  \begin{tikzpicture}
  \foreach \i in {1, ...,5} {
      \draw [gray] (-0.25, \i * 0.5) -- (2.75, \i * 0.5);
  }
  \foreach \i in {0, ...,5} {
      \draw [gray] (\i * 0.5, 0.25) -- (\i * 0.5, 2.75);
  }
  \draw [thick](0.5, 1.5)--(1.5, 1.5);   
  \draw [thick](1, 1)--(1, 2); 
  \draw [black] (1,1.5) circle (0.05); 
  \node[]at(0.8, 1.7){$v$};
  \draw [thick](1.5, 1)rectangle(2, 1.5);   
  \node[]at(1.75, 1.25){$p$};
  \end{tikzpicture}
\end{figure}

And the Hamiltonian is defined to be 
$$\mathsf{H}:=\sum_v (1-\mathsf{A}_v) + \sum_p (1-\mathsf{B}_p).$$
In this lattice model, we have 4 simple objects (topological excitations): $\one, \mathbf{e}, \mathbf{m}, \mathbf{f}$, where $\bfe$ is the $\bZ_2$-charge and $\bfm$ is the $\bZ_2$-flux.
Their fusion rules are $\bfe\ot\bfe\cong \bfm\ot\bfm\cong \bff\ot\bff\cong\one$, $\bfe\ot\bfm\cong\bff\cong\bfm\ot\bfe$.
The non-trivial braiding is generated by $\beta_{\bfe,\bfm}=-1$.
It is well-known that there are three 2d condensable algebras in $\TC$: the trivial condensable algebra $\one$, and two lagrangian algebras $\one\oplus\bfe$ and $\one\oplus\bfm$, which correspond to the rough boundary $\vect_{\bZ_2}$ and the smooth boundary $\rep(\bZ_2)$ respectively.
So we have $\one\oplus\bfe\widesim[3]{2-Morita}{}\one\oplus\bfm$.
Although $\TC$ is a simple example, it is clear enough to show the power of our methods.

We first use A$\overrightarrow{\text{rrow}}$ 3 in Trinity \ref{fig:alg_cycle} to classify the 2-Morita equivalent $E_2$ condensable algebras in $\TC$.
There are six 1d condensable algebras $\bC[H,\omega]$ in $\TC$: five correspond to five subgroups $H$ of $\bZ_2\times \bZ_2$, the other one corresponds to the non-trivial 2-cohomology class of $\bZ_2\times \bZ_2$.
For example, $\bC[\bZ_2^{\bff}]=\one\oplus \bff$ corresponds to subgroup $\bZ_2^{\bff}$ generated by $\one, \bff$.
We list all the six 1d condensable algebras in the first column of Table \ref{TC_table} below. 

\begin{table}[H]
  \centering
  \begin{tabular}{|c|c|c|c|}
          \hline
          $B_i\in \TC$&$Z_l(B_i)/Z_r(B_i)$ & Domain wall&  Lagrangian algebras $L_i\in\TC\btd\overline{\TC}$\\
          \hline
          $\one$& $ \one/\one $ & trivial wall & $ \one\overline{\one}\oplus \bfe\overline{\bfe}\oplus \bfm\overline{\bfm}\oplus \bff\overline{\bff} $ \\
          \hline
          $ \one\oplus \bff $ & $ \one/\one $ & $\bfe-\bfm$ exchange  &$\one\overline{\one}\oplus \bfm\overline{\bfe}\oplus \bfe\overline{\bfm}\oplus \bff\overline{\bff} $ \\
          \hline
          $\one\oplus \bfe $ & $ \one\oplus \bfe / \one\oplus \bfe  $& $\vect_{\bZ_2}\btd \vect_{\bZ_2}$ & $ \one\overline{\one}\oplus \bfe\overline{\one}\oplus \one\overline{\bfe}\oplus \bfe\overline{\bfe}$ \\
          \hline
          $ \one\oplus \bfm  $ & $ \one \oplus \bfm/\one \oplus \bfm $ &${\rep}(\bZ_2)\btd {\rep}(\bZ_2) $ &$ \one\overline{\one}\oplus \bfm\overline{\one}\oplus\one \overline{\bfm}\oplus \bfm\overline{\bfm} $\\
          \hline
          $ \one\oplus \bfe \oplus \bfm \oplus \bff  $ & $ \one \oplus \bfe/ \one \oplus \bfm $& $\vect_{\bZ_2}\btd \rep(\bZ_2)$ &$ \one\overline{\one}\oplus \bfe\overline{\one}\oplus \one\overline{\bfm}\oplus \bfe\overline{\bfm} $\\
          \hline
          $ \one\oplus \bfe \oplus \bfm \oplus \bff, \omega$&$ \one \oplus \bfm/\one \oplus \bfe $& $\rep(\bZ_2)\btd \vect_{\bZ_2} $ &$ \one\overline{\one}\oplus \bfm\overline{\one}\oplus \one\overline{\bfe}\oplus \bfm\overline{\bfe} $\\
          \hline
  \end{tabular}
  \caption{table of $\TC$, $i\in{1,2,3,4,5,6}$}\label{TC_table}
\end{table}

\begin{figure}[H]
    \centering
    \setcounter {subfigure} {0}
    \subfigure[]{
        \begin{minipage}[t]{0.14\linewidth}
        \centering
        \begin{tikzpicture}[scale=0.8]
            \draw[gray] (0,0) -- (0,4);
            \draw[gray] (-1,1) -- (1,1);
            \draw[gray] (-1,2) -- (1,2);
            \draw[gray] (-1,3) -- (1,3);
            \draw[draw=none](-0.1,-0.5)--(0.1, -0.8);
            \node at(0,4.4){$\one$};
            \node at(0,-0.5){$1$};
            \node at(0.35,2.15){\footnotesize $(0,0)$};
        \end{tikzpicture}
        \end{minipage}
    }
    \subfigure[]{
        \begin{minipage}[t]{0.14\linewidth}
        \centering
        \begin{tikzpicture}[scale=0.8]
            \draw[gray] (0,0) -- (0,4);
            \draw[dashed,very thick] (-1,2) -- (0,2);
            \draw[dashed,very thick] (-1,1) -- (0,1);
            \draw[gray] (-1,3) -- (0,3);
            \draw[gray] (-0,3.5) -- (1,3.5);
            \draw [dashed,very thick](0,1.5)--(1,1.5);
            \draw [dashed,very thick](0,2.5)--(1,2.5);
            \draw [dashed,very thick](-1,1)--(-1,2);
            \draw [dashed,very thick](1,1.5)--(1,2.5);
  
            \node at(0,4.4){$\one\oplus \bff$};
            \node at(0,-0.5){$\sC_{-\frac{1}{2},-\frac{1}{2}} + \sD_{\frac{1}{2},0}$};
        \end{tikzpicture}
        \end{minipage}
    }
    \subfigure[]{
        \begin{minipage}[t]{0.14\linewidth}
        \centering
        \begin{tikzpicture}[scale=0.8]
            \draw[gray] (0,0) -- (0,4);
            \draw[gray] (-1,1) -- (1,1);
            \draw[very thick] (-1,2) -- (1,2);
            \draw[very thick] (0,1) -- (0,3);
            \draw[gray] (-1,3) -- (1,3);
            \node at(0,4.4){$\one\oplus \bfe$};
            \node at(0,-0.5){$\sA_{v_{0,0}}$};
        \end{tikzpicture}
        \end{minipage}
    }
    \subfigure[]{
        \begin{minipage}[t]{0.14\linewidth}
        \centering
        \begin{tikzpicture}[scale=0.8]
            \draw[gray] (-0.5,0) -- (-0.5,4);
            \draw[gray] (0.5,0) -- (0.5,4);
            \draw[gray] (-1,1) -- (1,1);
            \draw[gray] (-1,2) -- (1,2);
            \draw[gray] (-1,3) -- (1,3);
            \draw [very thick](-0.5, 1)rectangle(0.5, 2);  
            \node at(0,4.4){$\one\oplus \bfm$};
            \node at(0,-0.5){$\sB_{p_{\frac{1}{2},-\frac{1}{2}}}$};
        \end{tikzpicture}
        \end{minipage}
    }
    \subfigure[]{
        \begin{minipage}[t]{0.14\linewidth}
        \centering
        \begin{tikzpicture}[scale=0.8]
            \draw[gray] (1,0) -- (1,4);
            \draw[gray] (0,0) -- (0,4);
            \draw[gray] (-1,1) -- (1.25,1);
            \draw[gray] (-1,2) -- (1.25,2);
            \draw [very thick](0, 1)rectangle(1, 2);  
            \draw [very thick](-1,2)--(0,2);
            \draw [very thick](0,2)--(0,3);
            \draw[gray] (-1,3) -- (1.25,3);
            \node at(0,4.4){$\one\oplus \bfe \oplus \bfm \oplus \bff$};
            \node at(0,-0.5){$\sA_{v_{0,0}} + \sB_{p_{\frac{1}{2},-\frac{1}{2}}}$};
        \end{tikzpicture}
        \end{minipage}
    }
    \subfigure[]{
        \begin{minipage}[t]{0.14\linewidth}
        \centering
        \begin{tikzpicture}[scale=0.8]
            \draw[gray] (-1,0) -- (-1,4);
            \draw[gray] (0,0) -- (0,4);
            \draw[gray] (-1.25,1) -- (1,1);
            \draw[gray] (-1.25,2) -- (1,2);
            \draw [very thick](-1, 1)rectangle(0, 2);  
            \draw [very thick](0,2)--(1,2);
            \draw [very thick](0,2)--(0,3);
            \draw[gray] (-1.25,3) -- (1,3);
            \node at(0,4.4){$\one\oplus \bfm \oplus \bfe \oplus \bff, \omega$};
            \node at(0,-0.5){$\sB_{p_{-\frac{1}{2},-\frac{1}{2}}} + \sA_{v_{0,0}}$};
        \end{tikzpicture}
        \end{minipage}
    }
\caption{Lattice realizations of 1d condensable algebra $B_i$ in the first column of Table \ref{TC_table} locally. 
Condensing $B_i$ on trivial wall $\TC$ is equivalent to removing these thick edges for all $k$ along the neighborhood of column $0$. 
See the left subfigures of fig. \ref{fig:B_3_B_4}-\ref{fig:e-m_exchange} below.
}\label{fig:1d_alg}
\end{figure}
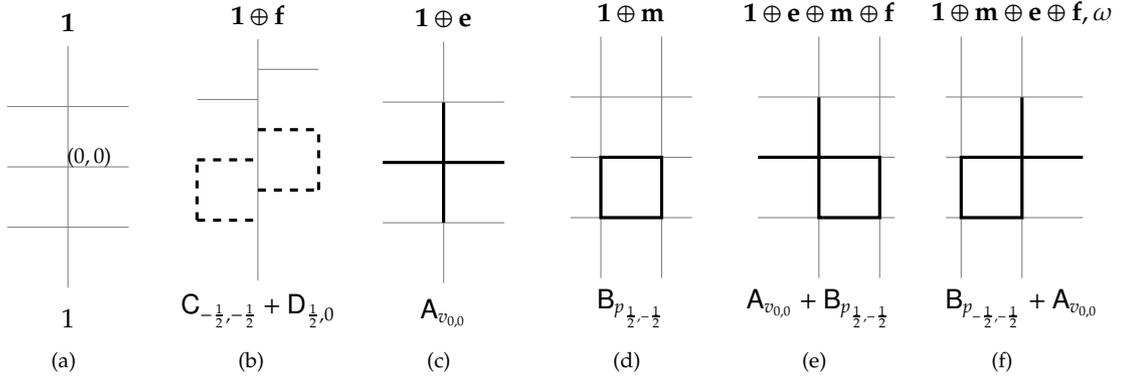

 
Inspired by the method that adding a trap $\sB_{p_0}$ results in a double degenerate state $\one\oplus \bfm$ in \cite{KZ22a}, we develop the lattice model depiction of 1d condensable algebras $B_i$. Suppose each vertex on lattice has a coordinate $(j,k)$ where the column is labeled by $j$ and the row is labeled by $k$, and suppose domain walls are sitting at column $0$. 
For example, given a plaquette $p_{\frac{1}{2},-\frac{1}{2}}$ and a vertex $v_{0,0}$ in the neighborhood of column 0 (see figure \ref{fig:1d_alg} (e)), we can add a local trap $\sA_{v_{0,0}}+\sB_{p_{\frac{1}{2},-\frac{1}{2}}}$ to the original Hamiltonian $\sH$, and obtain a new Hamiltonian:
\begin{align*}
    \sH':=\sH+\sA_{v_{0,0}}+\sB_{p_{\frac{1}{2},-\frac{1}{2}}}=\sum_{v\neq v_{0,0} }(1-\sA_v)+\sum_{p\neq p_{\frac{1}{2},-\frac{1}{2}}}(1-\sB_p)+2
\end{align*}
The new ground state subspace of $\sH'$ is 4-fold degenerate, which can be distinguished by the eigenvalues of $\sA_{v_{0,0}}=\pm 1$ and $\sB_{p_{\frac{1}{2},-\frac{1}{2}}}=\pm 1$.
The state with eigenvalues $\sA_{v_{0,0}}=1$ and $\sB_{p_{\frac{1}{2},-\frac{1}{2}}}=1$ is the ground state of the original Hamiltonian $\sH$ and generates the topological excitation $\one$; the state with $\sA_{v_{0,0}}=1$ and $\sB_{p_{\frac{1}{2},-\frac{1}{2}}}=-1$ generates $\bfm$; the state with $\sA_{v_{0,0}}=-1$ and $\sB_{p_{\frac{1}{2},-\frac{1}{2}}}=1$ generates $\bfe$; and the state with $\sA_{v_{0,0}}=-1$ and $\sB_{p_{\frac{1}{2},-\frac{1}{2}}}=-1$ generates $\bff$.
As a consequence, the topological excitation generated by the local trap $\sA_{v_{0,0}}+\sB_{p_{\frac{1}{2},-\frac{1}{2}}}$ is $\one\oplus \bfe\oplus\bfm\oplus \bff$.

\begin{figure}[H]
    \centering
    \begin{tikzpicture}
    \filldraw[gray!40] (1, 1) rectangle (1.5,1.5);
    \node[]at(1.25,1.25){$\bfm$};
    \node[]at(2.2,0.7){$\bfe$};
    \foreach \i in {1, ...,5} {
        \draw [gray] (-0.25, \i * 0.5) -- (2.75, \i * 0.5);
        \draw [very thick] (1, \i * 0.5) -- (2, \i * 0.5);
        \draw [gray] (\i * 0.5, 0.25) -- (\i * 0.5, 2.75);
    }
    \draw[gray](0,0.25)--(0, 2.75);
    \draw [very thick](1.5, 0.25)--(1.5, 2.75);    
    \draw [very thick](1, 0.5)--(1, 2.5);  
    \draw[latex-latex](4.5, 0.25) -- (4.5, 2.75);
    \node[rotate=90]at(4.2, 1.5){\small Condense};
    \node[rotate=90]at(4.8, 1.5){\small $\one\oplus \bfe \oplus \bfm \oplus \bff,\omega$};
    \node[]at(6.25, 3){\small $\TC_{Z_l(B_6)} \simeq \rep(\Zb_2)$};
    \node[]at(9, 3){\small $\vect_{\Zb_2} \simeq {}_{Z_r(B_6)} \TC$};
    \node[]at(0, 3){\small $\one \oplus \bfm \simeq Z_l(B_6)$};
    \node[]at(2.5, 3){\small $Z_r(B_6) \simeq \one \oplus \bfe$};
    \node[]at(1.25, 0){${}_{B_6} \TC_{B_6}$};
    \filldraw[] (2, 0.5) circle (0.075);
    \draw [](1.9, 0.6)--(1.1, 0.6);
    \draw [](1.1, 0.6)--(1.1, 1.9);
    \draw [-latex](1.1, 1.9)--(1.9, 1.9);
  
    \foreach \i in {1, ...,5} {
        \draw [gray] (6.25, \i * 0.5) -- (7.5, \i * 0.5);
        \draw [gray] (8, \i * 0.5) -- (9.25, \i * 0.5);
    }
    \foreach \i in {0, ...,1} {
        \draw [gray] (\i * 0.5 + 6.5, 0.25) -- (\i * 0.5 + 6.5, 2.75);
        \draw [gray] (\i * 0.5 + 8.5, 0.25) -- (\i * 0.5 + 8.5, 2.75);
    }
    \draw[gray](7.5,0.25)--(7.5, 2.75);
    \node[rotate=270] at(7.75, 1.5){\small $ \vect$};
    \node[]at(7.75, 0){$\rep(\bZ_2)\btd\vect_{\bZ_2}$};
    \node[]at(4.5, -0.9){\small $\sH_{wall} = \sH+ \sum\limits_k \sA_{v_{0,k}}+ \sum\limits_k \sB_{p_{-\frac{1}{2},k-\frac{1}{2}}}=\sum\limits_{v\neq v_{0,k} }(1-\sA_v)+\sum\limits_{p\neq p_{-\frac{1}{2},-\frac{1}{2}}}(1-\sB_p)+2N$};
    \end{tikzpicture}
    \caption{In the upper cases, 
    $B_5$ and $B_6$ located on the domain wall can not expand itself freely into the 2d bulk. 
    In case of $B_5$, $\one\oplus \bfe$ can be expanded to the left bulk of the domain wall. The half braiding $\beta_{\bfm,\bfe}=1$ from the left side of the wall is trivial. However, $\one \oplus\bfe$ is blocked from going to the right bulk due to the non-trivial braiding $\beta_{\bfe,\bfm}=-\id$ in $\TC$. Similarly, $\one\oplus \bfm$ has trivial braiding from the right bulk whence is blocked by the structure of wall from going to the right bulk. This shows $Z_l(B_5)$ should be $\one \oplus \bfe$ and $Z_r(B_5)$ should be $\one \oplus \bfm$. In case of $B_6$, we have a mirrored situation. Although the half braiding $\beta_{\bfe,\bfm}$ depicted in this case is $-1$, we need to multiply the non-trivial 2-cohomology class $\omega = -1$ given by $\beta_{\bfe,\bfm}$, which again results in commutativity of $\one\oplus \bfe$ with $B_6$ from right side.}\label{fig:wall_braiding}
\end{figure}
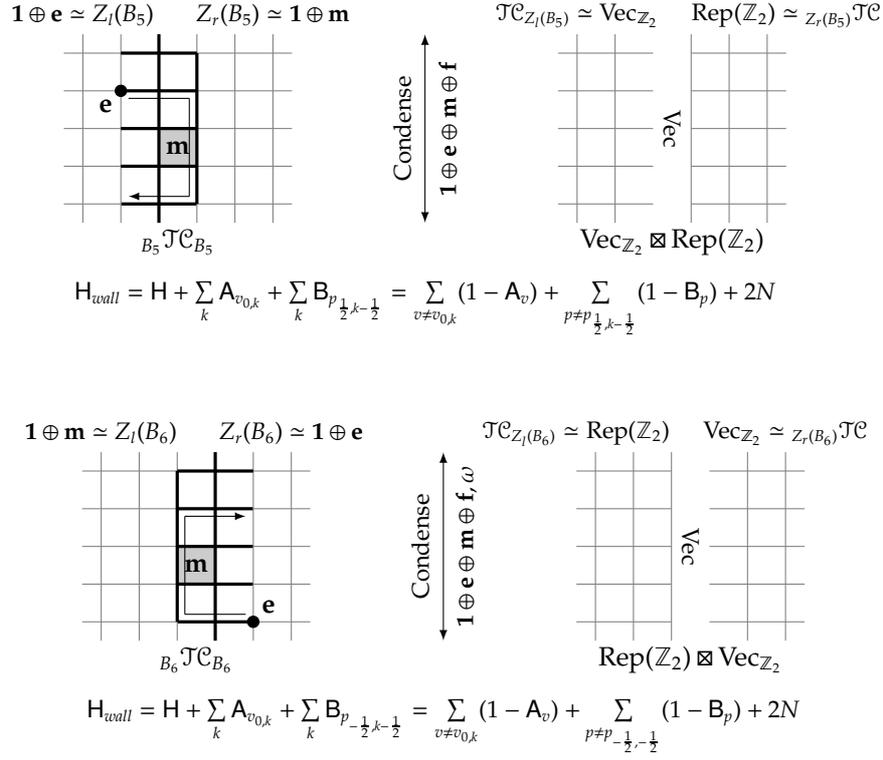

However, the relative position between $\sA_v$ and $\sB_p$ is not interchangeable. 
More precisely, changing trap $\sA_{v_{0,0}} + \sB_{p_{\frac{1}{2},-\frac{1}{2}}}$ to $\sB_{p_{-\frac{1}{2},-\frac{1}{2}}} + \sA_{v_{0,0}}$ would lead to a non-trivial half braiding $\beta_{\bfe,\bfm}=-1$ between $\bfe$ and $\bfm$, despite the excitations generated by these two traps remain the same (we illustrate the braiding distinctions on lattice model in figure \ref{fig:wall_braiding}). Mathematically, this non-trivial braiding is encoded in the non-trivial 2-cohomology class $\omega=-1$, which is due to a multiplication choice between $(\one\oplus\bfe)\ot(\one\oplus\bfm)$ and $(\one\oplus\bfm)\ot(\one\oplus\bfe)$. 
The non-interchangeability also shows that $B_5$ and $B_6$ are not commutative. 

We depict all six 1d condensable algebras $B_i$ in $\TC$ graphically in the lattice model in figure \ref{fig:1d_alg}. 
$B_3 = \one \oplus \bfe$ and $B_4 = \one \oplus \bfm$ can be obtained by adding traps similar to $\sH'$ directly. $\sC_{-\frac{1}{2},-\frac{1}{2}}$ and $ \sD_{\frac{1}{2},0}$ are 'interlocked' operators across each other according to \cite{KK12}, in which it translates an $\bfm$ to an $\bfe$ and vice versa.


Now we compute their left/right centers to obtain 2-Morita equivalent condensable algebras in $\TC$. 
Since $\TC$ is a $\TC$-$\TC$ bimodule category, we can use Davydov's right/left center to perform the calculation (see Definition \ref{defn:Dav} and Remark \ref{rmk:Dav_center}). 
Davydov's right/left center is usually more efficient because it only considers the maximal commutative subalgebra of $B$ such that the diagram \ref{diag:Dav_right_center} commutes.
It is clear that $B_1$, $B_3$ and $B_4$ are commutative algebras, their left and right centers are themselves. 
Since the maximal commutative subalgebra of $B_2 = \one\oplus \bff$ is $\one$, the left/right center of $B_2$ can only be $\one$.
We choose $B_5= \one\oplus \bfe \oplus \bfm\oplus \bff$ as a non-trivial example to display the calculation of left/right center. 

\begin{expl}\label{TC_cal}
    
    $B_5=\one\oplus \bfe \oplus \bfm\oplus \bff$ can be regarded as the tensor of two commutative algebras $A_\bfe:=\one\oplus \bfe$ and $A_\bfm:=\one\oplus\bfm$, i.e. $B_5\cong(\one\oplus\bfe)\ot(\one\oplus \bfm)$. So there are two candidates $A_\bfe$ and $A_\bfm$ to consider.
    Here we prove that the subalgebra $\one\oplus \bfm\cong \one\ot A_\bfm$ of $B_5\cong A_\bfe\ot A_\bfm$ is the Davydov's left center $C_l(B_5)$, and $\one\oplus \bfe\cong A_\bfe\ot \one$ is not.
    Indeed, the following diagram commutes:
    $$
    \begin{tikzcd}
        (\one\ot A_\bfm)\ot (A_\bfe \ot A_\bfm)  \ar[r, ""{name=s},"\beta_{A_\bfm,A_\bfe}"] \ar[dd,"\beta_{A_\bfm,A_\bfe}"']& (\one\ot A_\bfe)\ot (A_\bfm\ot A_\bfm) \ar[dr] & \\
        & & A_\bfe\ot A_\bfm \\
        (A_\bfe\ot A_\bfm)\ot (\one \ot A_\bfm) \ar[r,""{name=t},"\beta_{A_\bfm,\one}=\id"']& (A_\bfe\ot \one)\ot (A_\bfm \ot A_\bfm) \ar[ur] & \arrow[from=s, to=t, pos=.5, phantom, "\scalebox{1.5}{\Circlearrowright}"]
    \end{tikzcd}
    $$
    Hence we have $Z_r(B_5)\cong C_l(B_5)\cong A_\bfm=\one\oplus \bfm$. 
    In addition, the following diagram does not commute,
    $$
    \begin{tikzcd}
        (A_\bfe\ot \one)\ot (A_\bfe \ot A_\bfm)  \ar[r,""{name=s}, "\id"] \ar[dd,"\beta_{A_\bfe,A_\bfm}"']& (A_\bfe\ot A_\bfe)\ot (\one\ot A_\bfm) \ar[dr] & \\
        & & A_\bfe\ot A_\bfm \\
        (A_\bfe\ot A_\bfm)\ot (A_\bfe\ot \one) \ar[r,""{name=t},"\beta_{A_\bfm,A_\bfe}"']& (A_\bfe\ot A_\bfe)\ot (A_\bfm\ot\one ) \ar[ur] & \arrow[from=s, to=t, pos=.5, phantom, "\ncomm"]
    \end{tikzcd}
    $$
    since $\beta_{A_\bfe,A_\bfm}=\id\oplus\id\oplus\id\oplus \beta_{\bfe,\bfm}=\id\oplus\id\oplus\id\oplus-\id$ and $\beta_{A_\bfm,A_\bfe}=\id$.
    So $\one\oplus \bfe\cong A_\bfe\ot \one$ is not Davydov's left center of $B_5$.
    However, we can prove that $\one\oplus \bfe$ is Davydov's right center $C_r(B_5)$ of $B_5$ by a similar process.
    Hence, we have $Z_l(B_5)\cong C_r(B_5) = \one\oplus\bfe$.
    Therefore, $Z_l(B_5)\cong \one\oplus \bfe\widesim[3]{2-Morita}{}\one\oplus\bfm \cong Z_r(B_5)$.

\end{expl}


On the other hand, since $B_6\cong (\one\oplus \bfm)\ot (\one\oplus\bfe)$, the left/right centers of $B_6$ are mirrored to these of $B_5$. 
We list all left/right centers of $B_i$ in column 2 of Table \ref{TC_table}.
Based on this novel method, we recover the two 2-Morita equivalence classes of condensable algebras in the toric code model, namely $\one \widesim[3]{2-Morita}{\phi_{\bfe-\bfm}} \one$ 
and $\one \oplus \bfe \widesim[3]{2-Morita}{} \one \oplus \bfm$.

The role played by the left and right centers can also be realized in the lattice model:
Taking left/right center of $B_i$ is to expand the 1d condensable algebras $B_i$ on the trivial domain wall (fig. \ref{fig:1d_alg}) into the left and right bulks directly, such that they become 2-Morita equivalent 2d condensable algebras. 
By "expand directly", we mean to move the sub algebras of $B_i$ to left and right bulks and see which one commutes with $B_i$ from left and right, respectively. 
The bottom figures illustrate both the 1d condensation of $B_i$ on trivial domain wall and also the action of taking left and right centers, which shows that 1d condensation controlled by $B_i$ in $\CC$ is parallel to 2d condensations controlled by $Z_l(B_i)$ and $Z_r(B_i)$ in $\CC$. We also list their Hamiltonian below. $N$ represents the number of sites on a column under physical consideration.

  

\begin{figure}[H]
    \centering
    \begin{tikzpicture}
        \draw[latex-] (-0.25, 3.5) -- (1.25, 3.5);
        \draw[-latex] (1.75, 3.5) -- (3.25, 3.5);
        \node[]at(0.5, 3.8){left center};
        \node[]at(2.5, 3.8){right center};
        \node[]at(7.75, 3.8){$2d$ condensation};
        \draw[-latex] (6, 3.5) -- (7.5, 3.5);
        \draw[latex-] (8, 3.5) -- (9.5, 3.5);
        \draw[latex-latex](4.5, 0.25) -- (4.5, 2.75);
        \node[rotate=90]at(4.2, 1.5){\small Condense $\one \oplus \bfe$};

        \filldraw[] (1.5, 1.5) circle (0.075);
        \node[]at(1.7, 1.7){$\bfe$};
    \foreach \i in {1, ...,5} {
        \draw [gray] (0.25, \i * 0.5) -- (2.75, \i * 0.5);
        \draw [very thick] (1, \i * 0.5) -- (2, \i * 0.5);
        \draw [gray] (\i * 0.5, 0.25) -- (\i * 0.5, 2.75);
    }
    \draw [very thick](1.5, 0.25)--(1.5, 2.75);    
    \node[]at(6.5, 3){\small $\TC_{Z_l(B_3)} \simeq \vect_{\Zb_2}$};
    \node[]at(9, 3){\small $\vect_{\Zb_2} \simeq {}_{Z_r(B_3)} \TC$};
    \node[]at(0.5, 3){\small $B_3\simeq Z_l(B_3)$};
    \node[]at(2.5, 3){\small $Z_r(B_3) \simeq B_3$};
    \node[]at(1.5, 0){${}_{B_3} \TC_{B_3}$};
  
    \foreach \i in {1, ...,5} {
        \draw [gray] (6.25, \i * 0.5) -- (7.5, \i * 0.5);
        \draw [gray] (8, \i * 0.5) -- (9.25, \i * 0.5);
    }
    \foreach \i in {0, ...,1} {
        \draw [gray] (\i * 0.5 + 6.5, 0.25) -- (\i * 0.5 + 6.5, 2.75);
        \draw [gray] (\i * 0.5 + 8.5, 0.25) -- (\i * 0.5 + 8.5, 2.75);
    }
    \node[rotate=270] at(7.75, 1.5){\small $\vect$};
    \node[]at(7.75, 0){$\vect_{\bZ_2}\btd\vect_{\bZ_2}$};
    \node[]at(4.5, -0.8){\small $\sH_{wall}= \sH + \sum\limits_k \sA_{v_{0,k}} = \sum\limits_{v \neq v_{0,k}} (1-\sA_v) + \sum\limits_p (1-\sB_p)+N$};
    \end{tikzpicture}
  \end{figure}

  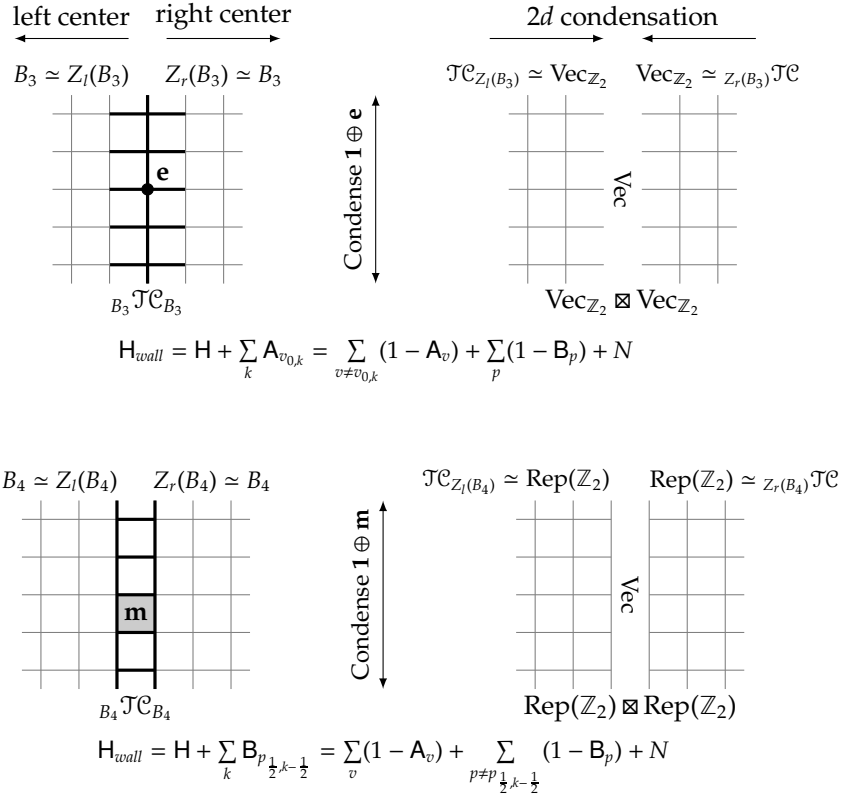
\begin{figure}[H]
    \centering
    \begin{tikzpicture}
        \filldraw[gray!40] (1, 1) rectangle (1.5,1.5);
        \node[]at(1.25,1.25){$\bfm$};
    \foreach \i in {1, ...,5} {
        \draw [gray] (-0.25, \i * 0.5) -- (2.75, \i * 0.5);
        \draw [very thick] (1, \i * 0.5) -- (1.5, \i * 0.5);
    }
    \foreach \i in {0, ...,5} {
        \draw [gray] (\i * 0.5, 0.25) -- (\i * 0.5, 2.75);
    }
    \draw [very thick](1, 0.25)--(1, 2.75);   
    \draw [very thick](1.5, 0.25)--(1.5, 2.75);    
    \draw[latex-latex](4.5, 0.25) -- (4.5, 2.75);
    \draw[draw=none] (0,0)--(-1,0);
    \node[rotate=90]at(4.2, 1.5){\small Condense $\one \oplus \bfm$};
    \node[]at(6.25, 3){\small $\TC_{Z_l(B_4)} \simeq \rep(\Zb_2)$};
    \node[]at(9.25, 3){\small $\rep(\Zb_2) \simeq {}_{Z_r(B_4)} \TC$};
    \node[]at(0.25, 3){\small $B_4\simeq Z_l(B_4)$};
    \node[]at(2.25, 3){\small $Z_r(B_4) \simeq B_4$};
    \node[]at(1.25, 0){${}_{B_4} \TC_{B_4}$};
    \foreach \i in {1, ...,5} {
        \draw [gray] (6.25, \i * 0.5) -- (7.5, \i * 0.5);
        \draw [gray] (8, \i * 0.5) -- (9.25, \i * 0.5);
    }
    \foreach \i in {0, ...,5} {
        \draw [gray] (\i * 0.5 + 6.5, 0.25) -- (\i * 0.5 + 6.5, 2.75);
    }
    \node[rotate=270] at(7.75, 1.5){\small $\vect$};
    \node[]at(7.75, 0){$\rep(\bZ_2)\btd\rep(\bZ_2)$};
    \node[]at(4.5, -0.8){\small $\sH_{wall}= \sH + \sum\limits_k \sB_{p_{\frac{1}{2},k-\frac{1}{2}}}= \sum\limits_v (1-\sA_v) + \sum\limits_{p \neq p_{\frac{1}{2},k-\frac{1}{2}}} (1-\sB_p)+N$};
    \end{tikzpicture}
    \caption{By directly expanding $B_3$ ad $B_4$ (fig \ref{fig:1d_alg} (c) (d)) in to the left and right bulk, we get their left and right centers as two pairs of 2-Morita equivalent $E_2$ condensable algebras. Since $B_3$ and $B_4$ are both commutative algebras, they moved from the 1d domain wall to the 2d bulk transparently, in which their left and right centers are just themselves.}\label{fig:B_3_B_4}
  \end{figure}

\begin{figure}[H]
    \centering
    \begin{tikzpicture}
    \filldraw[gray!40] (1, 1) rectangle (1.5,1.5);
    \node[]at(1.25,1.25){$\bfm$};
    \node[]at(0.3,1.8){$\bfe$};
    \foreach \i in {1, ...,5} {
        \draw [gray] (-0.25, \i * 0.5) -- (2.75, \i * 0.5);
        \draw [very thick] (0.5, \i * 0.5) -- (1.5, \i * 0.5);
        \draw [gray] (\i * 0.5, 0.25) -- (\i * 0.5, 2.75);
    }
    \draw[gray](0,0.25)--(0, 2.75);
    \draw [very thick](1, 0.25)--(1, 2.75);    
    \draw [very thick](1.5, 0.5)--(1.5, 2.5);  
    \draw[latex-latex](4.5, 0.25) -- (4.5, 2.75);
    \node[rotate=90]at(4.2, 1.5){\small Condense};
    \node[rotate=90]at(4.8, 1.5){\small $\one\oplus \bfe \oplus \bfm \oplus \bff$};

    \filldraw[] (0.5, 2) circle (0.075);
    \draw [](0.6, 1.9)--(1.4, 1.9);
    \draw [](1.4, 1.9)--(1.4, 0.6);
    \draw [-latex](1.4, 0.6)--(0.6, 0.6);
  
    \foreach \i in {1, ...,5} {
        \draw [gray] (6.25, \i * 0.5) -- (7.5, \i * 0.5);
        \draw [gray] (8, \i * 0.5) -- (9.25, \i * 0.5);
    }
    \foreach \i in {0, ...,1} {
        \draw [gray] (\i * 0.5 + 6.5, 0.25) -- (\i * 0.5 + 6.5, 2.75);
        \draw [gray] (\i * 0.5 + 8.5, 0.25) -- (\i * 0.5 + 8.5, 2.75);
    }
    \draw[gray](8,0.25)--(8, 2.75);
    \node[]at(6.5, 3){\small $\TC_{Z_l(B_5)} \simeq \vect_{\Zb_2}$};
    \node[]at(9.25, 3){\small $\rep(\Zb_2) \simeq {}_{Z_r(B_5)} \TC$};
    \node[]at(0, 3){\small $\one \oplus \bfe \simeq Z_l(B_5)$};
    \node[]at(2.5, 3){\small $Z_r(B_5) \simeq \one \oplus \bfm$};
    \node[]at(1.25, 0){${}_{B_5} \TC_{B_5}$};
    \node[rotate=270] at(7.75, 1.5){\small $\vect$};
    \node[]at(7.75, 0){$\vect_{\bZ_2}\btd \rep(\bZ_2)$};
    \node[]at(4.5, -0.9){\small $\sH_{wall} =\sH+ \sum\limits_k \sA_{v_{0,k}}+\sum\limits_k  \sB_{p_{\frac{1}{2},k-\frac{1}{2}}}=\sum\limits_{v\neq v_{0,k} }(1-\sA_v)+\sum\limits_{p\neq p_{\frac{1}{2},k-\frac{1}{2}}}(1-\sB_p)+2N$};
    \end{tikzpicture}
\end{figure}

The above figures with 1d condensations of $B_3,B_4, B_5$ and $B_6$ depict four non-invertible domain walls in $\CT \CC$. We can understand them by piecing together gapped boundaries ${\rep}(\bZ_2)$ and $\vect_{\bZ_2}$ two by two, see the right parts of the upper figures.

$B_1$ and $B_2$ correspond to two invertible domain walls, which are characterized by the group of braided auto-equivalence $\Aut_{E_2}(\TC)\cong \bZ_2$ of $\TC$. We omit the figure of the trivial domain wall $\TC = {}_{B_1}{\TC_{B_1}}$ and depict the situation related to $B_2$ below.



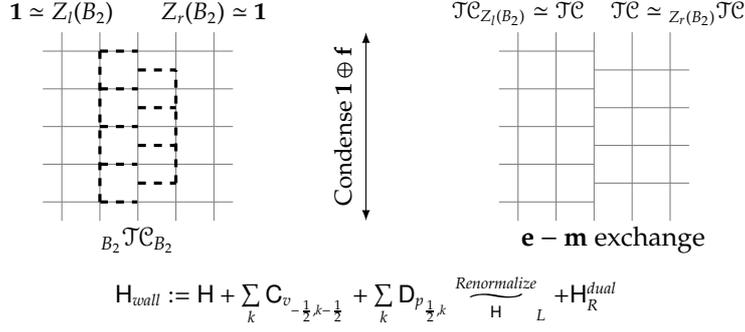
\begin{figure}[H]
    \centering
    \begin{tikzpicture}
    \foreach \i in {1, ...,5} {
        \draw [gray] (0.25, \i * 0.5) -- (2.75, \i * 0.5);
        \draw [gray] (\i * 0.5, 0.25) -- (\i * 0.5, 2.75);
        \draw [very thick, dashed] (1, \i * 0.5) -- (1.5, \i * 0.5);
    } 
    \foreach \i in {1, ...,4} {
      \draw [very thick,dashed] (1.5, \i * 0.5 + 0.25) -- (2, \i * 0.5 + 0.25);
    } 
    \draw [very thick, dashed] (1, 0.5) -- (1, 2.5);
    \draw [very thick, dashed] (2, 0.75) -- (2, 2.25);
    \draw[latex-latex](4.5, 0.25) -- (4.5, 2.75);
    \node[rotate=90]at(4.2, 1.5){\small Condense $\one\oplus \bff$};
    \node[]at(6.5, 3){\small $\TC_{Z_l(B_2)} \simeq \TC$};
    \node[]at(8.6, 3){\small $\TC \simeq {}_{Z_r(B_2)} \TC$};
    \node[]at(0.5, 3){\small $\one \simeq Z_l(B_2)$};
    \node[]at(2.5, 3){\small $Z_r(B_2) \simeq \one$};
    \node[]at(1.5, 0){${}_{B_2} \TC_{B_2}$};
    \foreach \i in {1, ...,5} {
        \draw [gray] (6.25, \i * 0.5) -- (7.5, \i * 0.5);
    }
    \foreach \i in {2, ...,5} {
        \draw [gray] (7.5, \i * 0.5-0.25) -- (8.75, \i * 0.5-0.25);
    }
    \foreach \i in {0, ...,4} {
        \draw [gray] (\i * 0.5 + 6.5, 0.25) -- (\i * 0.5 + 6.5, 2.75);
    }
    \node[]at(7.75, 0){$\bfe-\bfm$ exchange};
    \node[]at(4.5, -0.8){\small $\sH_{wall}:=\sH+ \sum\limits_k \sC_{v_{-\frac{1}{2},k-\frac{1}{2}}}+\sum\limits_k \sD_{p_{\frac{1}{2},k}}\widesim[4]{Renormalize}\sH_L+\sH_R^{dual}$};
    \end{tikzpicture}
    \caption{Since $\bfe\ot\bff \cong\bfm$ and $\bfm\ot \bff \cong \bfe$, $B_2 = \one\oplus \bff$ encodes an interchange between $\bfm$ and $\bfe$ (also called electromagnetic duality). Only $\one$ commutes with $B_2$ from left and right bulks. 1d condensing $B_2$ on the trivial wall leads to the $\bfe-\bfm$ exchange domain wall $\Phi_{\bfe-\bfm}$, we can realize this duality on lattice model by taking dual lattice on the right side.}\label{fig:e-m_exchange}
\end{figure}

\begin{rem}
    These six gapped domain walls correspond to six simple objects in the 2-category $\Sigma \TC$ of 1-codimensional topological defects of toric code model \cite{KZ22a,KZZZ24}. There fusion rules can be recognized directly through above lattice constructions. For example, 
    \begin{itemize}
        \item $(\vect_{\bZ_2}\btd \rep(\bZ_2)) \boxtimes_{\TC} (\vect_{\bZ_2}\btd \rep(\bZ_2)) \simeq \vect_{\bZ_2}\btd \rep(\bZ_2)$ since $\rep(\bZ_2) \boxtimes_{\TC} \vect_{\bZ_2}\simeq \vect$.
        \item $(\rep(\bZ_2)\btd \rep(\bZ_2))\btd_{\TC}(\rep(\bZ_2)\btd \vect_{\bZ_2})\simeq \mathrm{M}_2(\rep(\bZ_2)\btd \rep(\bZ_2))$ represents the superposition of two $\rep(\bZ_2)\btd \vect_{\bZ_2}$ domain walls, since $\rep(\bZ_2)\btd_{\TC}\rep(\bZ_2)\simeq \mathrm{M}_2(\vect)$ as multi-fusion categories 
        \item $\Phi_{\bfe-\bfm}\btd_{\TC}(\rep(\bZ_2)\btd \rep(\bZ_2))\simeq (\vect_{\bZ_2}\btd \rep(\bZ_2))$ since $\Phi_{\bfe-\bfm}$ has an $\bfe$-$\bfm$ exchange action on other walls. 
    \end{itemize}
\end{rem}

Through above figures we show how six 1d condensable algebras $B_i\in\TC$ correspond to six stable gapped domain walls ${}_{B_i}\TC_{B_i}$ in the lattice model of $\TC$ (see the third column of Table. \ref{TC_table}).
For more developments on this lattice model technique, see discussions of 1d condensable algebras in our future work.
We summarize above results in the following Table, where $H$ and $F$ are subgroups of $\bZ_2$ appearing in Davydov's classification of condensable algebras.

\begin{table}[H]
    \centering
    \begin{tabular}{|c|c|c|c|c|c|}
        \hline
        $H$ & $F$ &\makecell{2d condensable algebras \\ in $\TC$} &  \makecell{Condensed phase \\ $\TC^{loc}_A$} & Domain walls & Total: $6$ \\
        \hline
        \multirow{2}{*}{$\bZ_2$} & \multirow{2}{*}{$\bZ_2$} &\multirow{2}{*}{$\one\oplus \bfm$} &  \multirow{4}{*}{\text{$\vect$}} & \multirow{4}{*}{\begin{tikzpicture}[scale=0.75]
              \filldraw[fill=gray!40, draw=white] (-2,0) rectangle (-0.5,1.5);
              \filldraw[fill=gray!40, draw=white] (1,0) rectangle (2.5,1.5);
              \filldraw[fill=gray!0, draw=white] (-0.5,0) rectangle (1,1.5);
              \node at(0.25, 0.75){\ltiny $\vect$};
              \node at(-1.3, 0.75){\ltiny $\TC$};
              \node at(1.7, 0.75){\ltiny $\TC$};
              \draw[thick](-0.5,0)--(-0.5,1.5);
              \draw[thick](1,0)--(1,1.5);
              \node at(-0.4, -0.1){\tiny $2$};
              \node at(0.9, -0.1){\tiny $2$};
              \node at(0, 1.5){\quad};
          \end{tikzpicture}  }& \multirow{4}{*}{\makecell{\small non-invertible:\\ $2\times 2 =4$}} \\
          &&&&&\\
          $\{e\}$ &$\{e\}$ & $\one\oplus \bfe$  &   & & \\
          &&&&&\\
          \hline
          &&&&\multirow{3}{*}{             
              \begin{tikzpicture}[scale=0.75]
              \filldraw[fill=gray!40, draw=none] (-2,0) rectangle (1,1.5);
              \node at(-1.3, 0.75){\ltiny $\TC$};
              \node at(0.2, 0.75){\ltiny $\TC$};
              \draw[dashed](-0.5,0)--(-0.5,1.5);
              \node at(-0.5, 1.65){\tiny $2$};
              \node at(0, 1.4){\quad};
          \end{tikzpicture} 
          }&\\
          $\bZ_2$ &$\{e\}$ & $\one$ &  $\TC$ & &\small invertible: $2$ \\
        &&& \rule{0pt}{3ex}&&\\
        \hline
    \end{tabular}\label{TC_wall_table}
\end{table}

1d condensable algebras in $\TC$ can also be recovered by 2-Morita equivalent condensable algebras, i.e. A$\overrightarrow{\text{rrow}}$ 6 in Trinity \ref{fig:alg_cycle}.
When the condensed phase of the pair of 2-Morita equivalent 2d condensable algebras $(A_1,A_2)$ is $\vect$ (second row of above table), the extended tensor product of $A_1$ and $A_2$ is the tensor product of extension of $\phi$-algebra $B_{\Id}=\one \in\vect$ over $A_1$ and the extension of $\one$ over $A_2$, which is $A_1\ot A_2$.
Hence, 1d condensable algebras $B$ is an indecomposable subalgebra of $A_1\ot A_2$.
Precisely speaking, for $A_1=A_2=\one\oplus \bfe$, $A_1\ot A_2\cong \one\oplus \bfe\oplus\bfe\oplus\one$, so we pick the subalgebra $\one\oplus\bfe$ as $B_3$; for $A_1=\one\oplus\bfe$ and $A_2=\one\oplus\bfm$, $A_1\ot A_2\cong \one \oplus \bfe\oplus \bfm\oplus \bff$ is indecomposable, so $B_5$ is $\one \oplus \bfe\oplus \bfm\oplus \bff$.
Similarly, we can recover $B_4$ and $B_6$.

In the situation of the two invertible ones (third row of above table), we have two $\phi$-algebras in the condensed phase $\TC$ corresponding to the trivial domain wall and the $\bfe-\bfm$ exchange domain wall: $B_{\Id}=\one$ and $B_{\phi_{\bfe- \bfm}}=\one\oplus\bff$, which is an indecomposable subalgebra of tensor product of $\phi$-twisted lagrangian algebras $L_{\phi}=\bigoplus_{x\in\mathrm{Irr}}x\boxtimes \phi(x^*)$ in $\TC\boxtimes \overline{\TC}$ (in this case, $L_{\Id} =\one\overline{\one}\oplus \bfe\overline{\bfe}\oplus \bfm\overline{\bfm}\oplus \bff\overline{\bff}$ and $L_{\phi_{\bfe-\bfm}} =\one\overline{\one}\oplus \bfm\overline{\bfe}\oplus \bfe\overline{\bfm}\oplus \bff\overline{\bff}$).
The extended tensor product of 2d condensable algebra $A_1=\one=A_2$ 
is $B_{\phi}$ itself.
In the case of $B_{\Id}=\one$, the tensor product of $L_{\Id}$ is a direct sum of $\one$'s, so $B_1$ is just $\one$; for $B_{\phi_{\bfe- \bfm}}$, the tensor product of $L_{\phi_{\bfe-\bfm}}$ is $\one\oplus \bff\oplus \bff\oplus \one$, in which $B_2$ can only be $\one\oplus \bff$.

\vspace{+3ex}

Now we show the other method of classifying 2-Morita equivalent condensable algebras, namely A$\overrightarrow{\text{rrow}}$ 2 in Trinity \ref{fig:alg_cycle}.
We first classify lagrangian algebras in $\TC\btd\overline{\TC}\simeq \fZ(\vect_{\bZ_2\times \bZ_2})$.
By Theorem \ref{thm:classification_lag_alg_ZVecG}, the lagrangian algebras in $\fZ(\vect_{\bZ_2\times \bZ_2})\simeq\fZ(\rep(\bZ_2\times \bZ_2))$ are of the form $\fun(\bZ_2\times \bZ_2)\ot_{\bC[H]}\bC[H,\omega]$ for subgroups $H$ of $\bZ_2\times \bZ_2$. 
\begin{expl}
        For example, when $H=\{e\}$, $\omega$ must be trivial, the corresponding lagrangian is $\fun(\bZ_2\times \bZ_2)\ot_{\bC}\bC\simeq \fun(\bZ_2\times \bZ_2)$.
        Note that the function algebra $\fun(G)$, when forget to $\vect_G$, must have trivial grading and contains all irreducible representations $V$ of $G$ by $\dim(V)$ times.
        Since $\bfe_1$ and $\bfe_2$ are simple objects with trivial grading in $\fZ(\vect_{\bZ_2\times \bZ_2})$, $\fun(\bZ_2\times \bZ_2)$ can only be $\one\oplus \bfe_1\oplus \bfe_2\oplus \bfe_1\bfe_2 \in \fZ(\vect_{\bZ_2\times \bZ_2})$. Under the equivalence $\fZ(\vect_{\bZ_2\times \bZ_2})\to \TC\btd \overline{\TC}$ (namely, $\bfe_1\mapsto \bfe\boxtimes \overline{\bfe}$, $\bfe_2 \mapsto \bfm\boxtimes \overline{\bfm}$, $\bfm_1\mapsto \one\boxtimes \overline{\bfm}$ and $\bfm_2 \mapsto \bfe\boxtimes\overline{\one}$), we obtain $\one\overline{\one}\oplus \bfe\overline{\bfe}\oplus \bfm\overline{\bfm}\oplus \bff\overline{\bff}\in \TC\btd \overline{\TC}$\footnote{Here we use $\overline{x}$ to denote the object in $\overline{\TC}$, and we omit $\btd$ for the sake of simplicity.} which we denote by $L_1$. 

\end{expl}
Based on this method, we compute all lagrangian algebras in $\TC\btd \overline{\TC}$, results are listed in the column 4 of Table. \ref{TC_table}.
By intersecting $L_i$ with left and right components of $\TC\btd \overline{\TC}$, we can obtain pairs of 2-Morita equivalent condensable algebras immediately, which corresponds to column 2 of Table \ref{TC_table}.
For example, from $\one\overline{\one}\oplus \bfm\overline{\bfe}\oplus \bfe\overline{\bfm}\oplus \bff\overline{\bff} $, we can obtain $\one \widesim[3]{2-Morita}{\phi_{\bfe-\bfm}} \one$; from $ \one\overline{\one}\oplus \bfe\overline{\one}\oplus \one\overline{\bfm}\oplus \bfe\overline{\bfm} $, we can obtain $\one \oplus \bfe \widesim[3]{2-Morita}{} \one \oplus \bfm$.

\begin{rem}
    Finding lagrangian algebras of a pointed MTC $\FZ(\vect_{G})$ for an Abelian group $G$ can also be translated to the classification of isotropic subgroups of the corresponding metric group $(G\times \hat{G},q)$ \cite{DGNO10}.
\end{rem}

Also, lagrangian algebras in $\TC\btd\overline{\TC}$ are not hard to be reconstructed from these 2-Morita equivalent condensable algebras by using A$\overrightarrow{\text{rrow}}$ 1 in Trinity \ref{fig:alg_cycle}.
Due to Remark \ref{rmk:A_1boxA_2}, we can obtain $L_3$, $L_4$, $L_5$ and $L_6$ in column 4 of Table \ref{TC_table} by substituting $\one\oplus\bfe$ and $\one\oplus\bfm$ into $A^L_i\btd A^L_j$;
The canonical lagrangian algebra $L_1=\ot^{R}(\one)$ is given by $\bigoplus_{x\in\mathrm{Irr}(\TC)}x\btd x^*=\one\overline{\one}\oplus \bfe\overline{\bfe}\oplus \bfm\overline{\bfm}\oplus \bff\overline{\bff}$ \cite{DKR11}, where $\ot :\TC\btd \overline{\TC}\to \TC$ is the tensor functor;
Twist $L_1$ by the $\bfe$-$\bfm$ exchange braided autoequivalence $\phi_{\bfe-\bfm}$, i.e. $\bigoplus_{x\in\mathrm{Irr}(\TC)}x\btd \phi_{\bfe-\bfm}(x^*)$, we obtain $L_2$.

\vspace{+3ex}

We can also give the bijection between 1d condensable algebras in $\TC$ and lagrangian algebras in $\TC\btd \overline{\TC}$ through A$\overrightarrow{\text{rrow}}$ 4 and A$\overrightarrow{\text{rrow}}$ 5 in Trinity \ref{fig:alg_cycle}. 
This algebraic level bijection also induces a bijection between their module categories, which corresponds to the gapped domain walls in $\TC$ and the gapped boundaries of $\TC\btd \overline{\TC} \simeq \FZ(\TC)$, see the figure below.

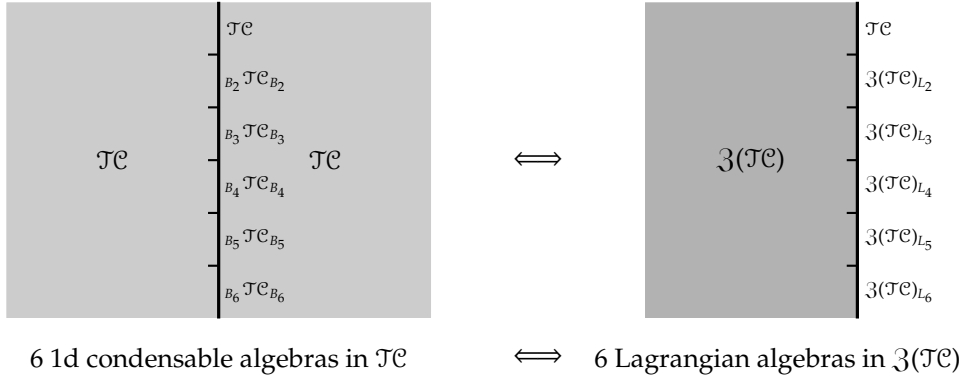
\begin{figure}[H]
    \centering
    \begin{tikzpicture}[scale=0.7]
        \filldraw[fill=gray!40, draw=white] (-8,6) rectangle (0,0);
        \filldraw[fill=gray!60, draw=white] (4,6) rectangle (8,0);
        \draw[very thick] (-4,6) -- (-4,0);
        \draw[very thick] (8,6) -- (8,0);
        \node at(2,3){$\Longleftrightarrow$};
        \node at (-6,3){$\TC$};
        \node at (-2,3){$\TC$};
        \node at (6,3){$\FZ(\TC)$};
        \draw[thick] (-4.2,5)--(-4,5);
        \node at(-3.6,5.5){\scriptsize $\TC$};
        \draw[thick] (-4.2,4)--(-4,4);
        \node at(-3.3,4.5){\scriptsize ${}_{B_2}{\TC_{B_2}}$};
        \draw[thick] (-4.2,3)--(-4,3);
        \node at(-3.3,3.5){\scriptsize ${}_{B_3}{\TC_{B_3}}$};
        \draw[thick] (-4.2,2)--(-4,2);
        \node at(-3.3,2.5){\scriptsize ${}_{B_4}{\TC_{B_4}}$};
        \draw[thick] (-4.2,1)--(-4,1);
        \node at(-3.3,1.5){\scriptsize ${}_{B_5} {\TC_{B_5}}$};
        \node at(-3.3,0.5){\scriptsize ${}_{B_6} {\TC_{B_6}}$};
  
        \draw[thick] (7.8,5)--(8,5);
        \node at(8.4,5.5){\scriptsize $\TC$};
        \draw[thick] (7.8,4)--(8,4);
        \node at(8.8,4.5){\scriptsize $\fZ(\TC)_{L_2}$};
        \draw[thick] (7.8,3)--(8,3);
        \node at(8.8,3.5){\scriptsize $\fZ(\TC)_{L_3}$};
        \draw[thick] (7.8,2)--(8,2);
        \node at(8.8,2.5){\scriptsize $\fZ(\TC)_{L_4}$};
        \draw[thick] (7.8,1)--(8,1);
        \node at(8.8,1.5){\scriptsize $\fZ(\TC)_{L_5}$};
        \node at(8.8,0.5){\scriptsize $\fZ(\TC)_{L_6}$};
        \node at(-4, -0.8){6 1d condensable algebras in $ \TC$ };
        \node at(6.5, -0.8){6 Lagrangian algebras in $\FZ(\TC)$};
        \node at(2,-0.75){$\Longleftrightarrow$};
  
    \end{tikzpicture}
  \end{figure}

  From lagrangian algebras $L_i$ to 1d condensable algebras $B_i$, we apply tensor functor $\ot:\TC\btd \overline{\TC}\to \TC$, the image is a direct sum of 1-Morita equivalent $B_i$ in $\TC$, e.g. $ L_1=\one\overline{\one}\oplus \bfe\overline{\bfe}\oplus \bfm\overline{\bfm}\oplus \bff\overline{\bff} $ becomes $\one\oplus \one\oplus \one\oplus \one$ under this functor, which is a direct sum of four $B_1$, and $L_5= \one\overline{\one}\oplus \bfe\overline{\one}\oplus \one\overline{\bfm}\oplus \bfe\overline{\bfm} $ becomes $\one\oplus \bfe\oplus \bfm\oplus \bff=B_5$.

On the other hand, $L_i$ can be obtained by taking full centers of 1d condensable algebras $B_i$. We can use internal hom $[B_i,B_i]_{\TC\boxtimes\overline{\TC}}$ to compute these full centers $Z(B_i)$ \cite{KYZ21}.
We compute $Z(\one)=[\one,\one]_{\TC\boxtimes\overline{\TC}}$ as an example.
Note that the $\TC\boxtimes\overline{\TC}$ action on $\TC$ is given by 
\begin{align*}
    \odot:\TC\boxtimes\overline{\TC}\times \TC&\to \TC\\
    (x\boxtimes \overline{y},c)&\mapsto (x\ot y)\ot c
\end{align*}
So by the following adjunction and Schur's Lemma \cite{EGNO15}:
\begin{align*}
    \hom_{\TC}((x\btd \overline{y})\odot \one,\one)\cong \hom_{\TC\boxtimes \overline{\TC}}(x\btd \overline{y},[\one,\one])
\end{align*}
we can see $[\one,\one]$ contains $x\btd \overline{y}$ if and only if $x\ot y$ contains $\one$.
Going through all the simple objects in $\TC\boxtimes \overline{\TC}$, it is not hard see $\one\overline{\one}$, $\bfe\overline{\bfe}$, $\bfm\overline{\bfm}$ and $\bff\overline{\bff}$ are tensored to $\one$.
Therefore, we have $Z(\one)\cong [\one,\one]\cong \one\overline{\one}\oplus \bfe\overline{\bfe}\oplus \bfm\overline{\bfm}\oplus \bff\overline{\bff}$.
For a detailed calculation of other cases, see \cite[Section 6.3]{YWL24}.

\vspace{3em}


\subsubsection{\texorpdfstring{$\fZ(\vect_{\bZ_4})$}{Z(vecZ4)} and abelian cases}\label{sec:Z4}

Now we step into a more complex case $\FZ(\vect_{G})$ for an abelian group $G$. 
Recall in the beginning of this section that it is more convenient to find 1d condensable algebras written as $\bC[H,\omega]$ in the case $\FZ(\vect_{G})$ for $G$ abelian.
By \cite{FRS04a}, we can use Kreuzer--Schellekens bicharacters to compute left/right center directly through the group data $(H,\omega)$, in which we perform in the second half of this subsection.

Before illustrating this method, we pick $\FZ(\vect_{\Zb_4}) \simeq \FZ(\rep(\bZ_4))$ as an example to perform A$\overrightarrow{\text{rrow}}$ 3 and A$\overrightarrow{\text{rrow}}$ 6 in Trinity \ref{fig:alg_cycle}. 
Lattice model of these kinds of 2d topological orders can be realized through Kitaev quantum double model \cite{Kit03}.
Data of MTC $\fZ(\vect_{\bZ_4})$ are listed as follows.
\bit
    \item The simple objects in $\fZ(\vect_{\bZ_4})$ can be written as $\{\bfe^{\alpha}\bfm^{\beta}\mid \alpha, \beta = 0,1,2,3 \}$, where $\bfe$ denotes the elementary $\bZ_4$-charge and $\bfm$ denotes the elementary $\bZ_4$-flux.
    \item The fusion rule of two simple objects $\bfe^{\alpha_1}\bfm^{\beta_1}$ and $\bfe^{\alpha_2}\bfm^{\beta_2}$ is given by $\bfe^{\alpha_1}\bfm^{\beta_1}\otimes \bfe^{\alpha_2}\bfm^{\beta_2}\simeq \bfe^{\alpha_1+\alpha_2}\bfm^{\beta_1+\beta_2}$.

    \item The braiding of $\bfe^{\alpha_1}\bfm^{\beta_1}$ and  $\bfe^{\alpha_2}\bfm^{\beta_2}$ is given by:
    \begin{align*}
        \beta_{\alpha_1,\beta_1;\alpha_2,\beta_2}:\bfe^{\alpha_1}\bfm^{\beta_1}\otimes \bfe^{\alpha_2}\bfm^{\beta_2}\xrightarrow{{\mathrm{i}}^{\alpha_1\beta_2}}\bfe^{\alpha_2}\bfm^{\beta_2}\otimes \bfe^{\alpha_1}\bfm^{\beta_1}.
    \end{align*}
\eit

\begin{rem}
    Data of MTC $\fZ(\rep(\bZ_N))$ for all $N\in\bZ_+$ can be generated by $\bfe$ and $\bfm$ through a similar process. 
\end{rem}

Similar to $\bZ_2$ case, we can first list all pairs $(H,\omega)$ where $H\subset \bZ_4\times \bZ_4$ and $\omega\in\mathrm{H}^2(H,\bC^{\times})$ (column 1 of Table \ref{table:Z4_alg} \footnote{Here we assume element $(\alpha,\beta)\in\bZ_4\times \bZ_4$ corresponds to $\bfe^{\alpha}\bfm^{\beta}$, and by $\langle (\alpha,\beta)\rangle$, we mean the subgroups generated by the elements $(\alpha,\beta)$.}),
then directly list all 1d condensable algebras $\bC[H,\omega]$ (see column 2 of Table \ref{table:Z4_alg}).
Again by taking the left and right centers of $B_i$ in $\fZ(\vect_{\bZ_4})$ similar to the algorithm in $\TC$, we can obtain all 2-Morita equivalent condensable algebras in $\fZ(\vect_{\bZ_4})$.
Results are listed in column 3 of Table \ref{table:Z4_alg}, 
we see there are four 2-Morita equivalence classes:
\bnu
    \item lagrangian algebras $\{\one \oplus\bfe\oplus\bfe^2\oplus\bfe^3,\one \oplus\bfm\oplus\bfm^2\oplus\bfm^3,\one\oplus \bfe^2\oplus\bfm^2\oplus\bff^2\}$ that condense to $\vect$;    
    
    \item $\one\oplus \bff^2$, which condense to double semion $\CD \CS := \vect_{\bZ_2}^{\alpha}\boxtimes \overline{\vect_{\bZ_2}^{\alpha}}$ \cite{HW14};
    
    \item $\{\one\oplus \bfe^2,\one\oplus\bfm^2\}$, which condense to the $\bZ_2$ topological order, i.e. $\fZ(\vect_{\bZ_4})^{loc}_{\one\oplus \bfe^2} \simeq \TC \simeq \fZ(\vect_{\bZ_4})^{loc}_{\one\oplus\bfm^2} $;
    
    \item trivial condensable algebra $\one$ that condenses to $\fZ(\vect_{\bZ_4})$ itself.    
\enu

Column 3 of table \ref{table:Z4_alg} contains all seven 2d condensable algebras in $\fZ(\vect_{\bZ_4})$, which is in accordance with Davydov's classification by Theorem \ref{thm:classification_lag_alg_ZVecG}.

\begin{rem}
    By dimension formula \cite{KO02}
\begin{align*}
  \dim(\CC_A^{loc})=\dfrac{\dim(\CC)}{\dim(A)^2}
\end{align*}
It is not hard to see that 2-Morita equivalent condensable algebras must have same dimension.
However, the converse is not true, $\bZ_4$ quantum double provides a counterexample: $\one\oplus \bff^2$ and $\one\oplus \bfe^2$ both have dimension 2, but the corresponding condensed phase $\CD\CS$ and $\TC$ are not equivalent. 
\end{rem}

\begin{rem}
    From 2-step condensation, by condensing $\one\oplus \bfe^2$ and $\one\oplus \bfm^2$, $\fZ(\vect_{\bZ_4})$ can condense to $\TC$.
    And from $\TC$, one can also condense $\one \oplus \bfe$ and $\one \oplus \bfm$ to $\vect$. This is equivalent to condense lagrangian algebras in $\fZ(\vect_{\bZ_4})$ directly.
    For example, $\Ext^R_{\one\oplus\bfe^2}(\one\oplus\bfe)=\one\oplus \bfe\oplus\bfe^2\oplus\bfe^3$, and $\Ext^R_{\one\oplus\bfm^2}(\one\oplus\bfe)=\one\oplus\bfe^2\oplus\bfm^2\oplus \bff^2$.
\end{rem}

\begin{table}[htp]
    \centering
    \caption{Results of $\fZ(\vect_{\bZ_4})$, $i=1,2,\dots,21,22$}
    \begin{tabular}{|m{1.85cm}<{\centering}|m{3.85cm}<{\centering}|m{5cm}<{\centering}|m{2.8cm}<{\centering}|}
        \hline
        $(H,\omega)$ & $B_i$& $A_l = Z_l(B_i)/A_r=Z_r(B_i)$& Domain Wall \\
        \hline
        \rowcolor{gray!40} \cellcolor{white} $\{e\}$ & $\one$ & $\one$, $\one$& $\fZ(\vect_{\bZ_4})$\\
        \hline
        \rowcolor{gray!20} \cellcolor{white}$\bZ_2\times \{e\}$ & $\one\oplus \bfe^2$ & $\one\oplus \bfe^2$, $\one\oplus \bfe^2$& $\CS^\bfe\btd_{\TC}\CS^\bfe$\\
        \hline
        \rowcolor{gray!20} \cellcolor{white}$\{e\}\times \bZ_2$ & $\one\oplus \bfm^2$ & $\one\oplus \bfm^2$, $\one\oplus \bfm^2$& $\CS^\bfm\btd_{\TC}\CS^\bfm$\\
        \hline
        \rowcolor{gray!20} \cellcolor{white}$\bZ_2=\langle (2,2)\rangle$ & $\one\oplus \bfe^2\bfm^2(\simeq \one\oplus \bff^2)$ & $\one\oplus \bff^2$, $\one\oplus \bff^2$& $\CN\btd_{\CD\CS}\CN$ \\
        \hline
        $\bZ_2\times \bZ_2$, $\omega=1$ & $\one\oplus \bfe^2\oplus \bfm^2\oplus \bff^2$ & $\one\oplus \bfe^2\oplus \bfm^2\oplus \bff^2$, $\one\oplus \bfe^2\oplus \bfm^2\oplus \bff^2$& $\CM\btd \CM$\\
        \hline
        \rowcolor{gray!40} \cellcolor{white}$\bZ_2\times \bZ_2$, $\omega=-1$ & $\one\oplus \bfe^2\oplus \bfm^2\oplus \bff^2$, $\omega=-1$ & $\one$, $\one$& $\Phi_{1-3}$ \\
        \hline
        $\bZ_4\times\{e\}$ & $\one\oplus \bfe\oplus \bfe^2\oplus \bfe^3$ & $\one\oplus \bfe\oplus \bfe^2\oplus \bfe^3$, $\one\oplus \bfe\oplus \bfe^2\oplus \bfe^3$& $\vect_{\bZ_4}\btd \vect_{\bZ_4}$\\
        \hline
        $\{e\}\times\bZ_4$ & $\one\oplus \bfm\oplus \bfm^2\oplus \bfm^3$ & $\one\oplus \bfm\oplus \bfm^2\oplus \bfm^3$, $\one\oplus \bfm\oplus \bfm^2\oplus \bfm^3$& $\rep(\bZ_4)\btd \rep(\bZ_4)$ \\
        \hline
        \rowcolor{gray!40} \cellcolor{white}$\bZ_4=\langle (1,1)\rangle$ & $\one\oplus \bff\oplus \bff^2\oplus \bff^3$ & $\one$, $\one$& $\Phi_{1-3}\circ \Phi_{\bfe-\bfm}^{\bZ_4}$\\
        \hline
        \rowcolor{gray!40} \cellcolor{white}$\bZ_4=\langle (1,3)\rangle$ & $\one\oplus \bfe\bfm^3\oplus \bfe^2\bfm^2\oplus \bfe^3\bfm$ & $\one$, $\one$& $\Phi_{\bfe-\bfm}^{\bZ_4}$\\
        \hline
        \rowcolor{gray!20} \cellcolor{white}$\bZ_4=\langle (1,2)\rangle$ & $\one\oplus \bfe\bfm^2\oplus \bfe^2\oplus \bfe^3\bfm^2$ & $\one\oplus \bfe^2$, $\one\oplus \bfe^2$& $\CS^\bfe\btd_{\TC}\CS^\bfe$, $\phi_{\bfe-\bfm}$ \\
        \hline
        \rowcolor{gray!20} \cellcolor{white}$\bZ_4=\langle (2,1)\rangle$ & $\one\oplus \bfe^2\bfm\oplus \bfm^2\oplus \bfe^2\bfm^3$ & $\one\oplus \bfm^2$, $\one\oplus \bfm^2$& $\CS^\bfm\btd_{\TC}\CS^\bfm$, $\phi_{\bfe-\bfm}$ \\
        \hline
        $\bZ_4\times \bZ_2$, $\omega=1$ & $(\one\oplus \bfe\oplus \bfe^2\oplus \bfe^3)\ot (\one\oplus \bfm^2)$ & $\one\oplus \bfe\oplus \bfe^2\oplus \bfe^3$, $\one\oplus \bfe^2\oplus \bfm^2\oplus \bff^2$ & $\vect_{\bZ_4} \btd \CM$\\
        \hline
        $\bZ_4\times \bZ_2$, $\omega=-1$ &  $(\one\oplus \bfm^2)\ot(\one\oplus \bfe\oplus \bfe^2\oplus \bfe^3) $ & $\one\oplus \bfe^2\oplus \bfm^2\oplus \bff^2$,$\one\oplus \bfe\oplus \bfe^2\oplus \bfe^3$ & $\CM \btd \vect_{\bZ_4}$\\
        \hline
        $\bZ_2\times \bZ_4$, $\omega=1$ & $(\one\oplus \bfe^2)\ot (\one\oplus \bfm\oplus \bfm^2\oplus \bfm^3)$ & $\one\oplus \bfe^2\oplus \bfm^2\oplus \bff^2$, $\one\oplus \bfm\oplus \bfm^2\oplus \bfm^3$& $\CM \btd \rep(\bZ_4)$ \\
        \hline
        $\bZ_2\times \bZ_4$, $\omega=-1$ & $(\one\oplus \bfm\oplus \bfm^2\oplus \bfm^3)\ot (\one\oplus \bfe^2)$ & $\one\oplus \bfm\oplus \bfm^2\oplus \bfm^3$, $\one\oplus \bfe^2\oplus \bfm^2\oplus \bff^2$& $\rep(\bZ_4)\btd \CM$ \\
        \hline
        \rowcolor{gray!20} \cellcolor{white}$\bZ_4^{(1,3)}\times \bZ_2$, $\omega=1$ & $(\one\oplus \bfe\bfm^3\oplus \bfe^2\bfm^2\oplus \bfe^3\bfm)\ot (\one\oplus \bfm^2)$ & $\one\oplus \bfe^2$, $\one\oplus \bfm^2$& $\CS^\bfe\btd_{\TC}\CS^\bfm$ \\
        \hline
        \rowcolor{gray!20} \cellcolor{white}$\bZ_4^{(1,3)}\times \bZ_2$, $\omega=-1$ & $(\one\oplus \bfe\bfm^3\oplus \bfe^2\bfm^2\oplus \bfe^3\bfm)\ot (\one\oplus \bfe^2)$ & $\one\oplus \bfm^2$, $\one\oplus \bfe^2$& $\CS^\bfm\btd_{\TC}\CS^\bfe$ \\
        \hline
        $\bZ_4\times \bZ_4$, $\omega=1$ & $(\one\oplus \bfe\oplus \bfe^2\oplus \bfe^3)\ot (\one\oplus \bfm\oplus \bfm^2 \oplus \bfm^3)$ & $\one\oplus \bfe\oplus \bfe^2\oplus \bfe^3$, $\one\oplus \bfm\oplus \bfm^2\oplus \bfm^3$& $\vect_{\bZ_4}\btd \rep(\bZ_4)$ \\
        \hline
        $\bZ_4\times \bZ_4$, $\omega=-1$ & $ (\one\oplus \bfm\oplus \bfm^2 \oplus \bfm^3)\ot(\one\oplus \bfe\oplus \bfe^2\oplus \bfe^3)$& $\one\oplus \bfm\oplus \bfm^2\oplus \bfm^3$, $\one\oplus \bfe\oplus \bfe^2\oplus \bfe^3$& $\rep(\bZ_4) \btd \vect_{\bZ_4}$ \\
        \hline
        \rowcolor{gray!20} \cellcolor{white}$\bZ_4\times \bZ_4$, $\omega=\mathrm{i}$ & $(\one\oplus \bfe\bfm^2\oplus \bfe^2\oplus \bfe^3\bfm^2)\ot (\one\oplus \bfe^2\bfm\oplus \bfm^2\oplus \bfe^2\bfm^3)$ & $\one\oplus \bfe^2$, $\one\oplus \bfm^2$& $\CS^\bfe\btd_{\TC}\CS^\bfm$, $\phi_{\bfe-\bfm}$ \\
        \hline
        \rowcolor{gray!20} \cellcolor{white}$\bZ_4\times \bZ_4$, $\omega=-\mathrm{i}$ & $(\one\oplus \bfe^2\bfm\oplus \bfm^2\oplus \bfe^2\bfm^3)\ot (\one\oplus \bfe\bfm^2\oplus \bfe^2\oplus \bfe^3\bfm^2)$ & $\one\oplus \bfm^2$, $\one\oplus \bfe^2$& $\CS^\bfm\btd_{\TC}\CS^\bfe$, $\phi_{\bfe-\bfm}$ \\
        \hline    
    \end{tabular}\label{table:Z4_alg}
\end{table}

Now we illustrate gapped domain walls associated to these 2-Morita equivalence classes of 2d condensable algebras (see fourth column of table \ref{table:Z4_alg}).
\bnu
    \item Denote $\CM$ as the gapped boundary of $\fZ(\vect_{\bZ_4})$ by condensing $\one\oplus \bfe^2\oplus \bfm^2\oplus \bff^2$; $\rep(\bZ_4)$ as the boundary by condensing $\one\oplus \bfm\oplus \bfm^2\oplus \bfm^3$; and $\vect_{\bZ_4}$ as the boundary by condensing $\one\oplus \bfe\oplus \bfe^2\oplus \bfe^3$. 
    There are nine gapped domain walls associated to these lagrangian algebras.

    \item Denote $\CN$ as the domain wall between $\bZ_4$ quantum double $\fZ(\vect_{\bZ_4})$ and the double semion $\CD\CS$.
    Since $\Aut_{E_2}(\CD\CS)=\{e\}$, there is only one domain wall $\CN\btd_{\CD\CS}\CN$ in $\fZ(\vect_{\bZ_4})$ associated to $\one\oplus \bff^2$.

    \item Denote $\CS^\bfe$ as the gapped domain wall between $\fZ(\vect_{\bZ_4})$ and $\TC$ by condensing $\one\oplus \bfe^2$; and $\CS^\bfm$ as the gapped domain wall between $\fZ(\vect_{\bZ_4})$ and $\TC$ by condensing $\one\oplus \bfm^2$.
    Note that $\Aut_{E_2}(\TC)\cong \bZ_2$, so there are totally $2\times 2\times 2=8$ domain walls associated to the 2-Morita class $\{\one\oplus \bfe^2,\one\oplus\bfm^2\}$\footnote{Here we implicitly choose $\phi$ as autoequivalence within $\CC_{A_1}^{loc}$ instead of braided equivalence between $\CC_{A_1}^{loc}$ and $\CC_{A_2}^{loc}$.}.

    \item Invertible domain walls are characterized by braided autoequivalence $\Aut_{E_2}(\fZ(\vect_{\bZ_4}))=\bZ_4^{\times}\times \bZ_2\cong \bZ_2\times\bZ_2$, where the first $\bZ_2$ is generated by $1-3$ order exchange $\phi_{1-3}$, i.e. $\bfe\mapsto \bfe^3, \bfm\mapsto\bfm^3$ and the second $\bZ_2$ is generated by $\bfe-\bfm$ exchange $\phi_{\bfe-\bfm}^{\bZ_4}$.
    So there are totally four invertible domain walls.
\enu

We have combined the above situations pictorially in table 
\ref{table:Z4}, in which there are 22 gapped domain walls in  $\fZ(\vect_{\bZ_4})$ in total.


\begin{table}[H]
    \centering
    \begin{tabular}{|c|c|c|c|c|c|}
            \hline
            $H$& $F$& \makecell{2d condensable algebras \\ in $\fZ(\vect_{\bZ_4})$}&  \makecell{Condensed phase \\ $\fZ(\vect_{\bZ_4})^{loc}_{A}$} & Domain walls & Total: 22\\
            \hline
            $\{e\}$ & $\{e\}$& $\one\oplus \bfe \oplus \bfe^2\oplus \bfe^3$ &  \multirow{3}{*}{\small $\vect$}  & \multirow{3}{*}{             
                \begin{tikzpicture}[scale=0.75]
                \filldraw[fill=gray!40, draw=white] (-2,0) rectangle (-0.5,1.5);
                \filldraw[fill=gray!40, draw=white] (1,0) rectangle (2.5,1.5);
                \node at(0.25, 0.75){\ltiny $\vect$};
                \node at(-1.3, 0.75){\ltiny $\fZ(\vect_{\bZ_4})$};
                \node at(1.7, 0.75){\ltiny $\fZ(\vect_{\bZ_4})$};
                \node at(-0.4, 0){\tiny $3$};
                \node at(0.9, 0){\tiny $3$};
                \draw[thick](-0.5,0)--(-0.5,1.5);
                \draw[thick](1,0)--(1,1.5);
                \node at(0, 1.5){\quad};
            \end{tikzpicture} 
            } & \multirow{3}{*}{\makecell{\small non-invertible:\\ $3\times 3 = 9$}} \\
            $\bZ_2$ & $\bZ_2$ & $\one\oplus \bfe^2 \oplus \bfm^2\oplus \bff^2$&& &\\
            $\bZ_4$ & $\bZ_4$ & $\one\oplus \bfm \oplus \bfm^2\oplus \bfm^3$&& & \\
            \hline
             & & & \multirow{3}{*}{\small $\CD \CS$}& \multirow{3}{*}{             
                \begin{tikzpicture}[scale=0.75]
                \filldraw[fill=gray!40, draw=white] (-2,0) rectangle (-0.5,1.5);
                \filldraw[fill=gray!40, draw=white] (1,0) rectangle (2.5,1.5);
                \filldraw[fill=gray!20, draw=white] (-0.5,0) rectangle (1,1.5);
                \node at(0.25, 0.75){\ltiny $\CD\CS$};
                \node at(-1.3, 0.75){\ltiny $\fZ(\vect_{\bZ_4})$};
                \node at(1.7, 0.75){\ltiny $\fZ(\vect_{\bZ_4})$};
                \draw[thick](-0.5,0)--(-0.5,1.5);
                \draw[thick](1,0)--(1,1.5);
                \node at(0, 1.5){\quad};
            \end{tikzpicture} 
            } & \\
            $\bZ_4$ & $\bZ_2$ & $\one\oplus \bff^2$&  && \small non-invertible: 1\\
            & & & && \\
            \hline
             & & &\multirow{4}{*}{\small $\TC$} &\multirow{4}{*}{             
                \begin{tikzpicture}[scale=0.75]
                \filldraw[fill=gray!40, draw=white] (-2,0) rectangle (-0.5,1.5);
                \filldraw[fill=gray!40, draw=white] (1,0) rectangle (2.5,1.5);
                \filldraw[fill=gray!20, draw=white] (-0.5,0) rectangle (1,1.5);
                \node at(0.2, 0.75){\ltiny $\TC$};
                \node at(-1.3, 0.75){\ltiny $\fZ(\vect_{\bZ_4})$};
                \node at(1.7, 0.75){\ltiny $\fZ(\vect_{\bZ_4})$};
                \draw[thick](-0.5,0)--(-0.5,1.5);
                \draw[thick](1,0)--(1,1.5);
                \draw[ dashed](0.25,0)--(0.25,1.5);
                \node at(0.25, 1.6){\tiny $2$};
                \node at(-0.4, -0.1){\tiny $2$};
                \node at(0.9, -0.1){\tiny $2$};
                \node at(0, 1.5){\quad};
            \end{tikzpicture} 
            } & \\
            $\bZ_4$ & $\bZ_2$ & $\one\oplus \bfm^2$& && \small \makecell{non-invertible: \\
            $2\times 2\times 2=8$ }\\
            $\bZ_2$ & $\{e\}$ & $\one\oplus \bfe^2$&  &&\\
            \hline
             & & &\multirow{3}{*}{\small $\fZ(\vect_{\bZ_4})$ } &\multirow{3}{*}{             
                \begin{tikzpicture}[scale=0.75]
                \filldraw[fill=gray!40, draw=white] (-2,0) rectangle (1,1.5);
                \node at(-1.3, 0.75){\ltiny $\fZ(\vect_{\bZ_4})$};
                \node at(0.2, 0.75){\ltiny $\fZ(\vect_{\bZ_4})$};
                \draw[dashed](-0.5,0)--(-0.5,1.5);
                \node at(-0.5, 1.65){\tiny $4$};
                \node at(0, 1.4){\quad};
            \end{tikzpicture} 
            } & \\
            $\bZ_4$ & $\{e\}$ & $\one$&  && \small invertible: 4\\
             & & & \rule{0pt}{3ex} && \\
            \hline    
        \end{tabular} 
      \caption{Seven 2d condensable algebras in $\fZ(\vect_{\bZ_4})$ are listed in the third column of the table. 
      There are four 2-Morita equivalence classes of  condensable algebra that condense to $\vect, \CD \CS, \TC, \fZ(\vect_{\bZ_4})$ respectively. 
      We have 22 domain walls in total (drawn in the fifth column), which can be written as the bimodule of twenty-two 1d condensable algebras $B_i$ in $\fZ(\vect_{\bZ_4})$ (see table \ref{table:Z4_alg}), i.e. ${}_{B_i} \fZ(\vect_{\bZ_4})_{B_i}$.
      Four invertible domain walls in  $\fZ(\vect_{\bZ_4})$ can be counted by $\Aut_{E_2}(\fZ(\vect_{\bZ_4}))$.}\label{table:Z4}
\end{table}

\begin{rem}
    One may use the module category $\CD(H,K)$ of modified quantum double to describe these domain walls \cite{BM07,HBJP23} clearly.
    The quantum double $\rep(D(G))\simeq \fZ(\vect_{G})$ can be written as $\CD(G,G)$, and the gapped domain wall produced by condensing 2d condensable algebra $A(H,F)$ in $\CD(G,G)$ is $\CD(G/F,H)$. 
\end{rem}

\vspace{2em}

We can also use A$\overrightarrow{\text{rrow}}$ 6 in Trinity \ref{fig:alg_cycle} to recover 1d condensable algebras from pairs of 2-Morita equivalent condensable algebras $A_1 \widesim[3]{2-Morita}{\phi} A_2$, here $\fZ(\vect_{\bZ_4})$ provides a non-trivial example of extended tensor product.

\bit
    \item When the condensed phase of $A_1$ and $A_2$ is $\vect$, the 1d condensable algebras are indecomposable subalgebras of $A_1\ot A_2$, this produce $B_i$ for $i=5,7,8,{13}, {14}, {15},{16},{19},{20}$. 
    
    \item When the condensed phase is double semion, $B_4$ is obtained by the extension of $B_{\Id}=\one\in\CD\CS$ over $\one\oplus\bff^2$, which is $\one\oplus\bff^2$ itself, this recovers $B_4$.
    
    \item When the condensed phase is $\bZ_2$ topological order $\TC$, here $A_1$/$A_2$ is either $\one \oplus\bfe^2$ or $\one\oplus\bfm^2$, we discuss each situation below:
    \bit
        \item When $A_1=A_2$ and the symmetry $\phi\in\Aut_{E_2}(\TC)$ is trivial, extension of $B_{\Id}=\one\in\TC$ over $A_1$ recovers $B_{2}$ and $B_3$,  which are just $\one\oplus \bfe^2$ and $\one\oplus \bfm^2$.
        
        \item When $A_1=A_2$ but the symmetry $\phi\in\Aut_{E_2}(\TC)$ is the non-trivial $\bfe-\bfm$ exchange in $\TC$, we have $B_{\phi_{\bfe-\bfm}}=\one\oplus\bff\in\TC$.
        To compute the extension of $B_{\phi_{\bfe-\bfm}}$ over $A_1$, we need to explicitly know how the condensation process from $\fZ(\vect_{\bZ_4})$ to $\TC$ is controlled by $A_1$.
        For example, by condensing $\one\oplus \bfe^2$, we have $\one\oplus \bfe^2 \mapsto \one$ and $\bfe\bfm^2\oplus \bfe^3\bfm^2\mapsto \bff$, so $\Ext^R_{\one\oplus \bfe^2}(\one\oplus \bff)=\one\oplus \bfe^2\oplus \bfe\bfm^2\oplus \bfe^3\bfm^2$.
        This calculation process can give $B_{11}$ and $B_{12}$.

        \item When $A_1\neq A_2$ but the symmetry $\phi\in\Aut_{E_2}(\TC)$ is trivial. 
        We can use $\bfe-\bfm$ exchange in $\fZ(\vect_{\bZ_4})$ to reduce this case to the case that $A_1=A_2$ with trivial inner symmetry.
        For example, when $A_1=\one\oplus\bfm^2$ and $A_2=\one\oplus\bfe^2$, the domain wall ${}_{B_{18}}\CC_{B_{18}}$ can be regarded as a fusion of $\bfe-\bfm$ exchange domain wall 
        with the domain wall ${}_{B_{2}}\CC_{B_{2}}$, see the following figure
        
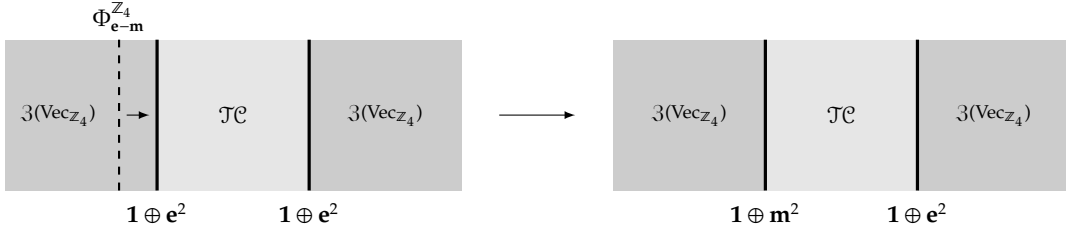
\begin{figure}[H]
            \centering
            \begin{tikzpicture}
                \filldraw[draw=none, fill=gray!40] (-7, 0)rectangle(-1,2);
                \filldraw[draw=none, fill=gray!40] (1, 0)rectangle(7,2);
                \filldraw[draw=none, fill=gray!20] (-5, 0)rectangle(-3,2);
                \filldraw[draw=none, fill=gray!20] (5, 0)rectangle(3,2);
                \draw[very thick](-5, 0)--(-5,2);
                \draw[very thick](5, 0)--(5,2);
                \draw[very thick](-3, 0)--(-3,2);
                \draw[very thick](3, 0)--(3,2);
                \draw[dashed, thick](-5.5, 0)--(-5.5,2);
                \draw[-latex](-5.4, 1)--(-5.1,1);
                \draw[-latex](-0.5, 1)--(0.5,1);
                \node[]at(-6.3, 1){\scriptsize $\fZ(\vect_{\bZ_4})$};
                \node[]at(-5.5, 2.3){\small $\Phi_{\bfe-\bfm}^{\bZ_4}$};
                \node[]at(-5, -0.3){\small $\one\oplus \bfe^2$};
                \node[]at(-3, -0.3){\small $\one\oplus \bfe^2$};
                \node[]at(-4, 1){\small $\TC$};
                \node[]at(-2, 1){\scriptsize $\fZ(\vect_{\bZ_4})$};
        
                \node[]at(6, 1){\scriptsize $\fZ(\vect_{\bZ_4})$};
                \node[]at(5, -0.3){\small $\one\oplus \bfe^2$};
                \node[]at(3, -0.3){\small $\one\oplus \bfm^2$};
                \node[]at(4, 1){\small $\TC$};
                \node[]at(2, 1){\scriptsize $\fZ(\vect_{\bZ_4})$};
                
            \end{tikzpicture}
        \end{figure}
        The corresponding 1d condensable algebra $B_{18}$ also should be the indecomposable subalgebra of the tensor product of $B_{10}$ and $B_2$ (by Proposition \ref{prp:fusing_rule_of_gapped_domain_walls}), i.e. $(\one\oplus \bfe\bfm^3\oplus \bfe^2\bfm^2\oplus \bfe^3\bfm)\ot (\one\oplus \bfe^2)$.
        By this method, we can also obtain $B_{17}$.

        \item When $A_1\neq A_2$ and the symmetry $\phi\in\Aut_{E_2}(\TC)$ is the non-trivial $\bfe-\bfm$ exchange in $\TC$, the corresponding 1d condensable algebra is the indecomposable subalgebra of $\Ext^R_{A_1}(\one\oplus \bff) \ot \Ext^L_{A_2}(\one\oplus \bff)$.
        Since $\Ext^R_{\one\oplus\bfe^2}(\one\oplus \bff)$ and $\Ext^L_{\one\oplus\bfm^2}(\one\oplus \bff)$ are $B_{11}$ and $B_{12}$, we obtain $B_{21}$ and $B_{22}$ immediately.
    \eit

    \item For $A_1=\one=A_2$, $B_{\phi}=\ot(L_{\phi})$ is indeed the 1d condensable algebra we need.
    By applying tensor functor on four $\phi$-twisted lagrangian algebras $L_{\phi}$, we obtain 1d condensable algebras $\one, \one\oplus \bfe^2\oplus \bfm^2\oplus \bff^2,\one\oplus\bff\oplus\bff^2\oplus\bff^3,\one\oplus\bfe\bfm^3\oplus\bfe^2\bfm^2\oplus\bfe^3\bfm$, which are indeed the last four 1d condensable algebras $B_1,B_6,B_9,B_{10}$.
\eit

\begin{rem}
    In general, for $\fZ(\vect_{\bZ_n})$, the 1d condensable algebra that corresponds to $\bfe-\bfm$ exchange has the form $\bigoplus_{i=0}^{n-1}\bfe^i\bfm^{n-i}$. And $\bigoplus_{i=0}^{n-1}\bff^i$ corresponds to $\bfe-\bfm$ exchange composing with $1 \Leftrightarrow n-1$ exchange.
\end{rem}



For a general abelian gauge symmetry $G$, we can use the metric group $(G\times \hat{G},q)$ to describe the MTC $\FZ(\vect_{G})$ \cite{EGNO15}.  Then we can use the KS bicharacter to compute the left/right center of a 1d condensable algebra $\bC[H,\omega]$.
\begin{defn}[\cite{KS94,FRS04a}]
    Let $(G,q)$ be a pre-metric group.
    For a subgroup $H\subset G$, we define a {\bf Kreuzer--Schellekens bicharacter} (abbreviated as KS bicharacter) on $H$ as a bicharacter 
    \begin{align*}
        \Xi :H\times H\to \bC^{\times}
    \end{align*} 
    such that $\Xi(g,g)=q(g)$ for each $g\in H$.
\end{defn}
A natural choice of a KS bicharacter is the braiding $\beta$ of the pointed braided fusion category $\CC(G,q)$, in which $\CC(G,q)$ one-to-one corresponds to pre-metric group $(G,q)$ up to equivalence.
However, since $\CC(G,q)$ represents a braided equivalence class, there does not exist a canonical choice of braiding.
So we fix a braiding $\beta$, then for a 2-cohomology class $\omega \in \mathrm{H}^2(H,\bC^{\times})$, we can obtain a KS bicharacter associated to $\omega$:
\begin{align}
    \Xi^{\omega}(g,h):=\beta_{g,h} \dfrac{\omega(g,h)}{\omega(h,g)}
\end{align}

\begin{thm}[\cite{KS94,FRS04a}]
    Let $B=\bC[H,\omega]$ be a 1d condensable algebra in $\FZ(\vect_{G})\simeq \CC(G\times \hat{G},q)$.
    Then its left/right center has the support 
    \begin{align*}
        K_l(H,\omega):=\{g\in H\mid \Xi^{\omega}(h,g)=1,\, \forall h\in H\}\\
        K_r(H,\omega):=\{g\in H\mid \Xi^{\omega}(g,h)=1,\, \forall h\in H\}
    \end{align*}
    i.e. $Z_l(B)\cong \bigoplus_{g\in K_l(H,\omega)} \bC g$ and $Z_r(B)\cong \bigoplus_{g\in K_r(H,\omega)} \bC g$.
\end{thm}
The above Theorem provides a general method to compute left and right centers directly from group theoretical data that can be applied to any finite abelian group $G$.

\vspace{2em}

\subsubsection{\texorpdfstring{$\fZ(\vect_{S_3})$}{Z(vecS3)} and non-abelian cases}\label{sec:expl_S3}
In this subsection, we discuss general finite gauge symmetry $G$, in which it might be non-abelian.
We can use \Ar{2} to find all 2-Morita equivalent condensable algebras. Lagrangian algebras in $\fZ(\vect_{G \times G})$ can be characterized based on characters.

We first review characters of $\FZ(\vect_{G})$ which are similar to characters in representation theory of finite groups.
\begin{defn}[\cite{Bantay94,Dav10a}]\label{dfn:2-char}
    Let $x$ be an object of $\FZ(\vect_{G})$, we define the {\bf character} $\chi_x$ associated to $x$ to be the  map from $C^2(G):=\{(g,h)\in G\times G\mid gh=hg\}$ to $\bC^{\times}$:
    \begin{align*}
        \chi_x:C^2(G)&\to\bC^{\times}\\
        (g,h)&\mapsto \tr(x_g(h)) 
    \end{align*}
    where $x_g$ is the $g$-grading component of $x$, which is a $G$-representation.
\end{defn}
We can compute characters for all simple objects in $\FZ(\vect_{G})$ to obtain a basis of space of characters.
We call them the {\bf irreducible} characters.

Next, we find all lagrangian algebras in $\fZ(\vect_{G \times G})$.
By Theorem \ref{thm:classification_lag_alg_ZVecG}, lagrangian algebras $L(H,\omega)$ in $\fZ(\vect_{G \times G})$ are uniquely determined by a subgroup $H\subset G\times G$ and a 2-cohomology class $\omega\in\mathrm{H}^2(H,\bC^{\times})$ up to conjugation.
And by the following theorem, we can directly write down the character associated to $L(H,\omega)$ from group-theoretical data.
\begin{thm}[\cite{Dav10a}]\label{thm:character}
    Let $L(H,\omega)$ be the lagrangian algebra in $\FZ(\vect_{G})$ associated to the pair $(H,\omega)$.
    We have 
    \begin{align*}
        \chi_{L(H,\omega)}(g,h)=\dfrac{1}{|H|}\sum_{x\in G,xgx^{-1},xhx^{-1}\in H}\dfrac{\omega(xgx^{-1},xhx^{-1})}{\omega(xhx^{-1},xgx^{-1})}
    \end{align*}
\end{thm}
Since $\{\chi_x\mid x\in\mathrm{Irr}(\fZ(\vect_{G \times G}))\}$ form a basis of space of characters, we can write the character $\chi_{L(H,\omega)}$ as a liner combination of these irreducible ones.
This decomposition tells us the support of any lagrangian algebra $L(H,\omega)$ in $\fZ(\vect_{G \times G})$.

\begin{expl}
    Consider $G=\bZ_2$. There are four irreducible characters $\chi_{\one}=(1,1,0,0)$, $\chi_{\bfe}=(1,-1,0,0)$, $\chi_{\bfm}=(0,0,1,1)$ and $\chi_{\bff}=(0,0,1,-1)$.
    The two subgroup $\bZ_2$ and $\{e\}$ determine two lagrangian algebras in $\fZ(\vect_{\bZ_2})$.
    By Theorem \ref{thm:character}, we have $\chi_{L(\bZ_2)}=(1,1,1,1)=\chi_{\one}+\chi_{\bfm}$ and $\chi_{L(\{e\})}=(2,0,0,0)=\chi_{\one}+\chi_{\bfe}$.
    Thus, we recover $L(\bZ_2)=\one\oplus\bfm$ and $L(\{e\})=\one\oplus \bfe$.
\end{expl}

After finding characters of all lagrangian algebras in $\fZ(\vect_{G \times G})\simeq \FZ(\vect_{G})\boxtimes \overline{\FZ(\vect_{G})}$ explicitly, we can write them as the sum over irreducible characters,
\begin{align*}
    \chi_{L}=\sum_{ij}Z_{ij}\chi_i\chi_j^*.
\end{align*}
Here the coefficient matrix $Z_{ij}$ take values in integers $\bZ$.
$\chi_L$ is also called the modular invariant {\it partition function} associated to $L$ \cite{BE00}.
The lagrangian algebra $L$ can be written as $L=\bigoplus_{i,j} Z_{ij} i\boxtimes \bar{j}$.
By intersecting all lagrangian algebras with left/right components of $\FZ(\vect_{G})\boxtimes \overline{\FZ(\vect_{G})}$, which is Arrow 2 in Trinity \ref{fig:alg_cycle}, we obtain all 2-Morita equivalent condensable algebras $A_l \widesim[3]{2-Morita}{} A_r$ in $\FZ(\vect_{G})$.
Indeed, $A_l=L\cap(\FZ(\vect_{G})\boxtimes \overline{\one})=\bigoplus_{i} Z_{i0} i\boxtimes \overline{\one}$ is determined by the first column $Z_{i0}$ of the coupling matrix $Z_{ij}$ and $A_r=L\cap(\one\boxtimes \overline{\FZ(\vect_{G})})=\bigoplus_{j} Z_{0j} \one\boxtimes \overline{j}$ is determined by the first row $Z_{0j}$ of the coupling matrix $Z_{ij}$.

\begin{rem}
    A finite group $G$ can be promoted to a finite 2-group $\CG$ canonically.
    Similar to the representation theory of finite group, 2-representations of 2-group can also be characterized by {\bf 2-characters} \cite{HXZ24}.
    Indeed, Definition \ref{dfn:2-char} is the {\bf joint} 2-character of $\CG$ defined on torus. where the commutative condition of the subset $C^2(G)$ is induced by the compatible condition on torus.
    The language of 2-characters of 2-groups provides a natural comprehension of Definition \ref{dfn:2-char} and the relation to partition functions.
\end{rem}

\vspace{1.5em}
Now we apply above method to the simplest example of non-Abelian group, i.e. the symmetric group $S_3$.
For a MTC with $S_3$ gauge symmetry, it can be described categorically by $\FZ(\vect_{S_3}) \simeq \fZ(\rep(S_3))$. 
The simple objects of $\fZ(\vect_{S_3})$ are characterized by the conjugacy class $C(g)$ and irreducible representations of its centralizer $Z(C(g))$ \cite{BK01}, in which we obtain 8 simple objects, we denote them by $\bfA$ to $\bfH$ \cite{CCW16}. 
See the table below.

\begin{table}[H]
    \centering
    \caption{$\fZ(\vect_{S_3})$}
    \begin{tabular}{|c|c|c|c|c|c|}
    \hline
    $C(g)$ & $Z(C(g))$ & $\rm IrrRep$ & Simple Obj & Dim &Character  \\
    \hline
    \multirow{3}{*}{$\{e\}$} & \multirow{3}{*}{$S_3$} & $\one$ & $\bfA$ & 1 & $\chi_{\bfA}$ \\
    \cline{3-6}
     &  & $\pi$ & $\bfB$ & 1 & $\chi_{\bfB}$ \\
    \cline{3-6}
     &  & $S$ & $\bfC$ & 2 & $\chi_{\bfC}$ \\
    \hline
    \multirow{3}{*}{$\{t, t^2\}$} & \multirow{3}{*}{$\bZ_3$} & $\one$ & $\bfF$ & 2 & $\chi_{\bfF}$ \\
    \cline{3-6}
     & & $\omega$ & $\bfG$ & 2 & $\chi_{\bfG}$ \\
    \cline{3-6}
     &  & $\omega^2$ & $\bfH$ & 2 & $\chi_{\bfH}$ \\
    \hline
    \multirow{2}{*}{$\{s, st, st^2\}$} & \multirow{2}{*}{$\bZ_2$} & $\one$ & $\bfD$ & 3 & $\chi_{\bfD}$ \\
    \cline{3-6}
     & & $E$ & $\bfE$ & 3 & $\chi_{\bfE}$ \\
    \hline
    \end{tabular}
\end{table}

\begin{itemize}
    \item The category $\rep(S_3)$ of $\bC$-linear representations of $S_3$ has three simple objects: the trivial representation $\one$, the sign representation $\pi$ and the standard representation $S$.
    The fusion rule of $\rep(S_3)$ is given by $\pi\ot \pi\cong \one$, $\pi\ot S\cong S\cong S\ot \pi$, $S\ot S\cong \one\oplus \pi\oplus S$.

    \item Simple objects in $\rep(\bZ_3)$ are denoted by $\one, \omega, \omega^2$.
    
    \item Simple object in $\rep(\bZ_2)$ are denoted by $\one$ and $E$.
\end{itemize}
We also list the corresponding quantum dimensions and the irreducible characters in last two columns, which coincide with Ostrik's and Davydov's notation \cite{Ost03a,Dav10a}.
The value of each irreducible character can be found in \cite[Section 5.3]{DS17}.
Fusion rules of $\fZ(\vect_{S_3})$ are listed in the following table.

\begin{table}[H]
    \centering
    \caption{Fusion rules of $\fZ(\vect_{S_3})$}
    \begin{tabular}{|c|c|c|c|c|c|c|c|c|}
    \hline
    $\ot$ & $\bfA$ & $\bfB$ & $\bfC$ & $\bfD$ & $\bfE$ & $\bfF$ & $\bfG$ & $\bfH$ \\
    \hline
    $\bfA$ & $\bfA$ & $\bfB$ & $\bfC$ & $\bfD$ & $\bfE$ & $\bfF$ & $\bfG$ & $\bfH$ \\
    \hline
    $\bfB$ & $\bfB$ & $\bfA$ & $\bfC$ & $\bfE$ & $\bfD$ & $\bfF$ & $\bfG$ & $\bfH$ \\
    \hline
    $\bfC$ & $\bfC$ & $\bfC$ & $\bfA\oplus \bfB\oplus \bfC$ & $\bfD\oplus \bfE$ & $\bfD\oplus \bfE$ & $\bfG \oplus \bfH$ & $\bfF\oplus \bfH$ & $\bfF\oplus \bfG$ \\
    \hline
    $\bfD$ & $\bfD$ & $\bfE$ & $\bfD\oplus \bfE$ & \makecell{$\bfA \oplus \bfC\oplus \bfF$ \\ $\oplus \bfG\oplus \bfH$}& \makecell{$\bfB \oplus \bfC\oplus \bfF$ \\ $\oplus \bfG\oplus \bfH$} & $\bfD \oplus \bfE$ & $\bfD\oplus \bfE$ & $\bfD\oplus \bfE$ \\
    \hline
    $\bfE$ & $\bfE$ & $\bfD$ & $\bfD\oplus \bfE$ & \makecell{$\bfB \oplus \bfC\oplus \bfF$ \\ $\oplus \bfG\oplus \bfH$}& \makecell{$\bfA \oplus \bfC\oplus \bfF$ \\ $\oplus \bfG\oplus \bfH$} & $\bfD \oplus \bfE$ & $\bfD\oplus \bfE$ & $\bfD\oplus \bfE$ \\
    \hline
    $\bfF$ & $\bfF$ & $\bfF$ & $\bfG \oplus \bfH$ & $\bfD \oplus \bfE$ & $\bfD \oplus \bfE$ & $\bfA \oplus \bfB\oplus \bfF$ & $\bfH \oplus \bfC$ & $\bfG\oplus \bfC$ \\
    \hline
    $\bfG$ & $\bfG$ & $\bfG$ & $\bfF\oplus \bfH$ & $\bfD \oplus \bfE$ & $\bfD \oplus \bfE$ & $\bfH \oplus \bfC$ & $\bfA \oplus \bfB\oplus \bfG$ & $\bfF\oplus \bfC$ \\
    \hline
    $\bfH$ & $\bfH$ & $\bfH$ & $\bfF\oplus \bfG$ & $\bfD \oplus \bfE$ & $\bfD \oplus \bfE$ & $\bfG\oplus \bfC$ & $\bfF\oplus \bfC$ & $\bfA \oplus \bfB\oplus \bfH$ \\
    \hline
    \end{tabular}
\end{table}

Then we can use Theorem \ref{thm:character} to compute characters for each pair $(H,\omega)$ where $H\subset S_3\times S_3$ to determine all lagrangian algebras in $\fZ(\vect_{S_3 \times S_3})$.
Since these characters have been calculated in different literatures \cite{CGR00,Ost03a,Dav10a}\footnote{Note that Davydov's Table miss two coefficients of $|\chi_4|^2$ and $|\chi_5|^2$ terms in $A(A_3\times A_3, \gamma )$ row \cite{Dav10a}.
}, we directly list the results in the first column of Table \ref{table:S3}.
Now by computing $Z_{i0}$ and $Z_{0j}$ of the coupling matrix $Z_{ij}$ of these partition functions, we obtain all 2-Morita equivalent condensable algebras in $\fZ(\vect_{S_3})$ listed in the second column of Table \ref{table:S3}.
There are four classes of 2-Morita equivalent condensable algebras:
\bnu
    \item lagrangian algebras $\{\bfA\oplus \bfF\oplus \bfD,\bfA\oplus \bfB\oplus 2\bfF, \bfA\oplus \bfC\oplus \bfD, \bfA\oplus \bfB\oplus 2\bfC\}$;
    
    \item $\{\bfA\oplus \bfC,\bfA\oplus \bfF\}$ which condense to the $\bZ_2$ topological order $\fZ(\vect_{\bZ_2}) =:\TC$;
    
    \item $\bfA\oplus \bfB$ which condense to the $\bZ_3$ topological order $\fZ(\vect_{\bZ_3})$;
    
    \item trivial condensable algebra $\bfA$.
\enu

\begin{table}
    \centering
    \caption{Results of $\fZ(\vect_{S_3})$}\label{table:S3}
    \begin{tabular}{|m{6.5cm}<{\centering}|m{3.8cm}<{\centering}|m{3.5cm}<{\centering}|}
        \hline
        $\chi_{L(H,\omega)}$ & $A_l /A_r$& Domain Wall \\
        \hline
        $|\chi_{\bfA}+\chi_{\bfF}+\chi_{\bfD}|^2$ & $\bfA\oplus \bfF\oplus \bfD$, $\bfA\oplus \bfF\oplus \bfD$ & $\rep(S_3)\btd \rep(S_3)$ \\
        \hline
        $(\chi_{\bfA}+\chi_{\bfF}+\chi_{\bfD})(\chi_{\bfA}+\chi_{\bfB}+2\chi_{\bfF})^*$ & $\bfA\oplus \bfF\oplus \bfD$, $\bfA\oplus \bfB\oplus 2\bfF$ & $\rep(S_3)\btd \vect_{S_3}^\bfF$ \\
        \hline
        $(\chi_{\bfA}+\chi_{\bfF}+\chi_{\bfD})(\chi_{\bfA}+\chi_{\bfC}+\chi_{\bfD})^*$ & $\bfA\oplus \bfF\oplus \bfD$, $\bfA\oplus \bfC\oplus \bfD$ & $\rep(S_3)\btd \rep(S_3)^\bfC$ \\
        \hline
        $(\chi_{\bfA}+\chi_{\bfF}+\chi_{\bfD})(\chi_{\bfA}+\chi_{\bfB}+2\chi_{\bfC})^*$ & $\bfA\oplus \bfF\oplus \bfD$, $\bfA\oplus \bfB\oplus 2\bfC$ & $\rep(S_3)\btd \vect_{S_3}$ \\
        \hline
        $|\chi_{\bfA}+\chi_{\bfB}+2\chi_{\bfF}|^2$ & $\bfA\oplus \bfB\oplus 2\bfF$, $\bfA\oplus \bfB\oplus 2\bfF$ & $\vect_{S_3}^\bfF\btd \vect_{S_3}^\bfF$ \\
        \hline
        $(\chi_{\bfA}+\chi_{\bfB}+2\chi_{\bfF})(\chi_{\bfA}+\chi_{\bfF}+\chi_{\bfD})^*$ & $\bfA\oplus \bfB\oplus 2\bfF$, $\bfA\oplus \bfF\oplus \bfD$ & $\vect_{S_3}^\bfF\btd \rep(S_3)$ \\
        \hline
        $(\chi_{\bfA}+\chi_{\bfB}+2\chi_{\bfF})(\chi_{\bfA}+\chi_{\bfC}+\chi_{\bfD})^*$ & $\bfA\oplus \bfB\oplus 2\bfF$, $\bfA\oplus \bfC\oplus \bfD$ & $\vect_{S_3}^\bfF\btd \rep(S_3)^\bfC$ \\
        \hline
        $(\chi_{\bfA}+\chi_{\bfB}+2\chi_{\bfF})(\chi_{\bfA}+\chi_{\bfB}+2\chi_{\bfC})^*$& $\bfA\oplus \bfB\oplus 2\bfF$, $\bfA\oplus \bfB\oplus 2\bfC$ & $\vect_{S_3}^\bfF\btd \vect_{S_3}$ \\
        \hline
        $|\chi_{\bfA}+\chi_{\bfC}+\chi_{\bfD}|^2$ & $\bfA\oplus \bfC\oplus \bfD$, $\bfA\oplus \bfC\oplus \bfD$ & $\rep(S_3)^\bfC\btd \rep(S_3)^\bfC$ \\
        \hline
        $(\chi_{\bfA}+\chi_{\bfC}+\chi_{\bfD})(\chi_{\bfA}+\chi_{\bfF}+\chi_{\bfD})^*$ & $\bfA\oplus \bfC\oplus \bfD$, $\bfA\oplus \bfF\oplus \bfD$ & $\rep(S_3)^\bfC\btd \rep(S_3)$ \\
        \hline
        $(\chi_{\bfA}+\chi_{\bfC}+\chi_{\bfD})(\chi_{\bfA}+\chi_{\bfB}+2\chi_{\bfF})^*$ & $\bfA\oplus \bfC\oplus \bfD$, $\bfA\oplus \bfB\oplus 2\bfF$ & $\rep(S_3)^\bfC\btd \vect_{S_3}^\bfF$ \\
        \hline
        $(\chi_{\bfA}+\chi_{\bfC}+\chi_{\bfD})(\chi_{\bfA}+\chi_{\bfB}+2\chi_{\bfC})^*$ & $\bfA\oplus \bfC\oplus \bfD$, $\bfA\oplus \bfB\oplus 2\bfC$ & $\rep(S_3)^\bfC\btd \vect_{S_3}$ \\
        \hline
        $|\chi_{\bfA}+\chi_{\bfB}+2\chi_{\bfC}|^2$ & $\bfA\oplus \bfB\oplus 2\bfC$, $\bfA\oplus \bfB\oplus 2\bfC$ & $\vect_{S_3}\btd \vect_{S_3}$ \\
        \hline
        $(\chi_{\bfA}+\chi_{\bfB}+2\chi_{\bfC})(\chi_{\bfA}+\chi_{\bfF}+\chi_{\bfD})^*$ & $\bfA\oplus \bfB\oplus 2\bfC$, $\bfA\oplus \bfF\oplus \bfD$ & $\vect_{S_3}\btd \rep(S_3)$ \\
        \hline
        $(\chi_{\bfA}+\chi_{\bfB}+2\chi_{\bfC})(\chi_{\bfA}+\chi_{\bfB}+2\chi_{\bfF})^*$ & $\bfA\oplus \bfB\oplus 2\bfC$, $\bfA\oplus \bfB\oplus 2\bfF$ & $\vect_{S_3}\btd \vect_{S_3}^\bfF$ \\
        \hline
        $(\chi_{\bfA}+\chi_{\bfB}+2\chi_{\bfC})(\chi_{\bfA}+\chi_{\bfC}+\chi_{\bfD})^*$ & $\bfA\oplus \bfB\oplus 2\bfC$, $\bfA\oplus \bfC\oplus \bfD$ &  $\vect_{S_3}\btd \rep(S_3)^\bfC$ \\
        \hline
        \rowcolor{gray!20} $|\chi_{\bfA}+\chi_{\bfF}|^2+|\chi_{\bfB}+\chi_{\bfF}|^2+|\chi_{\bfD}|^2+|\chi_{\bfE}|^2$ & $\bfA\oplus \bfF$, $\bfA\oplus \bfF$ & $\CM^\bfF\boxtimes_{\TC}\CM^\bfF$ \\
        \hline
        \rowcolor{gray!20} $|\chi_{\bfA}+\chi_{\bfF}|^2+(\chi_{\bfB}+\chi_{\bfF})\chi_{\bfD}^*+\chi_{\bfD}(\chi_{\bfB}+\chi_{\bfF})^*+|\chi_{\bfE}|^2$ & $\bfA\oplus \bfF$, $\bfA\oplus \bfF$ & $\CM^\bfF\boxtimes_{\TC}\CM^\bfF$, $\phi_{\bfe-\bfm}$ \\
        \hline
        \rowcolor{gray!20} $(\chi_{\bfA}+\chi_{\bfF})(\chi_{\bfA}+\chi_{\bfC})^*+(\chi_{\bfB}+\chi_{\bfF})(\chi_{\bfB}+\chi_{\bfC})+|\chi_{\bfD}|^2+|\chi_{\bfE}|^2$ & $\bfA\oplus \bfF$, $\bfA\oplus \bfC$ & $\CM^\bfF\boxtimes_{\TC}\CM^\bfC$ \\
        \hline
        \rowcolor{gray!20} $(\chi_{\bfA}+\chi_{\bfF})(\chi_{\bfA}+\chi_{\bfC})^*+(\chi_{\bfB}+\chi_{\bfF})\chi_{\bfD}^*+\chi_{\bfD}(\chi_{\bfB}+\chi_{\bfC})^*+|\chi_{\bfE}|^2$ & $\bfA\oplus \bfF$, $\bfA\oplus \bfC$ & $\CM^\bfF\boxtimes_{\TC}\CM^\bfC$, $\phi_{\bfe-\bfm}$ \\
        \hline
        \rowcolor{gray!20} $|\chi_{\bfA}+\chi_{\bfC}|^2+|\chi_{\bfB}+\chi_{\bfC}|^2+|\chi_{\bfD}|^2+|\chi_{\bfE}|^2$ & $\bfA\oplus \bfC$, $\bfA\oplus \bfC$ & $\CM^\bfC\boxtimes_{\TC}\CM^\bfC$ \\
        \hline
        \rowcolor{gray!20} $|\chi_{\bfA}+\chi_{\bfC}|^2+(\chi_{\bfB}+\chi_{\bfC})\chi_{\bfD}^*+\chi_{\bfD}(\chi_{\bfB}+\chi_{\bfC})^*+|\chi_{\bfE}|^2$ & $\bfA\oplus \bfC$, $\bfA\oplus \bfC$ & $\CM^\bfC\boxtimes_{\TC}\CM^\bfC$, $\phi_{\bfe-\bfm}$ \\
        \hline
        \rowcolor{gray!20} $(\chi_{\bfA}+\chi_{\bfC})(\chi_{\bfA}+\chi_{\bfF})^*+(\chi_{\bfB}+\chi_{\bfC})(\chi_{\bfB}+\chi_{\bfF})+|\chi_{\bfD}|^2+|\chi_{\bfE}|^2$   & $\bfA\oplus \bfC$, $\bfA\oplus \bfF$ & $\CM^\bfC\boxtimes_{\TC}\CM^\bfF$ \\
        \hline
        \rowcolor{gray!20} $(\chi_{\bfA}+\chi_{\bfC})(\chi_{\bfA}+\chi_{\bfF})^*+(\chi_{\bfB}+\chi_{\bfC})\chi_{\bfD}^*+\chi_{\bfD}(\chi_{\bfB}+\chi_{\bfF})^*+|\chi_{\bfE}|^2$ & $\bfA\oplus \bfC$, $\bfA\oplus \bfF$ & $\CM^\bfC\boxtimes_{\TC}\CM^\bfF$, $\phi_{\bfe-\bfm}$ \\
        \hline
        \rowcolor{gray!30} $|\chi_{\bfA}+\chi_{\bfB}|^2+2|\chi_{\bfC}|^2+2|\chi_{\bfF}|^2+2|\chi_{\bfG}|^2+2|\chi_{\bfH}|^2$ & $\bfA\oplus \bfB$, $\bfA\oplus \bfB$ & $\CN\boxtimes_{\fZ(\vect_{\bZ_3})}\CN$ \\
        \hline
        \rowcolor{gray!30} $|\chi_{\bfA}+\chi_{\bfB}|^2+2\chi_{\bfC}\chi_{\bfF}^*+2\chi_{\bfF}\chi_{\bfC}^*+2|\chi_{\bfG}|^2+2|\chi_{\bfH}|^2$ & $\bfA\oplus \bfB$, $\bfA\oplus \bfB$ & $\CN\boxtimes_{\fZ(\vect_{\bZ_3})}\CN$, $\phi_{\bfe-\bfm}^{\bZ_3}$ \\
        \hline
        \rowcolor{gray!60} $|\chi_{\bfA}|^2+|\chi_{\bfB}|^2+|\chi_{\bfC}|^2+|\chi_{\bfF}|^2+|\chi_{\bfG}|^2+|\chi_{\bfH}|^2+|\chi_{\bfD}|^2+|\chi_{\bfE}|^2$ & $\bfA$, $\bfA$ & $\fZ(\vect_{S_3})$\\
        \hline
        \rowcolor{gray!60} $|\chi_{\bfA}|^2+|\chi_{\bfB}|^2+\chi_{\bfC}\chi_{\bfF}^*+\chi_{\bfF}\chi_{\bfC}^*+|\chi_{\bfG}|^2+|\chi_{\bfH}|^2+|\chi_{\bfD}|^2+|\chi_{\bfE}|^2$ & $\bfA$, $\bfA$ & $\Phi_{\bfC-\bfF}$\\
        \hline    
    \end{tabular}
\end{table}

Now we illustrate gapped domain walls associated to these 2-Morita equivalence classes of 2d condensable algebras (see third column of table \ref{table:S3}).

\bnu
    \item 
    \bit
        \item Note that the boundaries condensed by $\bfA\oplus \bfF\oplus \bfD$ and $\bfA\oplus \bfC\oplus \bfD$ are both equivalent to $\rep(S_3)$ as fusion categories, this equivalence is provided by the $\bfC$-$\bfF$ (charge-flux) exchange symmetry.
        The difference between these two boundaries can only be seen by the different actions from the 2d bulk $\fZ(\vect_{S_3})$.
        To distinct them, we denote the boundary condensed by $\bfA\oplus \bfF\oplus \bfD$ by $\rep(S_3)$, and denote the boundary condensed by $\bfA\oplus \bfC\oplus \bfD$ by $\rep(S_3)^\bfC$.

        \item Unsurprisingly, the boundaries condensed by $\bfA\oplus \bfB\oplus 2\bfF$ and $\bfA\oplus \bfB\oplus 2\bfC$ are both equivalent to $\vect_{S_3}$.
        So we denote the boundary condensed by $\bfA\oplus \bfB\oplus 2\bfF$ by $\vect_{S_3}^\bfF$, and denote the boundary condensed by $\bfA\oplus \bfB\oplus 2\bfC$ by $\vect_{S_3}$.
    \eit
    By combining above four gapped boundaries two by two, we obtain sixteen gapped domain walls in $\fZ(\vect_{S_3})$, which are pictured in the second row of table \ref{table:S3_results}.

    \item Similarly, we denote the gapped domain walls between $\fZ(\vect_{S_3})$ and $\TC$ condensed through $\bfA\oplus \bfF$ and $\bfA\oplus \bfC$ by $\CM^\bfF$ and $\CM^\bfC$  respectively.
    Since $\Aut_{E_2}(\fZ(\vect_{\bZ_2}))\cong \bZ_2$, we have two invertible gapped domain walls in $\TC$ (see section \ref{section:toric_code}). Combining them together, we obtain total eight gapped domain walls in $\fZ(\vect_{S_3})$ (pictured in the third row of table \ref{table:S3_results}).

    \item We denote the domain wall between $\fZ(\vect_{S_3})$ and $\fZ(\vect_{\bZ_3})$ condensed by $\bfA\oplus \bfB$ as $\CN$.
    Note that $\Aut_{E_2}(\fZ(\vect_{\bZ_3}))\cong \bZ_3^{\times}\times \bZ_2\cong \bZ_2\times \bZ_2$ where the first $\bZ_2$ is the $1-2$ order exchange, and the second $\bZ_2$ is the $\bfe-\bfm$ exchange $\phi_{\bfe-\bfm}^{\bZ_3}$ in $\fZ(\vect_{\bZ_3})$. So in principle, we should have four invertible domain walls in $\fZ(\vect_{\bZ_3})$.
    However, there are only two (not four) gapped domain wall in $\fZ(\vect_{S_3})$ associated to $\bfA\oplus \bfB$. This is due to the braided autoequivalence $\phi_{1-2}$ is induced by a non-trivial algebra automorphism of $\bfA\oplus\bfB$ discussed in section \ref{sec:sym}.
    We will explain immediately after 4.
    
    \item Since $\Aut_{E_2}(\fZ(\vect_{S_3}))\cong \bZ_2$, there are two invertible domain walls, one is the trivial wall, another is the $\bfC$-$\bfF$ exchange wall, which can also be regarded as an $S_3$-version electromagnetic duality (pictured in the fifth row of table \ref{table:S3_results}).
\enu

Recall that in section \ref{sec:sym}, we discuss an algebra automorphism of a 2d condensable algebra $A\in\CC$ may induce a non-trivial braided autoequivalence in $\CC_A^{loc}$.
When we fuse domain walls in the 2-step condensation to obtain a direct condensation process, different condensable algebras in the intermediate phase may be extended to the same condensable algebra in the original phase.
This phenomenon happens in the case that the inner part is $\fZ(\vect_{\bZ_3})$. Different invertible domain walls in $\fZ(\vect_{\bZ_3})$ correspond to different lagrangian algebras in $\fZ(\vect_{\bZ_3\times \bZ_3})$, but they can be extended to the same lagrangian algebra in $\fZ(\vect_{S_3 \times S_3})$, which results the same gapped domain wall in $\fZ(\vect_{S_3})$.
  
The condensable algebra $\bfA\oplus\bfB$ admit a non-trivial $\bZ_2$ automorphism 
\begin{align*}
    \bfA\oplus \bfB \xrightarrow{1\oplus -1}\bfA\oplus \bfB
\end{align*}
By Theorem \ref{lem:phi_is_br_auto}, this non-trivial automorphism may induce a non-trivial braided autoequivalence in $\fZ(\vect_{S_3})_{\bfA\oplus\bfB}^{loc}$.
To see whether the induced braided autoequivalence is trivial or not, we compute the condensation process via $\bfA\oplus\bfB$.

By the following adjunction and Schur's Lemma (see appendix \ref{appendix:condensable_algebras})
\begin{align*}
    \hom_{\fZ(\vect_{S_3})_{\bfA\oplus \bfB}}(x\ot (\bfA\oplus \bfB),y\ot (\bfA\oplus \bfB))\cong \hom_{\fZ(\vect_{S_3})}(x,y\ot (\bfA\oplus \bfB)).
\end{align*}
We find $\bfA$ and $\bfB$ are mapped to the same object, which should be the tensor unit $\bfA\oplus \bfB$ in $\fZ(\vect_{S_3})_{\bfA\oplus\bfB}$.
When $x=y=\bfC$, we have $\bfC\ot(\bfA\oplus\bfB)\cong \bfC\oplus\bfC$ in $\fZ(\vect_{S_3})$, and hence $\dim(\hom_{\fZ(\vect_{S_3})_{\bfA\oplus\bfB}}(\bfC\ot(\bfA\oplus\bfB),\bfC\ot(\bfA\oplus\bfB)))=\dim(\hom_{\fZ(\vect_{S_3})}(\bfC,\bfC\oplus\bfC))=2$.
Thus, the free module $\bfC\ot(\bfA\oplus \bfB)$ consists of two inequivalent simple modules, and can only be two $\bfC$ with different $\bfA\oplus \bfB$-actions.
We choose one of them as the standard $\bfC$, and denote another simple module by $\bfC^{tw}$.
The $\bfA\oplus \bfB$-module action on $\bfC^{tw}$ can be induced by that on $\bfC$ through composing with the non-trivial automorphism of $\bfA\oplus\bfB$.
The condensation process of other simple objects are similar to $\bfC$, we summarize them as follows:
\begin{align*}
    -\ot(\bfA\oplus\bfB):\fZ(\vect_{S_3})&\to\fZ(\vect_{S_3})_{\bfA\oplus\bfB}\\
    \bfA\mapsto \bfA\oplus \bfB \quad \quad \bfB\mapsto &\bfA\oplus \bfB\quad \quad \bfC\mapsto \bfC\oplus \bfC^{tw}\\
    \bfD\mapsto \bfD\oplus \bfD^{tw} \quad & \quad \bfE\mapsto \bfE\oplus \bfE^{tw}\\
    \bfF\mapsto \bfF\oplus \bfF^{tw}\quad \quad \bfG\mapsto &\bfG\oplus \bfG^{tw}\quad \quad \bfH\mapsto \bfH\oplus \bfH^{tw}
\end{align*}
Local modules can be determined by computing $S$-matrix \cite{CGR00}, which encodes information of double braidings.
Results are listed as follows
\begin{align*}
    \fZ(\vect_{S_3})_{\bfA\oplus\bfB}^{loc}&\xrightarrow{\sim} \fZ(\vect_{\bZ_3})\\
    \bfA\oplus \bfB&\mapsto \one \\
    \bfC\mapsto \bfe\quad &\quad \bfC^{tw}\mapsto \bfe^2\\
    \bfF\mapsto \bfm\quad &\quad \bfF^{tw}\mapsto \bfm^2\\
    \bfG\mapsto \bfe\bfm\quad &\quad \bfG^{tw}\mapsto \bfe^2\bfm^2\\
    \bfH\mapsto \bfe^2\bfm\quad &\quad \bfH^{tw}\mapsto \bfe\bfm^2
\end{align*}
Some assignments between simple local $\bfA \oplus \bfB$-modules and simple objects in $\fZ(\vect_{\bZ_3})$ are based on the fact that $\bfC$ corresponds to charge $\bfe$ and $\bfF$ corresponds to flux $\bfm$, and $\bfF,\bfG,\bfH$ form the representations of $\bZ_3$.
Other assignments are based on fusion rules of $\fZ(\vect_{S_3})$.

It is clear that the non-trivial automorphism of $\bfA\oplus\bfB$ induce the $1-2$ exchange $\phi_{1-2}$ in $\fZ(\vect_{\bZ_3})$.
And two extended lagrangian algebras that are connected by $\phi_{1-2}$ must be isomorphic in $\fZ(\vect_{S_3})$ according to Theorem \ref{thm:phi_alg_iso}, so do extended lagrangian algebras in the folded phase $\fZ(\vect_{S_3})\btd \overline{\fZ(\vect_{S_3})}$.
As a consequence, the total number of gapped domain walls in $\fZ(\vect_{S_3})$ should be the number of invertible domain walls in $\fZ(\vect_{\bZ_3})$ quotient by the $\phi_{1-2}$ action, i.e. $4/2=2$.
(See the fourth row of table \ref{table:S3_results}).

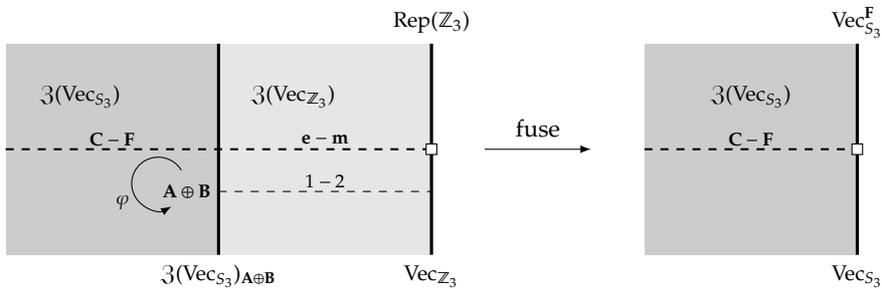
\begin{figure}[H]
    \footnotesize
    \centering
    \begin{tikzpicture}[scale=1.4]
        \filldraw[draw=none, fill=gray!40] (-4, 0)rectangle(-2,2);
        \filldraw[draw=none, fill=gray!20] (-2, 0)rectangle(0,2);
        \filldraw[draw=none, fill=gray!40] (2, 0)rectangle(4,2);
        \draw[very thick](-2, 0)--(-2,2);
        \draw[very thick](4, 0)--(4,2);
        \draw[very thick](0, 0)--(0,2);
        \draw[thick, dashed](-4, 1)--(0,1);
        \draw[dashed](-2, 0.6)--(0,0.6);
        \draw[thick, dashed](2, 1)--(4,1);

        \draw[-latex](0.5, 1)--(1.5,1);
        \node[]at(-2, -0.2){$\fZ(\vect_{S_3})_{\bfA \oplus \bfB}$};
        \node[]at(-3.3, 1.5){$\fZ(\vect_{S_3})$};
        \node[]at(3, 1.5){$\fZ(\vect_{S_3})$};
        \node[]at(4, -0.2){$\vect_{S_3}$};
        \node[]at(4, 2.2){$\vect_{S_3}^{\bfF}$};
        \node[]at(0, 2.2){$\rep(\Zb_3)$};
        \node[]at(0, -0.2){$\vect_{\Zb_3}$};
        \node[]at(-1.3, 1.5){$\fZ(\vect_{\bZ_3})$};
        \node[]at(-1, 1.1){\scriptsize $\bfe-\bfm$};
        \node[]at(-1, 0.7){\scriptsize $1-2$};
        \node[]at(-3, 1.1){\scriptsize $\bfC-\bfF$};
        \node[]at(3, 1.1){\scriptsize $\bfC-\bfF$};
        \node[]at(1, 1.2){\small fuse};
        \node[]at(-2.3, 0.6){\scriptsize $\bfA \oplus \bfB$};
        
        \draw[-latex] (-2.35,0.8) arc(30:300:0.25);
        \node[]at(-2.9, 0.5){\scriptsize $\varphi $};
        \filldraw[fill=white] (-0.05, 0.95)rectangle(0.05,1.05);
        \filldraw[fill=white] (3.95, 0.95)rectangle(4.05,1.05);
    \end{tikzpicture}
\end{figure}

The $\bfC-\bfF$ exchange domain wall in $\fZ(\vect_{S_3})$ can cross the $\bfA\oplus\bfB$-condensation and becomes the $\bfe-\bfm$ exchange domain wall in $\fZ(\vect_{\bZ_3})$.
    So $\bfA \oplus \bfB \widesim[3]{2-Morita}{\phi_{\bfe-\bfm}^{\bZ_3}} \bfA\oplus \bfB$ classifies the two gapped domain walls of $\fZ(\vect_{S_3})$ with inner phase $\fZ(\vect_{\bZ_3})$.
Different from the situation of Double Ising condenses to Toric code (Example \ref{proof:exp_DIS}, see also Example \ref{exp:exp_DIS} in next section), $1-2$ exchange induced by non-trivial automorphism in $\bfA\oplus\bfB$ does not generate distinguishable gapped boundaries after 2-step condensation. This is due to  $\fZ(\vect_{\bZ_3})$'s two gapped boundaries $\rep(\bZ_3)$ and $\vect_{\bZ_3}$ are obtained by condensing $\one\oplus\bfm\oplus\bfm^2$ and $\one\oplus\bfe\oplus\bfe^2$, respectively. 
And $1-2$ exchange acts trivially on these two lagrangian algebras. Extend these two lagrangian algebras in $\fZ(\vect_{\bZ_3})$ back to $\fZ(\vect_{S_3})$, we obtain lagrangian algebras $\bfA\oplus\bfB\oplus 2\bfF$ and $\bfA\oplus\bfB\oplus 2\bfC$, which tells us that the fused boundaries must be $\vect_{S_3}^{\bfF}$ and $\vect_{S_3}$ respectively.

\begin{rem}\label{rem:DNO}
    The fact that four invertible domain walls in $\fZ(\vect_{\bZ_3})$ produce only 2 different domain walls in $\fZ(\vect_{S_3})$ gives a counter example of Theorem 3.6 in \cite{DNO12}, which they use the term "pairwise non-isomorphic" to state that given different triples $(A_1,A_2,\phi)$, the lagrangian algebras $L(A_1, A_2, \phi)$ are non-isomorphic.
    
\end{rem}

\begin{rem}
Consider the MTC $\mathrm{Mod}_V$ for a VOA $V$, let $V'$ be a 2d condensable algebra in $\mathrm{Mod}_V$ and let $L$ be a lagrangian algebra in $\fZ(\mathrm{Mod}_V)$.
Condensing $L$ via condensable algebra $V'\ot \overline{V'}$ we obtain an incarnation $L'$, which is again a lagrangian in the "condensed" phase $\fZ(\mathrm{Mod}_{V'})$.
And the modular invariant $Z_{L'}$ corresponding to lagrangian $L'$ is indeed $Z_L$ written in terms of characters of objects in $\fZ(\mathrm{Mod}_{V'})$.
Physicists have noticed that phenomenon and conjectured that all block diagonal modular invariant can be obtained by some extensions of VOA $V$ (or equivalently, condensable algebras in $\mathrm{Mod}_V$, see Remark \ref{rem:ext_voa}) \cite{SY89}.
In \cite{DV88}, the authors find that the (braided) monoidal autoequivalence\footnote{In their paper and some related literatures, the notion "fusion algebra" is used to denote the (modular) fusion category $\mathrm{Mod}_V$ and "automorphisms of fusion algebra" is used to denote the (braided) monoidal autoequivalences of $\mathrm{Mod}_V$.} in $\mathrm{Mod}_{V'}$ would result the off-diagonal modular invariants.
Moreover, in \cite{MS89Natuality}, Moore and Seiberg show that all modular invariants should be determined by maximal extensions of some "chiral algebras" and braided autoequivalences.
Their result now has been rigorously proved by Theorem \ref{thm:classify_by_lag_alg}.
They also notice that not all braided autoequivalences would give new modular invariants, i.e., there are some redundancy in $\Aut_{E_2}(\mathrm{Mod}_{V'})$, whose reason is now clear through our analysis of algebra automorphisms.
\end{rem}

Above results are summarized in the following table. We also give the condensable algebras classified by Davydov \cite{Dav10} in the third column.
Here $A_3$ denotes the subgroup of order 3 that is isomorphic to $\Zb_3$, and $C_2$ denoted the one of order 2 that is isomorphic to $\Zb_2$.

\begin{table}[H]
    \centering
    \caption{Results in $\fZ(\vect_{S_3})$}\label{table:S3_results}
    \begin{tabular}{|c|c|c|c|c|c|}
        \hline
        $H$ & $F$ & \makecell{2d condensable algebras \\ in $\fZ(\vect_{S_3})$} &  \makecell{Condensed phase \\ $\fZ(\vect_{S_3})^{loc}_A$} & Domain walls & Total: $28$ \\
        \hline
        $S_3$ & $S_3$ & $\bfA\oplus \bfF\oplus \bfD$  &  \multirow{4}{*}{\text{$\vect$}} &\multirow{4}{*}{\begin{tikzpicture}[scale=0.75]
            \filldraw[fill=gray!60, draw=white] (-2,0) rectangle (-0.5,1.5);
            \filldraw[fill=gray!60, draw=white] (1,0) rectangle (2.5,1.5);
            \node at(0.25, 0.75){\ltiny $\vect$};
            \node at(-1.3, 0.75){\ltiny $\fZ(\vect_{S_3})$};
            \node at(1.7, 0.75){\ltiny $\fZ(\vect_{S_3})$};
            \node at(-0.4, 0){\tiny $4$};
            \node at(0.9, 0){\tiny $4$};
            \draw[thick](-0.5,0)--(-0.5,1.5);
            \draw[thick](1,0)--(1,1.5);
            \node at(0, 1.5){\quad};
        \end{tikzpicture} } & \multirow{4}{*}{\makecell{\small non-invertible: \\$4\times 4 =16$}}\\
        $A_3$ & $A_3$ & $\bfA\oplus \bfB\oplus 2\bfF$  &  & & \\
        $C_2$ & $C_2$ & $\bfA\oplus \bfC\oplus \bfD$  &   & & \\
        $\{e\}$ & $\{e\}$ &$\bfA\oplus \bfB\oplus 2\bfC$ & & & \\
        \hline
        \multirow{2}{*}{$S_3$} & \multirow{2}{*}{$A_3$} &\multirow{2}{*}{$\bfA\oplus \bfF$} & \multirow{4}{*}{\text{$\TC$}} & \multirow{4}{*}{\begin{tikzpicture}[scale=0.75]
            \filldraw[fill=gray!60, draw=white] (-2,0) rectangle (-0.5,1.5);
            \filldraw[fill=gray!60, draw=white] (1,0) rectangle (2.5,1.5);
            \filldraw[fill=gray!20, draw=white] (-0.5,0) rectangle (1,1.5);
            \node at(0.2, 0.75){\ltiny $\TC$};
            \node at(-1.3, 0.75){\ltiny $\fZ(\vect_{S_3})$};
            \node at(1.7, 0.75){\ltiny $\fZ(\vect_{S_3})$};
            \draw[thick](-0.5,0)--(-0.5,1.5);
            \draw[thick](1,0)--(1,1.5);
            \draw[ dashed](0.25,0)--(0.25,1.5);
            \node at(0.25, 1.6){\tiny $2$};
            \node at(-0.4, -0.1){\tiny $2$};
            \node at(0.9, -0.1){\tiny $2$};
            \node at(0, 1.5){\quad};
        \end{tikzpicture}  }& \multirow{4}{*}{\makecell{\small non-invertible: \\$2\times 2\times 2 =8$}} \\
        &&&&&\\
        $C_2$ &$\{e\}$ & $\bfA\oplus \bfC$  &   & & \\
        &&&&&\\
        \hline
        &&&&\multirow{3}{*}{\begin{tikzpicture}[scale=0.75]
            \filldraw[fill=gray!60, draw=white] (-2,0) rectangle (-0.5,1.5);
            \filldraw[fill=gray!60, draw=white] (1,0) rectangle (2.5,1.5);
            \filldraw[fill=gray!30, draw=white] (-0.5,0) rectangle (1,1.5);
            \node at(0.25, 0.75){\ltiny $\fZ(\vect_{\bZ_3})$};
            \node at(-1.3, 0.75){\ltiny $\fZ(\vect_{S_3})$};
            \node at(1.7, 0.75){\ltiny $\fZ(\vect_{S_3})$};
            \draw[thick](-0.5,0)--(-0.5,1.5);
            \draw[thick](1,0)--(1,1.5);
            \draw[dashed](0.25,0)--(0.25,1.5);
            \node at(0.25, 1.65){\tiny $4$};
            \node at(-0.4, -0.1){\tiny $1$};
            \node at(0.9, -0.1){\tiny $1$};
        \end{tikzpicture}  }&\\
        $A_3$ & $\{e\}$& $\bfA\oplus \bfB$ &  $\fZ(\vect_{\bZ_3})$ & & \makecell{\small non-invertible: \\$4/2=2$} \\
        &&&&&\\
        \hline
        &&&&\multirow{3}{*}{             
            \begin{tikzpicture}[scale=0.75]
            \filldraw[fill=gray!60, draw=none] (-2,0) rectangle (1,1.5);
            \node at(-1.3, 0.75){\ltiny $\fZ(\vect_{S_3})$};
            \node at(0.2, 0.75){\ltiny $\fZ(\vect_{S_3})$};
            \draw[dashed](-0.5,0)--(-0.5,1.5);
            \node at(-0.5, 1.65){\tiny $2$};
            \node at(0, 1.4){\quad};
        \end{tikzpicture} 
        }&\\
        $S_3$ &$\{e\}$ & $\bfA$ & $\fZ(\vect_{S_3})$ & &\small invertible: $2$ \\
        &&& \rule{0pt}{3ex}&&\\
        \hline
    \end{tabular}
    
\end{table}

\vspace{2em}
In principle, we can also find 1d condensable algebras in $\fZ(\vect_{\bZ_3})$ and use Arrow 3 to compute 2-Morita equivalent condensable algebras.
But it is not easy to write down 1d condensable algebras in general non-abelian cases.
Here we give a method to find 1d condensable algebras in fusion category $\CC$ based on the pre-knowledge of finite semisimple indecomposable left $\CC$-modules $\CP$ using internal hom, see appendix \ref{appendix:module_cat}:

\begin{thm}[\cite{KZ17}]\label{thm:indecomposable_left_module_and_1d_cond_alg}
      Let $\CC$ be a fusion category.
      Let $\CP$ be a finite semisimple indecomposable left $\CC$-module.    
      Then $\CP\simeq \CC_{[x,x]}$ for any simple object $x\in P$.
      And $[x,x]$ is a 1d condensable algebra in $\CC$.
\end{thm}
It is easy to see that $[x,x]\widesim[3]{1-Morita}{}[y,y]$ for any $x,y\in\mathrm{Irr}(\CC)$.

By Proposition 4.8 in \cite{DMNO13}, indecomposable left $\CC$-modules $\CP$ are also one-to-one corresponding to isomorphic classes of lagrangian algebras in $\fZ(\CC)$. 
Then we can use the following figure to find indecomposable semisimple module $\CP$ and then to compute internal hom $[x,x]$ using the following adjunction and Schur's Lemma:
\begin{align}\label{eq:adjunction}
    \hom_{\CP}(a\odot x,x)\cong \hom_{\CC}(a,[x,x]).
\end{align}

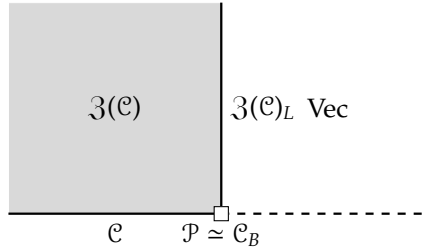
\begin{figure}[H]
    \begin{center}
        \begin{tikzpicture}[scale=0.933]
            \draw [fill=gray!30, draw=white] (-3,0) rectangle (0,3);
            \draw [fill=gray!0, draw=white] (0,0) rectangle (3,3);
            \draw [line width=0.3mm] (0,0)--(0,3);
            \draw [line width=0.3mm] (0,0)--(-3,0);
            \draw [line width=0.3mm,dashed] (0,0)--(3,0);            \draw [fill=gray!0,,draw] (-0.1,-0.1) rectangle (0.1,0.1);
            \node at (-1.5,1.5){$\fZ(\CC)$};
            \node at (1.5,1.5){$\vect$};
            \node [below] at (-1.5,0){$\CC$};
            \node [below] at (0,0){$\CP\simeq \CC_B$};
            \node [right] at (0,1.5){$\fZ(\CC)_L$};
        \end{tikzpicture}
    \end{center}  
    \caption[]{Correspondence between finite semisimple indecomposable left $\CC$-modules and lagrangian algebras in $\fZ(\CC)$}
    \label{fig:temp}
\end{figure}


In case of $S_3$,
an obvious choice of indecomposable $\fZ(\vect_{S_3})$-module is itself whose module action is the tensor product of $\fZ(\vect_{S_3})$.
Then the dual module can only be $\fZ(\vect_{S_3})$, which is a boundary  $\fZ(\fZ(\vect_{S_3}))_L$ of $\fZ(\fZ(\vect_{S_3}))$.
The corresponding lagrangian algebra $L$ is the canonical lagrangian algebra $\bigoplus_{x\in\mathrm{Irr}(\fZ(\vect_{S_3}))}x\btd x^*$.
Consider $[\bfA,\bfA]$ as an example, then equation \ref{eq:adjunction} becomes
\begin{align*}
      \hom_{\fZ(\vect_{S_3})}(x\ot \bfA,\bfA)\cong \hom_{\fZ(\vect_{S_3})}(x,[\bfA,\bfA])
\end{align*}
Since only for $x=\bfA$ we have $\bfA\ot \bfA\cong \bfA$, so $[\bfA,\bfA]\cong \bfA$.
Similarly, we can obtain $[\bfB,\bfB]\cong \bfA$, $[\bfC,\bfC]\cong \bfA\oplus \bfB\oplus \bfC$, $[\bfD,\bfD]\cong \bfA\oplus \bfC\oplus \bfF\oplus \bfG\oplus \bfH$, $[\bfE,\bfE]\cong \bfA\oplus \bfC\oplus \bfF\oplus \bfG\oplus \bfH$, $[\bfF,\bfF]\cong \bfA\oplus \bfB\oplus \bfF$, $[\bfG,\bfG]\cong \bfA\oplus \bfB\oplus \bfG$ and $[\bfH,\bfH]\cong \bfA\oplus \bfB\oplus \bfH$. They are all 1-Morita equivalent.

We can use \Ar{5} to check the above calculations give the correct 1d condensable algebras.
By tensoring the canonical lagrangian algebras, i.e.:
\begin{align*}
    & (\bfA \btd \bfA)\oplus (\bfB \btd \bfB) \oplus (\bfC\btd \bfC)\oplus (\bfD \btd \bfD) \oplus (\bfE \btd \bfE) \oplus (\bfF \btd \bfF) \oplus (\bfG \btd \bfG) \oplus (\bfH \btd \bfH)  \\
    & \qquad \qquad \qquad \qquad \qquad \qquad \qquad  \downarrow \ot  \\
    & \bfA \oplus \bfA \oplus (\bfA \oplus \bfB\oplus \bfC) \oplus (\bfA\oplus \bfC \oplus \bfF \oplus \bfG \oplus \bfH) \oplus (\bfA\oplus \bfC \oplus \bfF \oplus \bfG \oplus \bfH)\\
    &\oplus (\bfA \oplus \bfB\oplus \bfF) \oplus (\bfA\oplus \bfB \oplus \bfG) \oplus (\bfA\oplus \bfB \oplus \bfH) 
\end{align*}
We see above 1-Morita equivalent condensable algebras appear as direct summands.
Or to say, using this method, 1d condensable algebras can be recognized in the huge image of lagrangian algebras under the tensor functor.


If we choose $\fZ(\fZ(\vect_{S_3}))_L \simeq \rep(S_3)\btd \rep(S_3)$, recall that $\fZ(\fZ(\vect_{S_3}))\simeq \fZ(\vect_{S_3})\btd\overline{\fZ(\vect_{S_3})}$, then $\CP$ can be determined through the folding trick.
See the following figure.
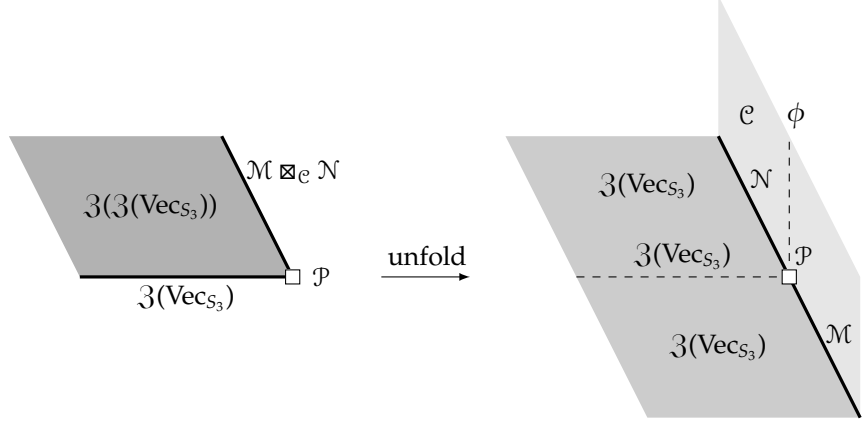
\begin{figure}[H]
    \centering
    \begin{tikzpicture}[scale=0.933]
        \filldraw[draw=none, fill=gray!60] (-3, 1)--(0,1)--(-1, 3)--(-4,3);
        \draw[very thick](-3, 1)--(0,1)--(-1, 3);
        \filldraw[rotate=0, fill=white] (-0.1, 0.9)rectangle(0.1,1.1);
        \draw[-latex](1.25,1)--(2.5,1);
        \node[]at(1.9, 1.3){unfold};
        \node[]at(-2, 2){$\FZ(\fZ(\vect_{S_3}))$};
        \node[]at(-1.5, 0.7){$\fZ(\vect_{S_3})$};
        \node[]at(0.4, 1){$\CP$};
        \node[]at(0, 2.5){$\CM\boxtimes_{\CC}\CN$};
  
        \filldraw[draw=none, fill=gray!40] (5, -1)--(8,-1)--(6, 3)--(3,3);
        \filldraw[draw=none, fill=gray!20] (8,-1)--(8,1)--(6, 5)--(6,3);
        \draw[very thick](8,-1)--(6,3);
        \draw[dashed](4,1)--(7,1);
        \draw[dashed](7,1)--(7,3);
        \filldraw[rotate=0, fill=white] (6.9, 0.9)rectangle(7.1,1.1);
        \node[]at(5.5, 1.3){$\fZ(\vect_{S_3})$};
        \node[]at(5, 2.3){$\fZ(\vect_{S_3})$};
        \node[]at(6, 0){$\fZ(\vect_{S_3})$};
        \node[]at(7.7, 0.2){$\CM$};
        \node[]at(6.6, 2.4){$\CN$};
        \node[]at(7.2, 1.3){$\CP$};
        \node[]at(6.4, 3.3){$\CC$};
        \node[]at(7.1, 3.3){$\phi$};
    \end{tikzpicture}
\end{figure}

In this case $\CM\simeq \CN\simeq \rep(S_3)$ and $\CC\simeq \vect$.
So a natural choice of $\CP$ is $\rep(S_3)$ itself.
The module action of $\fZ(\vect_{S_3})$ on $\rep(S_3)$ is given by first forgetting $\fZ(\vect_{S_3})$ to $\rep(S_3)$ then tensoring with $\rep(S_3)$.
By \cite{CCW16}, we have \footnote{Note that the assignment of $\bfF$ in \cite{CCW16} miss a $\pi$.}
\begin{align*}
    \fZ(\vect_{S_3})&\to \rep(S_3)\\
      \bfA\mapsto \one\quad \bfB &\mapsto \pi\quad \bfC\mapsto S\quad \\
      \bfD\mapsto \one\oplus S &\quad \bfE\mapsto \pi\oplus S\\
      \bfF\mapsto \one\oplus \pi\quad \bfG &\mapsto  S\quad \bfH\mapsto S
\end{align*}
Now by the adjunction \ref{eq:adjunction}, we can compute $[\one,\one]$ as follows
\begin{align*}
      \hom_{\rep(S_3)}(x\odot \one, \one)\cong \hom_{\fZ(\vect_{S_3})}(x,[\one,\one])
\end{align*}
Since $\bfA$, $\bfD$ and $\bfF$ forget to $\one$, $\one\oplus S$ and $\one\oplus \pi$ respectively, we have $[\one,\one]\cong \bfA\oplus \bfD\oplus \bfF$.
Similarly, we have $[\pi,\pi]\cong \bfA\oplus \bfD\oplus \bfF$ and $[S,S]\cong \bfA\oplus \bfB\oplus \bfC\oplus 2\bfD\oplus 2\bfE\oplus 2\bfF\oplus \bfG\oplus \bfH$.

Again consider the corresponding lagrangian algebra $(\bfA\oplus \bfF\oplus \bfD)\btd (\overline{\bfA}\oplus\overline{\bfF}\oplus \overline{\bfD})$.
By acting tensor functor, we get 
\begin{align*}
    &(\bfA\oplus \bfF\oplus \bfD)\ot (\bfA \oplus \bfF\oplus \bfD)\\
    =&\bfA\oplus (\bfA\oplus \bfB\oplus \bfF)\oplus (\bfA\oplus \bfC\oplus \bfF\oplus \bfG\oplus \bfH) \oplus \bfD\oplus \bfD\oplus \bfF\oplus \bfF\oplus (\bfD\oplus \bfE)\oplus (\bfD\oplus \bfE)
\end{align*}
It is clear that the image of the lagrangian algebra under the tensor functor is a direct sum of 1-Morita equivalent condensable algebras.

A more non-trivial case is $\fZ(\fZ(\vect_{S_3}))_L\simeq \rep(S_3)\btd \vect_{S_3}$.
Now $\CM\simeq \rep(S_3)$, $\CN\simeq \vect_{S_3}$ and $\CC\simeq \vect$.
A natural choice of the 0d defect $\CP$ is the invertible bimodule $\vect$.
For this case, $[\bC,\bC]\cong \bfA\oplus \bfB\oplus 2\bfC\oplus 3\bfD\oplus 3\bfE\oplus 2\bfF\oplus 2\bfG\oplus 2\bfH$ is computed to be the only 1d condensable algebra.

\subsection{Fusion category symmetries}\label{sec:expl_double_is}

Results of Trinity \ref{fig:alg_cycle} is not limited to the traditional topological orders with group gauge symmetries. In this section we perform some examples which are related to the fusion category symmetries \cite{TW19,JW20,KLWZZ20a}.

One example of fusion category symmetry is the non-chiral topological phases defined
by Levin-Wen models (or string-net models) \cite{LW05}, which is the Hamiltonian realizations of the Turaev-Viro 3D topological quantum field theory \cite{TV92}.
Indeed, Kitaev quantum double models $\FZ(\vect_{G})$ illustrated in section \ref{section:ZVec_G} cover a subset of the Levin-Wen models.

A 2+1D Levin-Wen model is associated to a unitary fusion category $\CS$. 
Consider a trivalent lattice, each edge admits some simple objects $\bfa,\bfb,\bfc,\dots\in\CS$.
And for each vertex $v$, we assign a Hilbert space $\cH_v:=\bigoplus_{\bfa,\bfb,\bfc\in\mathrm{Irr}(\CS)}\hom_{\CS}(\bfa\ot\bfb,\bfc)$ on it.
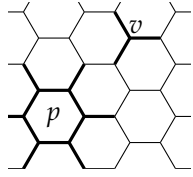
\begin{figure}[H]
    \centering
    \begin{tikzpicture}[scale=0.4]
        \clip (0.5,0) rectangle (6.5,5.5);
        \foreach \i in {0,...,2}{
            \foreach \j in {0,...,3}{
                \draw[](\i*3,{\j*sqrt(3)})--(\i*3+1,{\j*sqrt(3)});
                \draw[](\i*3+1.5,{\j*sqrt(3)+sqrt(3)/2})--(\i*3+2.5,{\j*sqrt(3)+sqrt(3)/2});
                \draw[](\i*3+1,{\j*sqrt(3)})--(\i*3+1.5,{\j*sqrt(3)+sqrt(3)/2});
                \draw[](\i*3+2.5,{\j*sqrt(3)+sqrt(3)/2})--(\i*3+3,{\j*sqrt(3)+sqrt(3)});
                \draw[](\i*3+1,{\j*sqrt(3)+sqrt(3)})--(\i*3+1.5,{\j*sqrt(3)+sqrt(3)/2});
                \draw[](\i*3+2.5,{\j*sqrt(3)+sqrt(3)/2})--(\i*3+3,{\j*sqrt(3)});
            }
        }
        \node[]at(2, 1.7){ $p$};
        \node[]at(4.7, 4.7){ $v$};
        \draw[very thick](1*3,{1*sqrt(3)})--(1*3+1,{1*sqrt(3)});
        \draw[very thick](0*3+1.5,{0*sqrt(3)+sqrt(3)/2})--(0*3+2.5,{0*sqrt(3)+sqrt(3)/2});
        \draw[very thick](0*3+1,{0*sqrt(3)})--(0*3+1.5,{0*sqrt(3)+sqrt(3)/2});
        \draw[very thick](0*3+2.5,{0*sqrt(3)+sqrt(3)/2})--(0*3+3,{0*sqrt(3)+sqrt(3)});
        \draw[very thick](0*3+1,{0*sqrt(3)+sqrt(3)})--(0*3+1.5,{0*sqrt(3)+sqrt(3)/2});
        \draw[very thick](0*3+2.5,{0*sqrt(3)+sqrt(3)/2})--(0*3+3,{0*sqrt(3)});

        \draw[very thick](0*3,{1*sqrt(3)})--(0*3+1,{1*sqrt(3)});
        \draw[very thick](0*3+1.5,{1*sqrt(3)+sqrt(3)/2})--(0*3+2.5,{1*sqrt(3)+sqrt(3)/2});
        \draw[very thick](0*3+1,{1*sqrt(3)})--(0*3+1.5,{1*sqrt(3)+sqrt(3)/2});
        \draw[very thick](0*3+2.5,{1*sqrt(3)+sqrt(3)/2})--(0*3+3,{1*sqrt(3)+sqrt(3)});
        \draw[very thick](0*3+1,{1*sqrt(3)+sqrt(3)})--(0*3+1.5,{1*sqrt(3)+sqrt(3)/2});
        \draw[very thick](0*3+2.5,{1*sqrt(3)+sqrt(3)/2})--(0*3+3,{1*sqrt(3)});

        \draw[very thick](1*3+1,{3*sqrt(3)})--(1*3+1.5,{2*sqrt(3)+sqrt(3)/2});
        \draw[very thick](1*3+1,{2*sqrt(3)})--(1*3+1.5,{2*sqrt(3)+sqrt(3)/2});
        \draw[very thick](1*3+1.5,{2*sqrt(3)+sqrt(3)/2})--(1*3+2.5,{2*sqrt(3)+sqrt(3)/2});

    \end{tikzpicture}
\end{figure}

The Hamiltonian can be written again as a combination of charge operators $\mathsf{Q}_v$ and flux operators $\sB_p$:
$$\mathsf{H}:=\sum_v (1-\mathsf{Q}_v) + \sum_p (1-\sB_p).$$
where the sums run over vertices $v$ and plaquettes $p$ of the honeycomb lattice.
The quasiparticle excitations in above model exhibits $\FZ(\CS)$-topological order.
\cite{KK12} outlines a construction of all possible boundaries and defects in Levin-Wen models.

Like the Toric code Example \ref{section:toric_code}, a local 1d condensable algebra should be a combination of charge and flux operators restrict on the neighborhood of a 1d region. 
And we propose that taking the left/right center to obtain the 2d condensable algebra are again to directly expand the 1d condensable algebra into the left/right bulk such the subalgebras' half braidings from left/right bulk are compatible with their algebraic multiplications.
It is possible to explicitly give a construction of 1d condensable algebras similar to the intertwiners constructed in \cite{LFHSV21}.
And show their left and right centers meet with the construction of 2d condensable algebras that give a 2d condensation of Levin-Wen systems \cite{CGHP23} based on the extended Levin-Wen models of \cite{HGW18}. 
The difficulty lies on how to define algebraic multiplications of 1d condensable algebras. 

Now we give some simple examples of fusion symmetry which can be realized by Levin-Wen model.

\begin{expl}[Double Fibonacci]
    The unitary fusion category $\CF ib$ has simple objects $\one$ and $\tau$, and the fusion rule is given by $\tau \otimes \tau = \one \oplus \tau$ \cite{Ost03classify,BD12}. This makes $\CF ib$ the smallest fusion category where the simple objects do not form a group.
    The double Fibonacci  $\FZ(\CF ib)$ contains a single nontrivial 2d condensable algebra $L=\one \overline{\one} \oplus \tau\overline{\tau}$, which is the canonical Lagrangian algebra. 
    Arrows are all trivial except \Ar{5}, i.e. $\ot(L)\cong \one\oplus\one\oplus\tau$ must be a direct sum of 1-Morita equivalent condensable algebras.
    Thus, we obtain $\one\widesim[3]{1-Morita}{}\one\oplus\tau$.
\end{expl}

\begin{expl}[Double Ising]\label{exp:exp_DIS}
    Consider the Ising topological order $\Is$ with anyons $\one$, $\psi$, $\sigma$ and the fusion rules are given by:
    $ \sigma \ot \sigma = \one \oplus \psi, \sigma \ot \psi = \sigma, \psi \ot \psi = \one.$
    Double Ising $\fZ(\Is)$ admits three 2d condensable algebras $A_0=\one\btd\overline{\one} $, $A_2=(\one \btd \overline{\one}) \oplus (\psi \btd \overline{\psi})$ and $A_L=(\one \btd \overline{\one}) \oplus (\sigma\btd \overline{\sigma})\oplus (\psi \btd \overline{\psi})$, which trivially condense to $\fZ(\Is)$ itself, $\TC$ and $\vect$ respectively \cite{CJKYZ20}.

    \begin{table}[H]
        \centering
        \begin{tabular}{|c|c|c|c|}
                \hline
                 \makecell{2d condensable algebras \\ in $\FZ(\Is)$}&  \makecell{Condensed phase \\ $\FZ(\Is)^{loc}_A$} & Domain walls & Total: 3\\
                \hline
                 & \multirow{3}{*}{$\vect$}& \multirow{3}{*}{             
                    \begin{tikzpicture}[scale=0.75]
                    \filldraw[fill=gray!40, draw=white] (-2,0) rectangle (-0.5,1.5);
                    \filldraw[fill=gray!40, draw=white] (1,0) rectangle (2.5,1.5);
                    \filldraw[fill=white, draw=white] (-0.5,0) rectangle (1,1.5);
                    \node at(0.25, 0.75){\ltiny $\vect$};
                    \node at(-1.3, 0.75){\ltiny $\FZ(\Is)$};
                    \node at(1.7, 0.75){\ltiny $\FZ(\Is)$};
                    \draw[thick](-0.5,0)--(-0.5,1.5);
                    \draw[thick](1,0)--(1,1.5);
                    \node at(-0.4, -0.1){\tiny $1$};
                    \node at(0.9, -0.1){\tiny $1$};
                    \node at(0, 1.5){\quad};
                \end{tikzpicture} 
                } & \\
                 $(\one\boxtimes \overline{\one}) \oplus (\psi\boxtimes \overline{\psi}) \oplus (\sigma\boxtimes \overline{\sigma})$&  && non-invertible: 1\\
                & && \\
                \hline
                  &\multirow{4}{*}{$\TC$} &\multirow{4}{*}{             
                    \begin{tikzpicture}[scale=0.75]
                    \filldraw[fill=gray!40, draw=white] (-2,0) rectangle (-0.5,1.5);
                    \filldraw[fill=gray!40, draw=white] (1,0) rectangle (2.5,1.5);
                    \filldraw[fill=gray!20, draw=white] (-0.5,0) rectangle (1,1.5);
                    \node at(0.2, 0.75){\ltiny $\TC$};
                    \node at(-1.3, 0.75){\ltiny $\FZ(\Is)$};
                    \node at(1.7, 0.75){\ltiny $\FZ(\Is)$};
                    \draw[thick](-0.5,0)--(-0.5,1.5);
                    \draw[thick](1,0)--(1,1.5);
                    \draw[ dashed](0.25,0)--(0.25,1.5);
                    \node at(0.25, 1.65){\tiny $2$};
                    \node at(0, 1.5){\quad};
                \end{tikzpicture} 
                } & \\
                $(\one\boxtimes \overline{\one}) \oplus (\psi\boxtimes \overline{\psi})$& && \makecell{non-invertible: \\
                $2 /2=1$ }\\
                &  &&\\
                \hline
                &\multirow{3}{*}{$\FZ(\Is)$ } &\multirow{3}{*}{             
                    \begin{tikzpicture}[scale=0.75]
                    \filldraw[fill=gray!40, draw=none] (-2,0) rectangle (1,1.5);
                    \node at(-1.3, 0.75){\ltiny $\FZ(\Is)$};
                    \node at(0.2, 0.75){\ltiny $\FZ(\Is)$};
                    \draw[dashed](-0.5,0)--(-0.5,1.5);
                    \node at(-0.5, 1.65){\tiny $1$};
                    \node at(0, 1.4){\quad};
                \end{tikzpicture} 
                } & \\
                $\one\boxtimes \overline{\one}$&  && invertible: 1\\
                &  \rule{0pt}{3ex}&& \\
                \hline    
            \end{tabular} 
    
    \end{table}

    Double Ising has three inequivalent gapped domain walls: $\fZ(\Is)\simeq \Is\btd\overline{\Is}$ is the trivial domain wall, and $\Is\btd \Is$ is the wall induced by condensing lagrangian algebra $A_L$. Since 2d condensable algebras can also be regarded as 1d condensable algebras, $A_L, A_2$ and $A_0$ are just the three $B_i \in \Alg_{E_1}^{cond}(\fZ(\Is))$ correspond to the gapped domain walls.

    Although $\TC$ has an $\bfe-\bfm$ exchange domain wall,
    $\fZ(\Is)_{A_2}\btd_{\TC}\fZ(\Is)_{A_2}$ is the unique domain wall associated to the 2-Morita class $(\one \btd \overline{\one}) \oplus (\psi \btd \overline{\psi})$.
    This is due to the $\bfe$-$\bfm$ exchange symmetry in $\fZ(\Is)_{A_2}^{loc}\simeq \TC$ is induced by the non-trivial algebra automorphism of $A_2$. 
    To explicitly see how the $\bfe$-$\bfm$ exchange is induced, we compute condensation process via $A_2$ more precisely.
    By the following adjunction
    \begin{align*}
        \hom_{\fZ(\Is)_{A_2}}(x\ot A_2, y\ot A_2)\cong \hom_{\fZ(\Is)}(x,y\ot A_2)
    \end{align*}
    when $x=y=\sigma\btd\overline{\sigma}$, since $(\sigma\btd\overline{\sigma})\ot A_2\cong (\sigma\btd\overline{\sigma})\oplus( \sigma\btd\overline{\sigma})$, so the free module $(\sigma\btd\overline{\sigma})\ot A_2$ must consists of two inequivalent simple modules and can only be $\sigma\btd\overline{\sigma}$ equipped with different $A_2$-actions.
    We denote one of them by $\sigma\btd\overline{\sigma}$ and another one by $(\sigma\btd\overline{\sigma})^{tw}$.
    Let $r:(\sigma\btd\overline{\sigma})\ot A_2 \to \sigma\btd\overline{\sigma}$ be the $A_2$-module action of $\sigma\btd\overline{\sigma}$.
    By composing $r$ with the non-trivial algebra automorphism $\varphi:A_2\to A_2$, we obtain another module action $r\circ(\id\ot\varphi)$.
    We can see the following diagram does not commute for any $\lambda\in\bC^{\times}$.
    \begin{figure}[H]
        \centering
        \begin{tikzcd}
            (\sigma\btd\overline{\sigma}) \ot A_2  \ar[r,""{name=s}, "\lambda\ot\id_{A_2}"] \ar[d,"r"'] & (\sigma\btd\overline{\sigma})\ot A_2 \ar[d,"r\circ (\id\ot\varphi)"] \\
            \sigma\btd\overline{\sigma} \ar[r,""{name=t},"\lambda"']& \sigma\btd\overline{\sigma} \arrow[from=s, to=t, pos=.5, phantom, "\ncomm"]
        \end{tikzcd}
    \end{figure}
    So these two modules are not isomorphic to each other. Thus, the module action on $(\sigma\btd\overline{\sigma})^{tw}$ must be $r\circ(\id\ot\varphi)$.
Then $\sigma\btd\overline{\sigma}$ is mapped to $\bfe$ and $(\sigma\btd\overline{\sigma})^{tw}$ is mapped to $\bfm$ (see \cite{CJKYZ20} for their braiding and twist structures), we find two non-free local $A_2$-modules $\bfe$ and $\bfm$ will exchange under the $\varphi$-action.
    This braided autoequivalence is indeed the electromagnetic duality in $\bZ_2$ topological order, the corresponding domain wall is the $\bfe$-$\bfm$-exchange domain wall. 
    See the left sub-figure below.

\begin{figure}[H]
    \centering
    \begin{tikzpicture}[scale=1.4]
        \filldraw[draw=none, fill=gray!40] (-4, 0)rectangle(-2,2);
        \filldraw[draw=none, fill=gray!20] (-2, 0)rectangle(0,2);
        \filldraw[draw=none, fill=gray!40] (2, 0)rectangle(4,2);
        \draw[very thick](-2, 0)--(-2,2);
        \draw[very thick](4, 0)--(4,2);
        \draw[very thick](0, 0)--(0,2);
        \draw[dashed](-2, 1)--(0,1);
        \filldraw[fill=white] (-0.05, 0.95)rectangle(0.05,1.05);
            
        \draw[-latex](0.5, 1)--(1.5,1);
        \node[]at(-2, -0.2){\small  $\FZ(\Is)_{A_2}$};
        \node[]at(-3.3, 1.5){\small $\FZ(\Is)$};
        \node[]at(3, 1){\small $\FZ(\Is)$};
        \node[]at(4, -0.2){\small $\FZ(\Is)_{A_L}$};
        \node[]at(0, 2.2){\small $\TC_{\one \oplus \bfe}$};
        \node[]at(0, -0.2){\small $\TC_{\one\oplus \bfm}$};
        \node[]at(-1.3, 1.5){\small $\TC$};
        \node[]at(-1, 0.8){\scriptsize $\bfe-\bfm$ exchange};
        \node[]at(1, 1.2){\small fuse};
        \node[]at(-2.3, 0.95){\small$A_2$};
        
        \draw[-latex] (-2.35,1.15) arc(30:300:0.25);
    
        \node[]at(-2.9, 1){\scriptsize $\varphi $};
    \end{tikzpicture}
    \caption{When double Ising $\FZ(\Is)$ condense to $\TC$.
    Inclusion $i:A_2\hookrightarrow A_L$ determines a 2-step condensation (see section \ref{sec:sym} for more mathematical details).
    The lagrangian algebra $A_L$ in $\FZ(\Is)$ can become either the lagrangian algebra $\one\oplus \bfe$ or $\one\oplus\bfm$ in $\TC$ depends on whether we compose the non-trivial automorphism $\varphi$ of $A_2$ to $i$ or not. 
   After fusion, we have $\fZ(\Is)_{A_2}\btd_{\TC}\TC_{\one\oplus\bfe}\simeq \FZ(\Is)_{\Ext^R_{A_2}(\one\oplus \bfe)}$ and $\fZ(\Is)_{A_2}\btd_{\TC}\TC_{\one\oplus\bfm}\simeq \FZ(\Is)_{\Ext^R_{A_2}(\one\oplus \bfm)}$. Since $\Ext^R_{A_2}(\one\oplus\bfe)\cong A_L\cong \Ext^R_{A_2}(\one\oplus\bfm)$, there is only one boundary of $\FZ(\Is)$.
    }
\end{figure}

Gapped domain walls $\Phi_{\varphi}$ in condensed phase $\CC_A^{loc}$ that are induced by automorphism $\varphi$ of 2d condensable algebra $A$ does not affect 1 codimensional defects (either boundaries or domain walls) related to the original phase $\CC$. 
We can also understand this triviality through calculating 1d condensable algebras in $\fZ(\Is)$ using \Ar{6}.
    When the condensed phase is $\TC$ via $A_2$, we have two 1d condensable algebras $B_{\Id}=\one$ and $B_{\phi_{\bfe-\bfm}}=\one\oplus\bff$ associated to the symmetry.
    The extension of $\one$ over $A_2$ is $A_2$ itself.
    And the extension $\Ext^R_{A_2}(\one\oplus\bff)$ of $\one\oplus\bff$ over $A_2$ is a direct sum of two $A_2$.
    So the extended 1d condensable algebras are 1-Morita equivalent, which both lead to $\fZ(\Is)_{A_2}\btd_{\TC}\fZ(\Is)_{A_2}$.
\end{expl}

Our method of classifying gapped domain walls can also be applied to chiral MTCs which are beyond the Levin-Wen models. We give an example of the simplest non-trivial anyon condensation happens in $\vect_G^{\alpha}$ for some $\alpha\in\mathrm{H}^3(G,\bC^{\times})$:

\begin{expl}[$\vect_{\bZ_8}^{\alpha}$]
    Consider the chiral MTC 
    $\vect_{\bZ_8}^{\alpha}$ with simple objects $\one, a, a^2...a^7$.
    There is only one condensable algebra given by $\one\oplus a^4$ such that the condensed phase is semion topological order $\vect_{\bZ_2}^{\alpha'}$.
    And there are two braided autoequivalences in $\vect_{\bZ_8}^{\alpha}$ given by identity and $a-a^5$ exchange.
    So $\vect_{\bZ_8}^{\alpha}$ has three gapped domain walls: two invertible ones and $(\vect_{\bZ_8}^{\alpha})_{\one\oplus a^4} \btd_{\vect_{\bZ_2}^{\alpha'}} (\vect_{\bZ_8}^{\alpha})_{\one\oplus a^4}$.
\end{expl}


\section{Generalizations and Outlooks}\label{section:outlook}
The study of condensable algebras related to 2d topological orders was initiated in \cite{BS09}, which developed a theoretical framework illustrating how the condensation of bosonic anyons induces transitions between topologically ordered phases, altering the fusion and braiding properties of the excitations in the system. \cite{Kon14} advanced the theoretical understanding of anyon condensation by formulating it within the framework of tensor categories.
\cite{Bur18} offers a review of anyon condensation, discussing its applications and potential implications for quantum computation. 
Anyon condensation gradually becomes a pivotal concept in understanding phase transitions between different topological phases.
However, there are still a couple of ingredients lack of discussions, for example:
\begin{itemize}
    \item A complete relation between 1-Morita equivalent $E_1$ condensable algebras and 2-Morita equivalent $E_2$ condensable algebras.
    \item The classification gapped domain walls from the perspective of 2-Morita equivalent $E_2$ algebras.
    \item Lattice model realizations of 1d condensable algebras and gapped domain walls.
    \item Fusion relations of gapped domain walls in two-step condensations.
    \item How symmetries induced by algebra automorphisms affect condensation process.
\end{itemize}

In this paper, we fill up these missing parts by studying 2-Morita equivalent condensable algebras in a MTC. Our study provides explicit examples of higher Morita equivalence in 1-category level.

We have accomplished the relation between the $E_2$-Morita equivalent 2d condensable algebras $A_i$ in a modular tensor category $\CC$ and the 1d condensable algebras $B_i$ in the spherical fusion category $\CC$, together with the lagrangian algebras $L_i$ in the Drinfeld center $\fZ(\CC)$ of $\CC$, which is summarized in the Trinity \ref{fig:alg_cycle} exhibited in preliminary.
Physically, by taking module categories of these algebras, we can also translate our result into a topological ordered version (a domain wall version), see the figure below. In which we have accomplished all the arrows proposed in the fusing or folding process of figure \ref{general_picture}.

Some examples are performed in Sec. \ref{section:lattice_model} to illustrate the interplay between these three, including 2d topological orders with abelian and non-abelian gauge group symmetries. We also explicitly write down the proper role of the left/right centers of the $E_1$ algebras in the toric code lattice model.

There are many future works related to our results that can be done. We would like to explicitly discuss some questions based on Witt equivalence in next subsection. 

\subsection{Witt equivalent MTCs} \label{sec:generalize}
Roughly speaking, two MTCs $\CC_1$ and $\CC_2$ are {\bf Witt equivalent} if they can be connected by a gapped domain wall $\CM$. 
More precisely, there exists condensable algebras $A_1 \in \CC_1, A_2 \in \CC_2$, and a braided equivalence $\phi$, such that $(\CC_1)^{loc}_{A_1}\stackrel{\phi}\simeq(\CC_2)^{loc}_{A_2}$ \cite{DMNO13}. Based on this concept, we propose the following definitions:
\begin{defn}
    Let $\CC_1$ and $\CC_2$ be two MTCs, $A_1\in \Alg_{E_2}^{cond}(\CC_1)$ and $A_2\in \Alg_{E_2}^{cond}(\CC_2)$ are {\bf generalized 2-Morita equivalent} if $(\CC_1)_{A_1}^{loc}\simeq (\CC_2)_{A_2}^{loc}$.
\end{defn}
Note that $\CC_1$ and $\CC_2$ are Witt equivalent if and only if there exists a pair of generalized 2-Morita equivalent condensable algebras:

\begin{align*}
    \exists \,\, A_1 \widesim[4]{g. \, 2-Morita}{} A_2 \, \, \, \Leftrightarrow\,\,\,   \CC_1\widesim[3]{Witt}{} \CC_2\\
    \text{($E_2$ algebras) }\,\,\,\,\,\,\,\,\,\,\,\,\,\,\,\,\,\, \,\text{ (MTCs) }
\end{align*}

We believe without proof that the following "generalized Trinity" is also true for two Witt-equivalent MTCs $\CC_1$ and $\CC_2$.

\begin{figure}[H]
    \centering
    \begin{tikzcd}[arrows=Rightarrow]
        \fbox{\begin{tabular}{@{}c@{}c@{}} Generalized 2-Morita \\equivalent condensable algebras\\ in $\CC_1$ and $\CC_2$
        \end{tabular}} \ar[rr, shift left=-0.5ex, "\text{6. generalized extended tensor}"'] \ar[ddr,"\text{1. Symmetry $\phi$ $+$ Extension}" ] && \fbox{\begin{tabular}{@{}c@{}c@{}}1-Morita class of 1d \\
            condensable algebras in $\CM$\end{tabular}} \ar[ll, shift left=-0.5ex,"\text{3. left and right center}"'] \ar[ddl,"\text{4. Full center}"' ]\\
        &&\\
        &\fbox{\begin{tabular}{@{}c@{}}Lagrangian algebras\\ in $\CC_1\boxtimes \overline{\CC_2}$\end{tabular}} \ar[uul, shift left=1.5ex, start anchor={[yshift=0.8ex,xshift=-1ex]},"\text{2. $\cap$ with components}"] \ar[uur, shift left=-1.5ex, start anchor={[yshift=0.9ex,xshift=0.95ex]},end anchor={[yshift=0.8ex,xshift=1ex]}, "\text{5. Forget}"']&
    \end{tikzcd}
    \caption{The results of Trinity \ref{fig:alg_cycle} should be also true when generalized to gapped domain walls between any two Witt equivalent MTCs $\CC_1$ and $\CC_2$. 
    }
\end{figure}
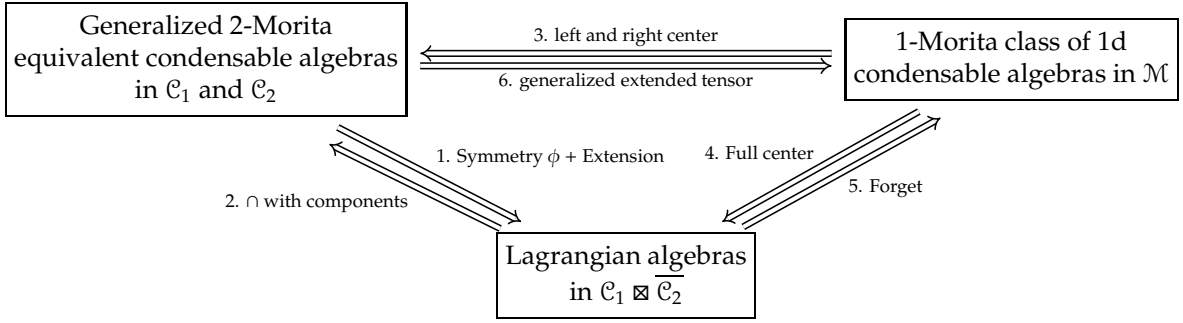

More things can be discussed based on this generalization. For example, Let $\CM$ be a gapped boundary wall of $\CC_1 \boxtimes \overline{\CC_2}$, or to say $\FZ(\CM) \simeq \CC_1 \boxtimes \overline{\CC_2}$, then all stable gapped domain walls between $\CC_1$ and $\CC_2$ can be classified by the ${}_{B_i} \CM_{B_i}$ with $B_i$ a 1d condensable algebra in $\CM$.
Then, according to Theorem \ref{thm:pull_open}, we propose that
\begin{thm}\label{thm:pull_open_generalized}
    Given a pair of topological orders $(\CC_1, \CC_2)$, any stable gapped domain wall $\CM$ between $\CC_1$ and $\CC_2$ such that $\FZ(\CM) \simeq \CC_1 \boxtimes \overline{\CC_2}$ can be written as $(\CC_1)_{A_1} \boxtimes_{(\CC_1)^{loc}_{A_1}} \Phi \boxtimes_{(\CC_2)^{loc}_{A_2}} \boxtimes {}_{A_2} \CC_2$ for some generalized 2-Morita equivalent condensable algebras $(A_1,A_2)$, where $\Phi$ is the invertible domain wall induced by the equivalence $\phi:(\CC_1)_{A_1}^{loc}\xrightarrow{\sim}(\CC_2)_{A_2}^{loc}$.
\end{thm}
The proof of above theorem is similar to the proof of Theorem \ref{thm:pull_open}.
The following conjecture is also straightforward:
\begin{conj}
The generalized 2-Morita equivalent condensable algebra pair $(A_1,A_2)$ can be written as $(Z_l(B) , Z_r(B))$ for $B$ a 1d condensable algebra in $\CM$.
\end{conj}

\begin{figure}[H]
    \centering
    \begin{tikzcd}
        \begin{tikzpicture}
            \centering
            \filldraw[fill=gray!40, draw=none] (-2,-1) rectangle (0,1);
            \filldraw[fill=gray!40, draw=none] (0,-1) rectangle (2,1);
            \draw[thick](0,-1)--(0,1);
            \node at(-1,-0.1){\small $\CC_1$};
            \node at(1,-0.1){\small $\CC_2$};
            \node at(0,-1.3){\small $\CM$};
        \end{tikzpicture}
        \ar[r, "\text{$\quad$ open $\quad$}","\simeq"'] &
        \begin{tikzpicture}
            \centering
            \filldraw[fill=gray!40, draw=none] (-3,-1) rectangle (-1,1);
            \filldraw[fill=gray!20, draw=none] (-1,-1) rectangle (2,1);
            \filldraw[fill=gray!40, draw=none] (2,-1) rectangle (4,1);
            \draw[thick](-1,-1)--(-1,1);
            \draw[thick](2,-1)--(2,1);
            \draw[dashed,thick](0.5,-1)--(0.5,1);
            \node at(-2,-0.1){\small $\CC_1$};
            \node at(3,-0.1){\small $\CC_2$};
            \node at(0.5,-0.1){\small $(\CC_1)^{loc}_{Z_l(B)}\simeq (\CC_2)^{loc}_{Z_r(B)}$};
            \node at(-1,-1.3){\small $(\CC_1)_{Z_l(B)}$};
            \node at(2,-1.3){\small ${}_{Z_r(B)} \CC_2$};
            \node at(.50,-1.3){\small $\Phi$};
        \end{tikzpicture}
    \end{tikzcd}
    \caption{Theorem \ref{thm:pull_open} tells us any stable gapped domain wall in a MTC $\CC$ can be 'pulled open'. Theorem \ref{thm:pull_open_generalized} further tells us any gapped domain wall between Witt equivalent topological orders can be opened to contain an inner condensed topological order.
    }
\end{figure}
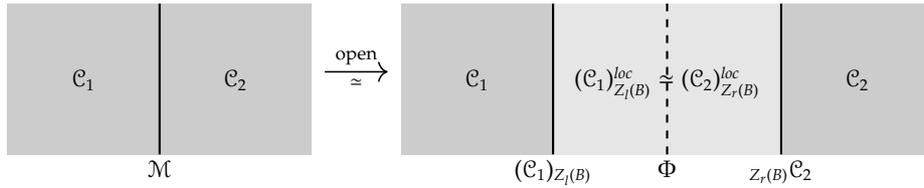

Two MTCs $\CC_1$ and $\CC_2$ are Witt equivalent also means that there exists a larger MTC $\CB$ and two condensable algebras $A_1$ and $A_2$ in $\CB$ such that $\CB_{A_1'}^{loc}\simeq \CC_1$ and $\CB_{A_2'}^{loc}\simeq \CC_2$ \cite{DMNO13}.

For two phases $\CC_1$ and $\CC_2$, we define their common condensed phase that has maximal quantum dimension to be the greatest common divisor.
Similarly, we define the common original phase $\CB$ that has the minimal quantum dimension to be their least common multiple.

Note that the least common multiple may not be unique.
For example, consider $\fZ(\vect_{\bZ_2})$ and $\fZ(\vect_{\bZ_3})$, it is clear that $\fZ(\vect_{\bZ_6})\simeq \fZ(\vect_{\bZ_2})\boxtimes \fZ(\vect_{\bZ_3})$ is a least common multiple.
Also, $\fZ(\vect_{S_3})$ is another common multiple of $\bZ_2$ and $\bZ_3$ topological orders. 
Since it has the same quantum dimension as $\fZ(\vect_{\bZ_6})$, they are both the least common multiples of $\fZ(\vect_{\bZ_2})$ and $\fZ(\vect_{\bZ_3})$.


Similar process of figure \ref{general_picture} happens. We can unfold a (probably unstable) gapped boundary (which is a multi-fusion category) of $\fZ(\CC) \simeq \CC\boxtimes \overline{\CC}$ to be a gapped domain wall in $\CC$ (\ref{general_picture} (c) to \ref{general_picture} (b)), and any (probably unstable) gapped domain wall can be opened to contain an interlayer MTC (probably have larger quantum dimension) (\ref{general_picture} (b) to \ref{general_picture} (a)).

On the other hand, since any pair of MTCs ($\CC_1, \CC_2$) in the same Witt class can be obtained from a single 2d phase $\CB$ via two different 2d condensations, we propose the following proposition by Theorem 1.4.8 in \cite{DR18}:

\begin{prop}
    Any gapped domain wall that can be written as  ${}_{A_1} \CB \boxtimes_{\CB}\CB_{A_2}$, in which ${\CB}_{A_1}^{loc} \simeq \CC_1$ and ${\CB}_{A_2}^{loc} \simeq \CC_2$, 
    must be a direct sum of indecomposable gapped domain walls between $\CC_1$ and $\CC_2$.
\end{prop}


\begin{figure}[H]
    \centering
    \begin{tikzcd}
        \begin{tikzpicture}
            \centering
            \filldraw[fill=gray!40, draw=none] (-3,-1) rectangle (-1,1);
            \filldraw[fill=gray!60, draw=none] (-1,-1) rectangle (1,1);
            \filldraw[fill=gray!40, draw=none] (1,-1) rectangle (3,1);
            \draw[thick](-1,-1)--(-1,1);
            \draw[thick](1,-1)--(1,1);
            \node at(-2,-0.1){\small $\CC_1 \simeq \CB^{loc}_{A_1}$};
            \node at(2,-0.1){\small $\CC_2\simeq\CB_{A_2}^{loc}$};
            \node at(0,-0.1){\small $\CB$};
            \node at(-1,-1.3){\small ${}_{A_1} \CB$};
            \node at(1,-1.3){\small $\CB_{A_2}$};
        \end{tikzpicture}
        =\bigoplus
        \begin{tikzpicture}
            \centering
            \filldraw[fill=gray!40, draw=none] (-3,-1) rectangle (-1,1);
            \filldraw[fill=gray!20, draw=none] (-1,-1) rectangle (1,1);
            \filldraw[fill=gray!40, draw=none] (1,-1) rectangle (3,1);
            \draw[thick](-1,-1)--(-1,1);
            \draw[thick](1,-1)--(1,1);
            \node at(-2,-0.1){\small $\CC_1 $};
            \node at(2,-0.1){\small $\CC_2$};
            \node at(0,-0.1){\footnotesize $(\CC_1)_{A_1'}^{loc}\simeq (\CC_2)_{A_2'}^{loc}$};
            \node at(-1,-1.3){\small $(\CC_1)_{A_1'}$};
            \node at(1,-1.3){\small ${}_{A_2'} \CC_2$};
        \end{tikzpicture}
    \end{tikzcd}
\end{figure}

$\CB$ drawn in the left sub-figure that can condense to $\CC_1$ and $\CC_2$, is related to the phase $(\CC_1)_{A_1'}^{loc}$ that condensed from $\CC_1$ and $\CC_2$ via "direct sum" of gapped domain walls. 
However, the precise decomposition has not been studied.
Here we provide an inspiration via the 'splitting channel"(see the following figure), which tells us that $\CM$ controls the decomposition.
This may give us a method to re-construct the bigger original phase from the smaller condensed phase in a 2-step condensation process.

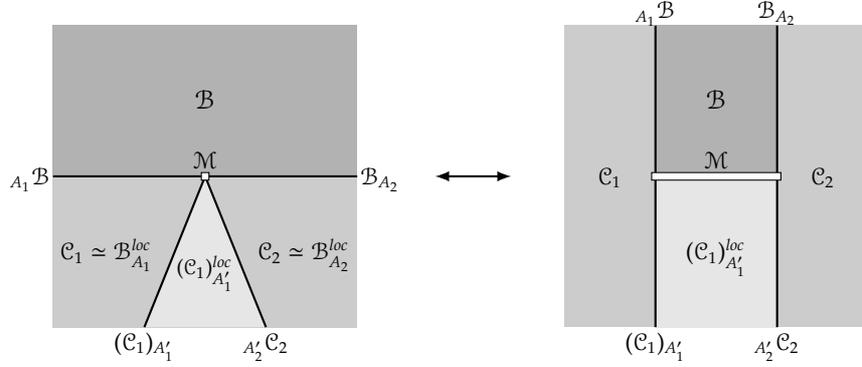
\begin{figure}[H]
    \centering
    \begin{tikzcd}
        \begin{tikzpicture}
            \centering
            \filldraw[fill=gray!60, draw=none] (-2,1) rectangle (2,3);
            \filldraw[fill=gray!40, draw=none] (-2,-1) rectangle (2,1);
            \filldraw[fill=gray!20, draw=none] (-0.8,-1)--(0.8,-1)--(0,1)--(-0.8,-1);
            \draw[thick](-2,1)--(2,1);
            \draw[thick](-0.8,-1)--(0,1);
            \draw[thick](0.8,-1)--(0,1);
            \node at(-1.3,-0.1){\small $\CC_1 \simeq \CB^{loc}_{A_1}$};
            \node at(1.3,-0.1){\small $\CC_2\simeq\CB_{A_2}^{loc}$};
            \node at(0,1.9){$\CB$};
            \node at(0,-0.3){\footnotesize $(\CC_1)_{A_1'}^{loc}$};
            \node at(0,1.1){\small ${\CM}$};
            \node at(-2.3,0.9){\small ${}_{A_1} \CB$};
            \node at(2.3,0.9){\small ${\CB}_{A_2}$};
            \node at(-0.8,-1.3){\small $(\CC_1)_{A_1'}$};
            \node at(0.8,-1.3){\small ${}_{A_2'} \CC_2$};
            \filldraw[fill=white] (-0.05, 0.95)rectangle(0.05,1.05);
        \end{tikzpicture}
        \begin{tikzpicture}
            \centering
            \filldraw[fill=white, draw=none] (-1,-1) rectangle (1,3);
            \draw[latex-latex, thick] (-0.7,1) -- (0.3,1);
        \end{tikzpicture}
        \begin{tikzpicture}
            \centering
            \filldraw[fill=gray!40, draw=none] (-2,-1) rectangle (2,3);
            \filldraw[fill=gray!60, draw=none] (-0.8,1) rectangle (0.8,3);
            \filldraw[fill=gray!20, draw=none] (-0.8,-1) rectangle (0.8,1);
            \draw[thick](-0.8,-1)--(-0.8,3);
            \draw[thick](0.8,-1)--(0.8,3);
            \node at(-0.8,3.1){\small ${}_{A_1} \CB$};
            \node at(0.8,3.1){\small $\CB_{A_2}$};
            \node at(-1.4,0.9){\small $\CC_1$};
            \node at(1.4,0.9){\small $\CC_2$};
            \node at(0,1.9){$\CB$};
            \node at(0,1.1){\small ${\CM}$};
            \node at(0,-0.1){\small $(\CC_1)_{A_1'}^{loc}$};
            \node at(-0.8,-1.3){\small $(\CC_1)_{A_1'}$};
            \node at(0.8,-1.3){\small ${}_{A_2'} \CC_2$};
            \filldraw[fill=white] (-0.85, 0.95)rectangle(0.85,1.05);
        \end{tikzpicture}
    \end{tikzcd}
    \caption{The upper part of the right panel corresponds to the decomposable gapped domain wall, and the bottom part corresponds to the indecomposable gapped domain walls. 0d defect $\CM$ controls the splitting channel between ${}_{A_1} \CB \boxtimes_{\CB}\CB_{A_2}$ and $(\CC_1)_{A_1'} \boxtimes_{(\CC_1)_{A_1'}^{loc}}{}_{A_2'} \CC_2$, thus give the relation from $\CB$ to $(\CC_1)_{A_1'}^{loc} \simeq (\CC_1)_{A_2'}^{loc}$}
\end{figure}





In particular, if we consider only one condensable algebra $A\in \CB$, then, ${}_A \CB_A$, as the indecomposable domain wall in $\CB$, just corresponds to the decomposable domain wall ${\Fun_{\CC}(\CB_A,\CB_A)}$ in $\CC$.
The following figure depicts this situation, which can be understood as a deformed configuration of a normal anyon condensation process.
Since we can bend domain wall $\CB_A$ to different directions without losing information, it is natural to conclude that anyon condensation process is reversible if we consider all condensation descendants.


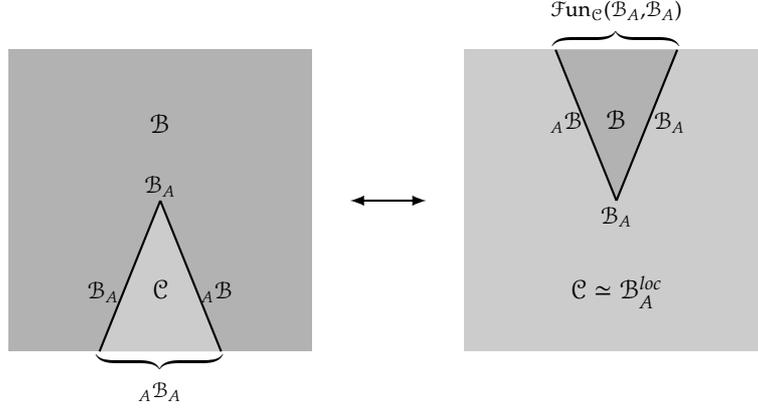
\begin{figure}[H]
    \centering
    \begin{tikzcd}
        \begin{tikzpicture}
            \centering
            \filldraw[fill=gray!60, draw=none] (-2,-1) rectangle (2,3);
            \filldraw[fill=gray!40, draw=none] (-0.8,-1)--(0.8,-1)--(0,1)--(-0.8,-1);
            \draw[thick](-0.8,-1)--(0,1);
            \draw[thick](0.8,-1)--(0,1);
            \node at(0,1.9){$\CB$};
            \node at(0,-0.3){$\CC$};
            \node at(0,1.1){\small ${\CB}_{A}$};
            \node at(-0.75,-0.3){\small ${\CB}_{A}$};
            \node at(0.75,-0.3){\small ${}_{A} \CB$};
            \node[]at(0, -0.9){\large $\underbrace{\qquad \qquad }_{{}_A\CB_{A}}$};
        \end{tikzpicture}
        \begin{tikzpicture}
            \centering
            \filldraw[fill=white, draw=none] (-1,-1) rectangle (1,3);
            \draw[latex-latex, thick] (-0.5,1) -- (0.5,1);
        \end{tikzpicture}
        \begin{tikzpicture}
            \centering
            \filldraw[fill=gray!40, draw=none] (-2,-1) rectangle (2,3);
            \filldraw[fill=gray!60, draw=none] (-0.8,3)--(0.8,3)--(0,1)--(-0.8,3);
            \draw[thick](-0.8,3)--(0,1);
            \draw[thick](0.8,3)--(0,1);
            \node at(0,1.95){$\CB$};
            \node at(0,-0.3){$\CC \simeq \CB_A^{loc}$};
            \node at(0,0.7){\small ${\CB}_{A}$};
            \node at(-0.65,1.95){\small ${}_{A} \CB$};
            \node at(0.7,1.95){\small ${\CB}_{A}$};
            \node[]at(0, 2.9){\large $\overbrace{\qquad \qquad }^{\Fun_{\CC}(\CB_A,\CB_A)}$};
        \end{tikzpicture}
    \end{tikzcd}
\end{figure}
It is also natural to consider the fusion of gapped domain walls in $\CC$ through the multiplication algorithms of 1d condensable algebras $B$.
We prove that (see appendix \ref{appendix:Inv_monoidal_bimod} for details) 
\begin{prop}\label{B_fusion_over_C}
    ${}_{B_1}\CC_{B_1}\boxtimes_{\CC} {}_{B_2}\CC_{B_2}\simeq {}_{B_1\ot B_2}\CC_{B_1\ot B_2}$
\end{prop}
up to 1-Morita equivalence. 
\begin{figure}[H]
    \centering
    \begin{tikzpicture}
        \filldraw[draw=none, fill=gray!40] (-7, 0)rectangle(-1,2);
        \filldraw[draw=none, fill=gray!20] (-5.8, 0)rectangle(-4.6,2);
        \filldraw[draw=none, fill=gray!20] (-3.4, 0)rectangle(-2.2,2);
        \draw[very thick](-5.8, 0)--(-5.8,2);
        \draw[very thick](-4.6, 0)--(-4.6,2);
        \draw[very thick](-3.4, 0)--(-3.4,2);
        \draw[very thick](-2.2, 0)--(-2.2,2);
        \draw[-latex](-4.5, 1.5)--(-4.1,1.5);
        \draw[-latex](-3.5, 1.5)--(-3.9,1.5);
        \node[]at(-6.4, 1){\small $\CC$};
        \node[]at(-4, 1){\small $\CC$};
        \node[]at(-1.6, 1){\small $\CC$};


        \node[rotate = 180] at (-5.2, 2.2) {$\underbrace{\hspace{1cm}}$};
        \node[rotate = 180] at (-2.8, 2.2) {$\underbrace{\hspace{1cm}}$};
        \node[]at(-5.2, 2.5){${}_{B_1}\CC_{B_1}$};
        \node[]at(-2.8, 2.5){${}_{B_2}\CC_{B_2}$};
        \node[] at (-4, -0.2) {$\underbrace{\hspace{3.5cm}}$};
        \node[]at(-4, -0.5){${}_{B_1\ot B_2}\CC_{B_1\ot B_2}$};

    \end{tikzpicture}
\end{figure}

\begin{expl}
    For $\CC:=\TC$, 
    \begin{itemize}
        \item choosing $B_1:=\one\oplus \bff=B_2$, we have ${}_{\one\oplus\bff}\TC_{\one\oplus\bff}=\Phi_{\bfe-\bfm}$.
        Consider ${}_{\one\oplus\bff}\TC_{\one\oplus\bff} \btd_{\TC} {}_{\one\oplus\bff}\TC_{\one\oplus\bff}$, this should be equivalent to ${}_{(\one\oplus\bff)\ot(\one\oplus\bff)}\TC_{(\one\oplus\bff)\ot (\one\oplus\bff)}$.
        Note that $(\one\oplus \bff) \ot (\one\oplus \bff)$ as an algebra is not a direct sum of two $\one \oplus \bff$, but a matrix algebra
            $\begin{pmatrix}
                \one &\bff\\
                \bff &\one
            \end{pmatrix}$,
        which is 1-Morita equivalent to trivial algebra $\one$. So ${}_{(\one\oplus\bff)\ot(\one\oplus\bff)}\TC_{(\one\oplus\bff)\ot (\one\oplus\bff)}\simeq \TC$ which coincides with $\Phi_{\bfe-\bfm}\circ \Phi_{\bfe-\bfm}\simeq \Phi_{\Id}$.

        \item choosing $B_1=\one\oplus \bfe=B_2$, we have ${}_{\one\oplus\bfe}\TC_{\one\oplus\bfe}=\vect_{\bZ_2}\btd \vect_{\bZ_2}$.
        Consider ${}_{\one\oplus\bfe}\TC_{\one\oplus\bfe} \btd_{\TC} {}_{\one\oplus\bfe}\TC_{\one\oplus\bfe}$, this should be equivalent to ${}_{(\one\oplus\bfe)\ot(\one\oplus\bfe)}\TC_{(\one\oplus\bfe)\ot (\one\oplus\bfe)}$.
        Note that $(\one\oplus\bfe)\ot(\one\oplus\bfe)$ as an algebra is the direct sum of two $\one\oplus \bfe$, so we have 
        ${}_{\one\oplus\bfe}\TC_{\one\oplus\bfe} \btd_{\TC} {}_{\one\oplus\bfe}\TC_{\one\oplus\bfe}\simeq {}_{(\one\oplus\bfe)\oplus(\one\oplus\bfe)}\TC_{(\one\oplus\bfe)\oplus (\one\oplus\bfe)}\simeq M_2(\vect_{\bZ_2})$.
    \end{itemize}
\end{expl}

More generally, we can consider fusing two 1d domain walls ${}_{B_1}(\CC_A)_{B_1}$ and ${}_{B_2}({}_A \CC)_{B_2}$, which should be a gapped domain wall in $\CC$. 
Namely ${}_{B_1}(\CC_A)_{B_1} \btd_{\CC_A^{loc}}{}_{B_2}({}_A \CC)_{B_2} \simeq {}_{B_{?}}\CC_{B_{?}}$, $B_?$ may not be indecomposable. Since $B_{?}$ only depends on $B_1, B_2$ and $A$,
we propose the following conjecture:

\begin{conj}
    ${}_{B_1}(\CC_A)_{B_1}\boxtimes_{\CC_A^{loc}} {}_{B_2}({}_A \CC)_{B_2}\simeq {}_{\Ext_A^R(B_1)\ot_A \Ext_A^L(B_2)}\CC_{\Ext_A^R(B_1)\ot_A \Ext_A^L(B_2)}$, where $\Ext_A^R:\Alg_{E_1}(\CC_A)\to \Alg_{E_1}(\CC)$ and $\Ext_{A}^L:\Alg_{E_1}({}_A \CC)\to \Alg_{E_1}(\CC)$.
\end{conj}

In particular, for $A=\one$, we recover Proposition \ref{B_fusion_over_C} since $\Ext^{L,R}_{\one}(B)=B$.
We depict the fusion of domain walls in the 
following figure:

\begin{figure}[H]
    \centering
    \begin{tikzpicture}
        \filldraw[draw=none, fill=gray!40] (1, 0)rectangle(7,2);
        \filldraw[draw=none, fill=gray!20] (5, 0)rectangle(3,2);
        \draw[very thick](5, 0)--(5,2);
        \draw[very thick](3, 0)--(3,2);

        \node[]at(6, 1){\small $\CC$};
        \node[]at(2, 1){\small $\CC$};

        \node[]at(5,-0.2){\small ${}_{B_2} ({}_A \CC)_{B_2}$};
        \node[]at(3,-0.2){\small ${}_{B_1} (\CC_A)_{B_1}$};
        \node[]at(4,1){\small $\CC_A^{loc}$};

        \node[rotate = 180] at (4, 2.2) {$\underbrace{\hspace{2cm}}$};
        \node[]at(4, 2.5){${}_{\Ext_A^R(B_1)\ot_A \Ext_A^L(B_2)}\CC_{\Ext_A^R(B_1)\ot_A \Ext_A^L(B_2)}$};
    \end{tikzpicture}
\end{figure}

\begin{expl}
    Consider $\CC:=\TC$ and $A=\one \oplus \bfe$. $\CC_A \simeq \vect_{\bZ_2}$, and the non-trivial 1d condensable algebra in $\CC_A$ is given by $\one \oplus M$.
    \begin{itemize}
        \item Choosing $B_1= \one=B_2$, we have $\Ext^R_{\one\oplus \bfe}(B_1)=\one\oplus \bfe$ and $\Ext^L_{\one\oplus\bfe}(\one)=\one\oplus\bfe$.
        Hence, $\Ext_A^R(B_1)\ot_A \Ext_A^L(B_2)= \one\oplus \bfe$, and the corresponding gapped domain wall is indeed $\vect_{\bZ_2}\btd \vect_{\bZ_2}$ (see Sec. \ref{section:toric_code} for 1d condensable algebras in $\TC$).

        \item Choosing $B_1= \one\oplus M$ and $B_2=\one$, we have $\Ext^R_{\one\oplus \bfe}(B_1)=(\one\oplus\bfm)\ot (\one\oplus \bfe)$ and $\Ext^L_{\one\oplus\bfe}(\one)=\one\oplus\bfe$.
        Hence, $\Ext_A^R(B_1)\ot_A \Ext_A^L(B_2)=(\one\oplus\bfm)\ot (\one\oplus \bfe)$ which corresponds to the gapped domain wall $\rep(\bZ_2)\btd \vect_{\bZ_2}$.

        \item Choosing $B_1= \one\oplus M=B_2$, we have $\Ext^R_{\one\oplus \bfe}(B_1)=(\one\oplus\bfm)\ot (\one\oplus \bfe)$ and $\Ext^L_{\one\oplus\bfe}(B_2)=(\one\oplus\bfe)\ot(\one\oplus\bfm)$.
        Hence, $\Ext_A^R(B_1)\ot_A \Ext_A^L(B_2)=(\one\oplus\bfm)\ot (\one\oplus \bfe)\ot (\one\oplus\bfm)\cong\begin{pmatrix}
            \one\oplus \bfm &\bfe\oplus \bff\\
            \bfe\oplus\bff &\one\oplus \bfm
        \end{pmatrix}\widesim[3]{1-Morita}{} \one\oplus \bfm$, in which the corresponding the gapped domain wall should be $\rep(\bZ_2)\btd \rep(\bZ_2)$.
    \end{itemize}
\end{expl}

There are also several promising directions emerge for future research.
For example: 
clarifying the relationships among different definitions of 2-Morita equivalence;
generalizing the Trinity framework (Figure \ref{fig:alg_cycle}) to include 0d defects;
or extending Witt equivalence to the algebraic level 
\cite{JMPP21,Dec22center} are all important and handy projects. 
Moreover, it is interesting to construct 1d condensable algebras and their centers in concrete models. 
We can also consider topological Wick "rotating" the spatial bulk phase to the temporal direction and describing these gapped domain walls under category symmetries
\cite{KWZ22,XZ22}. These could enhance our understanding of 2-Morita equivalences in physical systems. 

\appendix 
\appendixpage

\section{Condensable Algebras in MTCs}\label{appendix:condensable_algebras}
\subsection{Basic definition and results}
\begin{defn}
    Let $\CC$ be a MTC, an {\bf algebra} $A$ in $\CC$ is an object equipped with two morphisms $m:A\ot A\to A$ and $h: \one\to A$ satisfying
    \begin{align*}
        m\circ(m\ot\id_A)=m\circ(\id_A\ot m),\\
        m\circ(h\ot\id_A)=\id_A=m\circ(\id_A\ot h).
    \end{align*}
    An algebra $A$ is called 
    \begin{itemize}
        \item {\bf $E_2$} or {\bf commutative} if $m=m\circ \beta_{A,A}$.
        \item {\bf separable} if $m:A\ot A\to A$ splits as a $A$-$A$-bimodule homomorphism. 
        \item {\bf connected} if $\dim \hom_{\CC}(\one,A)=1$;
        \item  $E_2$-{\bf condensable} if $A$ is commutative connected separable;
        \item {\bf lagrangian} if $A$ is $E_2$-condensable and $\dim(A)^2=\dim(\CC)$.
    \end{itemize}
\end{defn}

\begin{defn}
    Let $A$ be an algebra in $\CC$.
    A {\bf right} $A$-{\bf module} in $\CC$ is a pair $(M,r_M)$, where $M$ is a object in $\CC$ and $r_M:M\ot A\to M$ is a morphism in $\CC$ such that 
    \begin{align*}
        r_M\circ(r_M\ot\id_A)=r_M\circ(\id_M\ot m),\\
        \id_A=r_M\circ(\id_M\ot h).
    \end{align*}
\end{defn}

\begin{thm}
    Let $A$ be an algebra in a monoidal category $\CC$.
    There is an adjunction $-\ot A\dashv U$,where $-\ot:A:\CC\to \CC_A$ sends any objects $x\in\CC$ to the free module $x\ot A$ and $U:\CC_A\to \CC$ is the forgetful functor.
    The adjunction can be written more explicitly
    \begin{align*}
        \hom_{\CC_A}(x\ot A,M)\cong \hom_{\CC}(x,U(M))
    \end{align*}
    for any $x\in \CC$ and $M\in\CC_A$.
\end{thm}


\begin{defn}
    A right $A$-module $M$ in $\CC$ is called a {\bf local} $A$-module if $r_M\circ \beta_{A,M}\circ \beta_{M,A}=r_M$.
\end{defn}
We denote the category of local $A$-modules in $\CC$ as $\CC_A^{loc}$. 
Mathematically, we can prove that 
\begin{thm}[\cite{BEK00,KO02}]
    Let $\CC$ be a MTC, $A$ be a condensable algebra in $\CC$.
    Then $\CC_A$ is a SFC and $\CC_A^{loc}$ is a MTC. 
\end{thm}

\begin{defn}
    Let $\CC$ be a braided monoidal category and $\CM$ be a monoidal category.
    Let $F:\CC\to \CM$ monoidal functor, a {\bf central functor structure} of $F$ is a braided monoidal functor $F':\CC\to \fZ(\CM)$ such that the following diagram commutes
    \begin{center}
        \begin{codi}[tetragonal]
            \obj{
                |(C)|\CC & |(Z)|\fZ(\CM)\\
                & |(M)|\CM\\
            };
            \mor[swap]: C F:-> M;
            \mor: * "F'":-> Z "U":-> *; 
        \end{codi}
    \end{center}
    where $U:\fZ(\CM)\to\CM$ is the forgetful functor.
\end{defn}

\begin{lem}[\cite{DMNO13}]
    Let $F: \CC \to \CM$ be a central functor, then $F^{R}(\one_{\CM})$ is a condensable algebra in $\CC$, and $\CC_{F^{R}(\one_{\CM})}$ is monoidal equivalent to the image of $F$.
\end{lem}

\subsection{Condensable algebras in \texorpdfstring{$\fZ(\vect_{G})$}{Z(vec(G))}}
In this subsection we briefly review the classification of $E_1$ and $E_2$ condensable algebras in $\fZ(\vect_{G})$.

An explicit description of the category $\FZ(\vect_{G})$ is given in \cite{BK01,Dav10a}. 
\begin{itemize}
    \item Its objects are pairs $(X, \rho_X)$, where $X$ is a $G$-graded vector spaces, i.e. $X=\oplus_{g\in G}X_g$, and $\rho_X: G\times X \rightarrow X$ is a compatible $G$-action, which means for $f, g\in G$, $(fg)(v)=f(g(v))$, $e(v)=v$ for all $v \in X$ and $f(X_g)=X_{fgf^{-1}}$.
    \item The tensor product of $(X, \rho_X)$ and $(Y, \rho_Y)$
    is just usual tensor product of $G$-graded
    vector spaces with the $G$-action $\rho_{X\ot Y}$ defined by $g(x\ot y)=g(x)\ot g(y)$ for $x\in X, y\in Y$.
    \item The tensor unit is $\bC$ which is viewed as a $G$-graded vector space supported only on the unit $e$ and equipped with a trivial $G$-action. 
    \item The braiding is given by
    \[\beta_{X,Y}(x\ot y) = f(y)\ot x, \qquad x\in X_f, y \in Y, f\in G.\]
    \item The dual object $X^{\vee} = \oplus_{g\in G}(X^{\vee})_g$ is given by
    \[(X^{\vee})_g = (X_{g^{-1}})^{\vee} = \text{hom}(X_{f^{-1}}, \bC)\]
    with action $g(l)(x)=l(g^{-1}(x))$ for $l \in \text{hom}(X_{f^{-1}}, \bC), x \in X_{gf^{-1}g^{-1}}$.
    \item The twist is given by $\theta_X(x)=f^{-1}(x)$ for $ x \in X_f$.
    \item The quantum dimension dim $X$ is just the usual vector space dimension.
\end{itemize}

\begin{thm}[\cite{Dav10a}]
    An $E_2$ condensable algebra $A = A(H, F,\omega, \epsilon ):=\fun(G)\otd_{\bC[H]}\bC[F,\omega,\epsilon]$ in $\FZ(\vect_{G})$ is determined by a subgroup $H \subset G$, a normal subgroup $F$ in $H$, a 2-cocycle $\omega \in Z^2(F,\bC^{\times})$ and $\epsilon : H \times F \rightarrow  \bC^{\times}$ satisfying the following conditions:
    \bnu
    \item (Action axiom)
    \begin{equation}{\label{eq:A1}}
        \epsilon_{gh}(f)=\epsilon_g(hfh^{-1})\epsilon_h(f), \qquad \forall g,h\in H, f\in F
    \end{equation}
    \item (Multiplicativity)
    \begin{equation}{\label{eq:A2}}
        \omega(f,g)\epsilon_h(fg)=\epsilon_h(f)\epsilon_h(g)\omega(hff^{-1}, hgh^{-1})\epsilon(f), \qquad \forall h \in H, f,g \in F
    \end{equation}
    \item (Commutativity)
    \begin{equation}{\label{eq:A3}}
        \omega(f,g)=\epsilon_f(g)\omega(fgf^{-1},f), \qquad f,g\in F.
    \end{equation}
    \enu
\end{thm}

\begin{itemize}
    \item The underlying vector space of algebra $A = A(H, F, \omega, \epsilon)$ is spanned by $\{\delta_g\ot_{\bC[H]} e_f\mid g \in G, f \in F\}$, where $\{\delta_g\mid g\in G\}$ is the standard basis of the regular algebra $\fun(G)$ and $\{e_f\mid f\in F\}$ is the standard basis of the group algebra $\bC[F,\omega]$.
    Equivalently, this basis can also be written as $\{\delta_g\ot e_f\mid g\in G,f\in F\}$ modulo the relations
    \begin{equation}{\label{eq:A4}}
        \delta_{gh}\ot e_f=\epsilon_h(f)\cdot \delta_g\ot e_{hfh^{-1}}, \qquad \forall h \in H,
    \end{equation}

    \item The $G$-grading on basis is $\delta_g\ot_{\bC[H]}e_f\in A_{gfg^{-1}}$ and $G$-action is $h(a_{hg,f})$. 
    
    \item The multiplication is given by 
    \begin{align*}
        (\delta_g\ot_{\bC[H]} e_f)\cdot (\delta_{g'}\ot_{\bC[H]} e_{f'})=\delta_{gg'}\omega(f,f')\cdot \delta_g\ot_{\bC[H]} e_{ff'}.
    \end{align*}
\end{itemize}

The algebra $A(H,F,\omega, \epsilon)$ is lagrangian if and only if $F=H$. 
In this case, $\epsilon$ is uniquely determined by $\omega$ by Eq.\ref{eq:A3}. 
Therefore, a lagrangian algebra is determined by a pair $(H,\omega)$.

\subsection{Extension of algebras}
\begin{lem}\label{lem:ext}
    A separable algebra $(B,m: B\otimes_A B \to B, h: A\to B)$ in $\CC_{A}$ can be extended to a separable algebra $\Ext^R_{A}(B):=(U(B),m^{ext},h^{ext})$ in $\CC$, where 
    \bit
        \item $U(B)$ is the image of $B$ under the forgetful functor $U:\CC_{A}\to \CC$,
        \item $m^{ext}: U(B)\ot U(B) \to U(B)\otimes_A U(B)\to U(B\ot_A B) \xrightarrow{U(m)} U(B)$, 
        \item and $h^{ext}: \one\to A\to U(A) \xrightarrow{U(h)} U(B)$.
    \eit
\end{lem}
For algebras in the category ${}_A \CC$ of left $A$-modules in $\CC$, we also denote the extension functor by $\Ext^L_A:\Alg_{E_1}({}_A \CC)\to \Alg_{E_1}(\CC)$.

By composing with the inclusion $i:\CC_A^{loc}\hookrightarrow \CC_A$, we have 
\begin{thm}
    A condensable algebra $(B,m: B\otimes_A B \to B, h: A\to B)$ in $\CC^{loc}_{A}$ can be extended to a condensable algebra $\Ext^R_{A}(B):=(U(B),m^{ext},h^{ext})$ in $\CC$.
    In particular, if $B$ is commutative in $\CC_A^{loc}$, then the extended algebra $U(B)$ is commutative in $\CC$.
\end{thm}

\section{Module Categories and Centers} \label{appendix:morita}
\subsection{Module categories}\label{appendix:module_cat}

\begin{defn}[\cite{KZ18}]
    Let $\CC_1$, $\CC_2$ be braided monoidal categories.
    \begin{itemize}
        \item A monoidal left $\CC_1$-module is a monoidal category $\CM$ equipped with a braided monoidal functor $F:\CC_1\to\fZ(\CM)$;
        \item A monoidal right $\CC_2$-module is a monoidal category $\CM$ equipped with a braided monoidal functor $\overline{\CC_2}\to\fZ(\CM)$;
        \item A monoidal $\CC_1$-$\CC_2$-bimodule is a monoidal category $\CM$ equipped with a braided monoidal functor $\CC_1\btd \overline{\CC_2}\to\fZ(\CM)$.
    \end{itemize}
\end{defn}

\begin{defn}
    Let $\CM$ be a left $\CC$-module.
    An \textbf{internal hom} in $\CM$ is a functor
    \begin{align}
        [-,-]:\CM^{op}\times\CM\to\CC
    \end{align} 
    such that for every object $x\in\CM$, we have a pair of adjoint functors
    \begin{align*}
        -\odot x\dashv [x,-]
    \end{align*}

\end{defn}

\begin{defn}
    Let $\CD$ be a braided fusion category and let $\CE$ be a monoidal right $\CD$-module with module action $\odot :\CE\times \CD\to \CE$.
    Consider an algebra $(A,m_A,h_A)$ in $\CD$.
    A {\bf right} $A$-module in $\CE$ is a pair $(M,r_M)$ where 
    \bit
        \item $M$ is an object in $\CE$;
        \item $r_M:M\odot A\to M$ is a morphism in $\CE$.
    \eit
    such that the following diagrams commute
    \begin{align*}
        \xymatrix{
             &(M \odot A)\odot A\ar[r]^{r_M\odot \id_A}\ar[dl] &M \odot A \ar[dd]^{r_M}\\
            M\odot (A\otimes_{\CD}A)\ar[d]_{\id_M\odot m_A} & & \\
            M\odot A\ar[rr]_{r_M} & &M
        }
        \quad \quad \quad 
        \xymatrix{
            M\odot \one_{\CD}\ar[r]^{\id_M\odot h_A}\ar[dr] & M\odot A\ar[d]^{r_M}\\
             &M
        }
    \end{align*}
\end{defn}

\subsection{Invertible monoidal bimodules}\label{appendix:Inv_monoidal_bimod}
\begin{prop}
    Let $\CC$ be a braided fusion category. Let $\mathrm{Mod}(\CC)$ denote the 2-category of finite semisimple $\CC$-modules.
    Then it admits a monoidal structure given by relative tensor product $\btd_{\CC}$.
\end{prop}

\begin{defn}[\cite{ENO10}]
    Let $\CC$ be a braided fusion category.
    Then the {\bf Picard group} $\mathrm{Pic}(\CC)$ is the group consists of all invertible objects in $\mathrm{Mod}(\CC)$ with respect to the relative tensor product $\btd_{\CC}$.
\end{defn}

Since finite semisimple $\CC$-module are characterized by $\CC_B$ for some separable algebra $B$ in $\CC$.
\begin{prop}[\cite{DN21}]\label{prp:fusing_rule_in_Mod(C)}
    There is an equivalence $\CC_B \btd_{\CC}\CC_{B'}\simeq \CC_{B\ot B'}$.
\end{prop}

For each finite semisimple $\CC$-module $\CM$, we have a finite semisimple monoidal $\CC$-$\CC$-bimodule $\Fun_{\CC}(\CM,\CM)$.
In particular $\CM\simeq \CC_B$, then we have $\Fun_{\CC}(\CC_B,\CC_B)\simeq {}_B \CC_B$ \cite{KZ18,DSPS19}.

\begin{prop}[Proposition \ref{B_fusion_over_C}]\label{prp:fusing_rule_of_gapped_domain_walls}
    There is an equivalence
    \begin{align*}
        {}_B \CC_B \btd_{\CC} {}_{B'} \CC_{B'}\simeq {}_{B\ot B'}\CC_{B\ot B'}
    \end{align*}
    as monoidal $\CC$-$\CC$-bimodule.
\end{prop}
\begin{proof}
    \[
    {}_B \CC_B \btd_{\CC} {}_{B'} \CC_{B'}\simeq \Fun_{\CC}(\CC_B,\CC_B)\btd_{\CC} \Fun_{\CC}(\CC_{B'},\CC_{B'})\simeq \Fun_{\CC}(\CC_B\btd_{\CC} \CC_{B'},\CC_B\btd_{\CC}\CC_{B'})\simeq \Fun_{\CC}(\CC_{B\ot B'},\CC_{B\ot B'})\simeq {}_{B\ot B'}\CC_{B\ot B'}
    \] 
\end{proof}

\begin{defn}\label{dfn:E_1_BrPic}
    Let $\CC$ be a braided fusion category.
    We define the $E_1$ {\bf Brauer-Picard group} $\mathrm{BrPic}_{E_1}(\CC)$ to be the set of all invertible $E_1$-monoidal $\CC$-$\CC$-bimodules with the multiplication given by Deligne's tensor product $\btd_{\CC}$ over $\CC$.
\end{defn}

\begin{thm}\label{thm:E_1_BrPic}
    There is an isomorphism $\mathrm{Pic}(\CC)\cong \mathrm{BrPic}_{E_1}(\CC)$ as groups.
\end{thm}
\begin{proof}
     We first prove that the map
    \begin{align*}
        f:\mathrm{Pic}(\CC)&\to \mathrm{BrPic}_{E_1}(\CC)\\
        \CC_B&\mapsto {}_B \CC_B
    \end{align*}
    is bijective.
    Indeed, $\CC_B\in \mathrm{Pic}(\CC)$ if and only if there exists $\CC_{B'}$ such that $\CC_B \btd_{\CC} \CC_{B'}\simeq \CC$ and $\CC_{B'}\btd_{\CC}\CC_B\simeq \CC$. By Proposition \ref{prp:fusing_rule_in_Mod(C)}, we have $\CC_{B\ot B'}\simeq \CC\simeq \CC_{B'\ot B}$, which implies $B\ot B' \widesim[3]{1-Morita}{} \one \widesim[3]{1-Morita}{} B'\ot B$.
    Hence, we have ${}_{B\ot B'}\CC_{B\ot B'}\simeq \CC \simeq {}_{B'\ot B}\CC_{B'\ot B}$.
    By Proposition \ref{prp:fusing_rule_of_gapped_domain_walls}, we have ${}_B \CC_B \btd_{\CC} {}_{B'} \CC_{B'}\simeq \CC \simeq {}_{B'} \CC_{'B} \btd_{\CC} {}_{B} \CC_{B}$, which is equivalent to say the monoidal $\CC$-$\CC$-bimodule ${}_B \CC_B \in \mathrm{BrPic}_{E_1}(\CC)$ is invertible.

    $f$ preserves group multiplication can be easily derived from Proposition \ref{prp:fusing_rule_in_Mod(C)} and Proposition \ref{prp:fusing_rule_of_gapped_domain_walls}.
\end{proof}

\subsection{Centers for algebras} \label{appendix:algebra_center}
Now we use a concept called unital action to define the left/right center \cite{KYZ21}.
\begin{defn}
    Let $\CM$ be a monoidal left $\CC$-module  with $\CC$-module action $\odot :\CC\times \CM\to \CM$, and let $A\in\Alg_{E_1}(\CC)$, $M\in\Alg_{E_1}(\CM)$.
    A {\bf unital} $A$-action is a morphism $f:A\odot M\to M$ such that the composition $M\simeq \one_{\CC}\odot M\to A\odot M\to M$ is identity $\id_M$.
\end{defn}

\begin{defn}[Center by Davydov]\label{defn:Dav}
    Let $\CC$ be an $E_2$-monoidal 1-category and let $B$ be an $E_1$-algebra in $\CC$.
    The {\bf Davydov's right center} $C_r(B)$ is an object in $\CC$ equipped with a morphism $\iota_l:C_r(B)\to B$, such that for any object $X\in \CC$ with a morphism $f: X\to B$ satisfying the following commutative diagram
    \begin{align}\label{diag:Dav_right_center}
        \xymatrix{
            B\ot X\ar[r]\ar[dd]_{\beta_{X,B}} &B\ot B\ar[dr]\\
             & &B\\
            X\ot B\ar[r]  &B\ot B\ar[ur]
        }
    \end{align}
    there is a morphism $g: X\to C_r(B)$ such that $f=\iota_l \circ g$.
\end{defn}

\begin{rem}

    For a fusion category ${}_{B}\CC_B$, the full center $Z(B) \in \FZ({}_{B}\CC_B)$ coincides with $L(B)$, where
 $L:{}_{B}\CC_B \to \FZ({}_{B}\CC_B)$ is the adjoint to the forgetful functor $F : \FZ({}_{B}\CC_B) \to {}_{B}\CC_B$.

    $$
    \begin{tikzcd}
        \FZ({}_{B}\CC_B) \ar[dr, "F"] \ar[d]  & \\
        \FZ({}_{B}\CC_B)_{Z(B)} \ar[r, "\sim"] & {}_{B}\CC_B
    \end{tikzcd}
    $$
\end{rem}

Compatible with bulk-to-wall map.

\section{(Higher) Morita Equivalence} \label{morita_equvi}
\subsection{1-Morita equivalence}

The original definition of Morita equivalence is to say that two $E_1$-algebras have equivalent module categories.
\begin{defn}\label{defn:1-Morita}
    Let $\CC$ be an $E_1$-monoidal $n$-category.
    Two $E_1$-algebras $A$ and $B$ in $\CC$ are {\bf Morita equivalent} if $\CC_A\simeq \CC_B$ as $n$-categories.
\end{defn}

Different 1-Morita equivalences can be unified by above definition, they differ by the choice of the $n$-category $\CC$.
For instance, when $\CC$ is a monoidal $1$-category, Definition \ref{defn:1-Morita} is the usual definition of 1-Morita equivalent algebras; when $\CC$ is $\Cat$, the 2-category of 1-categories, algebras in $\Cat$ are monoidal categories, then Definition \ref{defn:1-Morita} characterizes 1-Morita equivalence between monoidal categories.

\subsubsection{Ordinary 1-Morita equivalence in 1-categories}
In particular, let $\CC$ be the category $\Ab$ of Abelian groups.
It is clear that $\Ab$ is an $E_{\infty}$-monoidal 1-category (i.e. a symmetric monoidal 1-category).
And an $E_1$-algebra in $\Ab$ is a ring.
For two rings $R,S$, we can prove that 
\begin{thm}[Eilenberg-Watts]
    For $R, S\in\Alg_{E_1}(\Ab)$, the functor
    \begin{align*}
        {}_R\Ab_S&\to\Fun^{coc}_{\Ab}(\Ab_R,\Ab_S)\\
        M&\mapsto -\otimes_R M
    \end{align*}
    to the category of cocontinuous and additive functors is an equivalence of categories.
\end{thm}

\begin{cor}\label{crl:Morita_rings}
    Two rings $R,S$ are Morita equivalent if and only if there are $R$-$S$-bimodule $M$ and $S$-$R$-bimodule $N$ such that $ M\otd\limits_S N\cong R$ and $N\otd\limits_R M\cong S$.
\end{cor}
This is a famous result in the ordinary Morita theory, which concentrates on the category of modules over rings.

The Elienberg-Watts Theorem still holds in a more general setup.
\begin{thm}[Generalized Eilenberg-Watts Theorem]\label{thm:generalized_Eilenberg-Watts}
    Let $\CC$ be a cocomplete $E_1$-monoidal 1-category.
    For two $E_1$-algebras $A$, $B\in\Alg_{E_1}(\CC)$, there is an equivalence
    \begin{align*}
        {}_A\CC_B&\to\Fun^{coc}_{\CC}(\CC_A,\CC_B)\\
        M&\mapsto -\ot_A M
    \end{align*}
    of categories.
    In particular, the equivalence
    \begin{align*}
        {}_A\CC_A\to\Fun_{\CC}(\CC_A,\CC_A)
    \end{align*}
    is a monoidal equivalence.
\end{thm}
Above theorem assures that Corollary \ref{crl:Morita_rings} can be generalized to $E_1$-algebras in a more general category $\CC$.
That is to say, two $E_1$-algebras are 1-Morita equivalent if and only if there exists an invertible bimodule between them.
We believe this 1-categorical theorem can also be promoted to n-categories.

Indeed, above characterization will be more natural if we consider the Morita category $\Mrt_{E_1}(\CC)$.
\begin{defn}
    Let $\CC$ be a cocomplete $E_1$-monoidal 1-category.
    The bicategory $\Mrt_{E_1}(\CC)$ consists of 
    \begin{itemize}
        \item objects are $E_1$-algebras in $\CC$;
        \item for two $E_1$-algebras $A$ and $B$, 1-morphisms between them are $A$-$B$-bimodules;
        \item for two bimodules $M$, $N\in\Mor(A,B)$, 2-morphisms between them are $A$-$B$-bimodule homomorphisms.
        \item the composition of 2-morphisms are composition of bimodule homomorphisms.
        \item the composition of 1-morphisms are the relative tensor product: for $A$-$B$-bimodule $M$ and $B$-$C$-bimodule $N$, their composition is the $A$-$C$-bimodule $M\otimes_B N$.
    \end{itemize}
\end{defn}

In bicategory $\Mrt_{E_1}(\CC)$, an invertible bimodule is indeed an invertible morphism. Hence, two $E_1$ algebras are 1-Morita equivalent is the same as there is an equivalence between them in $\Mrt_{E_1}(\CC)$.
\begin{cor}
    Two rings are 1-Morita equivalent if and only if they are equivalent objects in the bicategory $\Mrt_{E_1}(\Ab)$.   
\end{cor}
Thus, we can also use the equivalences in the bicategory $\Mrt_{E_1}(\Ab)$ as the definition of the ordinary Morita theory.

Theorem \ref{thm:generalized_Eilenberg-Watts} also leads to another characterizations of 1-Morita equivalence.
First we notice that,
\begin{prop}
    Let $A_1$ and $A_2$ be two $E_1$ algebras in $\CC$.
    Then $A_1\widesim[3]{1-Morita}{} A_2$ implies ${}_{A_1}\CC_{A_1}$ is $E_1$-monoidal equivalent to $ {}_{A_2}\CC_{A_2}$.
\end{prop}
\begin{proof}
    By Definition \ref{defn:1-Morita}, we have $\CC_{A_1}\simeq \CC_{A_2}$.
    By Theorem \ref{thm:generalized_Eilenberg-Watts}, we have ${}_{A_1}\CC_{A_1}\simeq \Fun_{\CC}(\CC_{A_1},\CC_{A_1})\simeq \Fun_{\CC}(\CC_{A_2},\CC_{A_2})\simeq {}_{A_2}\CC_{A_2}$.
\end{proof}

Since a (local) $E_1$-module over an $E_1$-algebra is an $A$-$A$-bimodule, i.e., $\Mod_{A}^{E_1}(\CC)\simeq {}_A\CC_A$, so we have 
\begin{cor}
    Let $A_1$ and $A_2$ be two $E_1$-algebras in $\CC$.
    Then $A_1\widesim[3]{1-Morita}{} A_2$ implies $\Mod_{A_1}^{E_1}(\CC)\widesimeq[2]{E_1}\Mod_{A_2}^{E_1}(\CC)$.
\end{cor}
The converse might be true, that is, 
\begin{conj}
    for two $E_1$-algebras $A_1$ and $A_2$, if we have $\Mod_{A_1}^{E_1}(\CC)$ is $E_1$-monoidal equivalent to $\Mod_{A_2}^{E_1}(\CC)$, then $A_1$ is $E_1$-Morita equivalent to $A_2$.
\end{conj}

If above conjecture holds, we can use the following definition which everything is $E_1$ to equivalently define 1-Morita equivalence.
\begin{defn}
    Two $E_1$-algebras $A_1$ and $A_2$ are {\bf 1-Morita equivalent} if their $E_1$-module categories $\Mod_{A_1}^{E_1}(\CC)$ and $\Mod_{A_2}^{E_1}(\CC)$ are $E_1$-monoidal equivalent.
\end{defn}

\subsubsection{1-Morita equivalence in modular fusion categories}
It is well known that two rings are 1-Morita equivalent implies their centers are isomorphic to each other.
In general, Davydov shows that the full center is a Morita invariant in any monoidal 1-category \cite{Dav10}.
\begin{thm}\label{thm:Morita_inv}
    Let $\CC$ be a $E_1$-monoidal 1-category. 
    Two $E_1$-algebras $A_1$ and $A_2$ are 1-Morita equivalent implies their full center $Z(A_1)$ and $Z(A_2)$ are isomorphic in the Drinfeld center $\fZ(\CC)$ of $\CC$. 
\end{thm} 

As a consequence of above theorem, if we consider $E_2$-algebras in $\vect$, we have the following corollary:
\begin{cor}
    Two $E_2$-algebras are 1-Morita equivalent if and only if they are isomorphic.
\end{cor}

The converse of Theorem \ref{thm:Morita_inv} need not be true in general. For example, consider real numbers and quaternions in $\vect_{\Rb}$.
However, if we consider $E_1$-algebras in a MTC, it was proved that two simple algebras with non-degenerate trace pairing are 1- Morita equivalent if and only if their full centers are isomorphic as algebras \cite{KR08}.
More precisely,
\begin{thm}
    Let $\CC$ be a modular fusion category and let $A, B$ be simple normalized special Frobenius algebras in $\CC$. 
    Then the following two statements are equivalent.
    \begin{itemize}
        \item $A$ and $B$ are 1-Morita equivalent.
        \item The full centers of $Z(A)$ and $Z(B)$ are isomorphic as algebras.
    \end{itemize}
\end{thm}

\begin{rem}
    For a separable indecomposable algebra $A$ in a modular category $\CC$ the full center $Z(A)$ is a Lagrangian algebra in $\CC \boxtimes \overline{\CC}$. Moreover, the full center construction establishes an isomorphism between the set of Morita equivalence classes of separable indecomposable algebras in $\CC$ and isomorphism classes of Lagrangian algebras in $\CC \boxtimes \overline{\CC}$.
\end{rem}

\subsubsection{1-Morita equivalence in 2-category $\Cat^{Fin}_{\bk}$}
Let $\Cat^{Fin}_{\bk}$ denote the 2-category of finite $\bk$-linear categories.
\begin{defn}
    Let $\CC$ and $\CD$ be two $E_1$-algebras in $\Cat^{Fin}_{\bk}$, and $\CM$ a $\CC$-$\CD$-bimodule in $\Cat^{Fin}_{\bk}$.
    We say $\CC$ and $\CD$ is {\bf Morita equivalent} if $\CM$ is invertible, i.e. there exists $\CD$-$\CC$-bimodule $\CN$, such that $\CM\boxtimes_{\CD}\CN\simeq \CC$ and $\CN\boxtimes_{\CC}\CM\simeq \CD$.
\end{defn} 

There are equivalent characterizations of Morita equivalence between multi-fusion 1-categories. 
Let $\CC$ and $\CD$ be two multi-fusion 1-categories over an algebraically closed field of characteristic zero. 
Categorifying the classical notion of Morita equivalence for algebras, we say that $\CC$ and $\CD$ are Morita equivalent if the (linear) 2-categories $\RMod(\CC)$ and $\RMod(\CD)$ are equivalent ($\RMod(\CC)$ for the 2-category of finite semisimple right $\CC$-module 1-categories). 

Alternatively, given $\CM$ a finite semisimple left $\CC$-module 1-category, we can consider $\Endo_{\CC}(\CM)$, the multi-fusion 1-category of left $\CC$-module endofunctor of $\CM$. 
Following \cite{EO04}, we use $\CC^{*}_{\CM}$ to denote $\Endo_{\CC}(\CM)$, and call it the dual tensor 1-category to $\CC$ with respect to $\CM$. 
Then, we say that $\CC$ and $\CD$ are Morita equivalent if there exists a faithful finite semisimple left $\CC$-module 1-category $\CM$ together with a monoidal equivalence between $\CC^{*}_{\CM}$ and $\CD^{mop}$, that is $\CD$ equipped with the opposite monoidal structure. 
It follows from \cite{ENO10} that this coincides with the notion of Morita equivalence recalled above.
Moreover, it follows from \cite{Ost03a} that there exists an algebra $A$ in $\CC$ such that $\CM$ is equivalent to $\CC_A$, the 1-category of right $A$-modules in $\CC$. 
This implies that there is a monoidal equivalence between $\Endo_{\CC}(\CM)$ and $\BMod_{A}(\CC)^{mop}$, the monoidal 1-category of $A$-$A$-bimodules in $\CC$. 

Let us also note that, by \cite{ENO05}, the algebra $A$ is necessarily separable, i.e. $A$ is a special Frobenius algebra. 
It then follows that $\CC$ and $\CD$ are Morita equivalent if and only if there exists a faithful separable algebra $A$ in $\CC$ together with an equivalence $\CD \simeq  \BMod_{A}(\CC)$ of monoidal 1-categories. This recovers the notion of Morita equivalence introduced in \cite{FRS02a} and \cite{Mueg03}. 

In \cite{KZ18} the authors also proved that 
\begin{prop}
    Let $\CC$, $\CD$ be finite monoidal categories and $\CM$ an invertible $\CC$-$\CD$-bimodule. 
    Then $\fZ(\CC) \simeq \fZ(\CD)$ as braided monoidal categories.
\end{prop}

Since fusion categories are finite monoidal categories, we can restrict the above results to fusion categories.
Hence, two fusion categories are Morita equivalence is equivalent to say that their Drinfeld centers are braided equivalent, or they share the same bulk.

\subsection{2-Morita equivalence}
We define 2-Morita equivalence iteratively.
\begin{defn}
    Let $\CC$ be an $E_2$-monoidal $n$-category.
    Two $E_2$-algebras $A,B$ are {\bf 2-Morita equivalent} if $\CC_A$ and $\CC_B$ are 1-Morita equivalent, i.e. $\RMod_{\CC_A}(n\Cat)\simeq \RMod_{\CC_B}(n\Cat)$ as $(n+1)$-categories. 
\end{defn}

\begin{expl}
    Consider the $E_{\infty}$-monoidal 1-category $\Ab$.
    An $E_2$-algebra in $\Ab$ is a commutative ring.
    For two commutative rings $R$ and $S$, they are $E_2$-Morita equivalent if $\Ab_R$ and $\Ab_S$ are $E_1$-Morita equivalent.
    Since $E_1$-Morita equivalence implies isomorphic center, we have $\fZ_1(\Ab_R)\simeq \fZ_1(\Ab_S)$ as $E_2$-monoidal categories.
    However, since $\fZ_1(\Ab_R)\simeq \Ab_R$ for any commutative ring, we have $\Ab_R\simeq \Ab_S$, which implies $R$ and $S$ are $E_1$-Morita equivalent.
    Hence, again, we must have $Z(R)\cong Z(S)$.
    But $R$ and $S$ are both commutative, we have $R\cong S$.
    As a consequence, the higher Morita equivalence in classical algebras is trivial, which is equivalent to algebra isomorphisms.
\end{expl}

\begin{prop}
    For a MTC $\CC$, two $E_2$-algebra $A_1$ and $A_2$ are $E_2$-Morita equivalent if and only if their local modules categories are $E_2$-monoidal equivalent.
\end{prop}
\begin{proof}
    By definition, we have $\CC_{A_1}$ and $\CC_{A_2}$ are $E_1$-Morita equivalent, which means their centers are equivalent, i.e. $\fZ(\CC_{A_1})\simeq \fZ(\CC_{A_2})$.
    Since $\fZ(\CC_A)\simeq \CC\boxtimes \overline{\CC_{A}^{loc}}$, we have $\CC\boxtimes \overline{\CC_{A_1}^{loc}}\simeq \CC\boxtimes \overline{\CC_{A_2}^{loc}}$.
    By Muger's Prime Decomposition Theorem, we have 
    $\CC_{A_1}^{loc}\simeq \CC_{A_2}^{loc}$.
\end{proof}

Since $E_2$-modules are local modules, i.e. $\Mod^{E_2}_A(\CC)\simeq \CC_{A}^{loc}$.
Thus, we have 
\begin{cor}
    For a MTC $\CC$, two $E_2$-algebra $A_1$ and $A_2$ are $E_2$-Morita equivalent if their $E_2$ module categories $\Mod^{E_2}_{A_1}(\CC)$ and $\Mod^{E_2}_{A_2}(\CC)$ are $E_2$-monoidal equivalent.
\end{cor}

For MTCs, we can use $E_2$-monoidal equivalence between $E_2$-module categories to define $E_2$-Morita equivalence.
\begin{conj}
    This definition can be promoted to braided fusion categories, NOT need non-degeneracy.
\end{conj}
If above conjectures holds, we can use the following definition in which everything is $E_2$ to define 2-Morita equivalence. 
\begin{defn}
    Two $E_2$-algebras $A_1$ and $A_2$ are $E_2$-{\bf Morita equivalent} if their $E_2$-module categories $\Mod_{A_1}^{E_2}(\CC)$ and $\Mod_{A_2}^{E_2}(\CC)$ are $E_2$-monoidal equivalent.
\end{defn}

For 2-Morita equivalences in 2-categories, for instance, 2-Morita equivalence of nondegenerate braided fusion categories is just the Witt equivalence \cite{Dec22center}.

\newpage
\bibliography{Top}

\end{document}